\pdfoutput=1

\documentclass{amsart}

\usepackage{mathtools,bm,comment,amssymb,enumitem,hyperref,cite,xcolor,array,float,subcaption,float,pgfplots}
\usepackage[capitalise,nameinlink]{cleveref}
\usepackage[greek.polutoniko,english]{babel}
\usepackage[margin=1in]{geometry}
\usepackage[foot]{amsaddr}
\usepackage[linesnumbered,ruled]{algorithm2e}
\usepackage[noend]{algpseudocode}
\usepackage[locale=DE]{siunitx}
\usepackage{textcomp}
\usepackage{placeins}
\usetikzlibrary{shapes,arrows}
\usepackage{graphicx}

\usepackage[utf8x]{inputenc} 

\numberwithin{equation}{section}
\crefname{subsection}{Subsection}{Subsections}
\crefformat{equation}{(#2#1#3)}
\crefformat{enumi}{(#2#1#3)}
\crefname{figure}{Figure}{Figures}

\newcommand{\lambdap}{\lambda^{\!\textup{pro}}}
\newcommand{\lambdad}{\lambda^{\!\textup{deg}}}
\newcommand{\old}{{\text{old}}}

\newcommand{\eps}{\varepsilon}

\newcommand{\ECM}{{E\hspace{-.05em}C\hspace{-.15em}M}}
\newcommand{\WSS}{{\text{WSS}}}
\newcommand{\MDE}{{M\hspace{-.15em}D\hspace{-.1em}E}}
\newcommand{\TAF}{{T\hspace{-.15em}A\hspace{-.1em}F}}
\newcommand{\ecm}{\phi_{\ECM}}
\newcommand{\taf}{\phi_{\TAF}}
\newcommand{\mde}{\phi_{\MDE}}
\newcommand{\mean}{\textup{mean}}
\newcommand{\new}{\textup{new}}

\newcommand{\dd}{\textup{d}}

\renewcommand{\t}{\tilde}
\newcommand{\p}{\partial}
\newcommand{\pt}{\p_t}
\newcommand{\ov}{\overline}

\renewcommand{\div}{\textup{div}}

\newcommand{\A}{\mathcal{A}}

\renewcommand{\H}{\mathcal{H}}
\newcommand{\R}{\mathbb{R}}

\newcommand{\N}{\mathbb{N}}

\newcommand{\RD}{{\mathcal{R}\mathcal{D}}}
\newcommand{\phib}{\bm{\phi}}

\newtheorem{remark}{Remark}

\DeclareMathOperator{\sign}{sign}

\newcommand{\bbR}{\mathbb{R}}

\newcommand{\bphi}{{\bm{\phi}}}

\newcommand{\bn}{\bm{n}}

\newcommand{\bv}{\bm{v}}

\newcommand{\bx}{\bm{x}}
\newcommand{\by}{\bm{y}}

\newcommand{\phit}{\tilde{\phi}}
\newcommand{\mut}{\tilde{\mu}}

\newcommand{\CV}{{CV}}
\newcommand{\knew}{{{k+1}}}
\newcommand{\knewO}{{{k+1}}}

\newcommand{\surfInt}[1]{\int_{\sigma} {#1} \cdot \bn\, \dd S}
\newcommand{\surfIntSum}[1]{\sum_{\sigma \in \partial \CV}\int_{\sigma} {#1} \cdot \bn \, \dd S}
\newcommand{\elemInt}[1]{\int_{\CV} {#1} \, \dd V}
\newcommand{\weakDot}[2]{\left({#1}, {#2}\right)}
\newcommand{\plusOp}[1]{\left(#1\right)^{+}}

\makeatletter
\@namedef{subjclassname@2020}{\textup{2020} Mathematics Subject Classification}
\makeatother

\makeatletter
\renewcommand{\email}[2][]{%
	\ifx\emails\@empty\relax\else{\g@addto@macro\emails{,\space}}\fi%
	\@ifnotempty{#1}{\g@addto@macro\emails{\textrm{(#1)}\space}}%
	\g@addto@macro\emails{#2}%
}
\makeatother

\makeatletter
\pgfmathdeclarefunction{erf}{1}{%
	\begingroup
	\pgfmathparse{#1 > 0 ? 1 : -1}%
	\edef\sign{\pgfmathresult}%
	\pgfmathparse{abs(#1)}%
	\edef\x{\pgfmathresult}%
	\pgfmathparse{1/(1+0.3275911*\x)}%
	\edef\t{\pgfmathresult}%
	\pgfmathparse{%
		1 - (((((1.061405429*\t -1.453152027)*\t) + 1.421413741)*\t 
		-0.284496736)*\t + 0.254829592)*\t*exp(-(\x*\x))}%
	\edef\y{\pgfmathresult}%
	\pgfmathparse{(\sign)*\y}%
	\pgfmath@smuggleone\pgfmathresult%
	\endgroup
}
\makeatother

\let\originalleft\left
\let\originalright\right
\renewcommand{\left}{\mathopen{}\mathclose\bgroup\originalleft}
\renewcommand{\right}{\aftergroup\egroup\originalright}

\title[Modeling and simulation of vascular tumors]{Modeling and simulation of vascular tumors \\ embedded in evolving capillary networks}

\author[M. Fritz, P. K. Jha, T. K\"oppl, J. T. Oden, A. Wagner, and B. Wohlmuth]{Marvin Fritz$^1$, Prashant K. Jha$^{2}$, Tobias K\"oppl$^{1,*}$,  J. Tinsley Oden$^2$, \\ Andreas Wagner$^1$, and Barbara Wohlmuth$^{1,3}$}
\subjclass[2020]{65M08, 65M60, 76S05, 76Z05, 92C17, 92C42.}
\keywords{tumor growth, 3D-1D coupled blood flow models, angiogenesis, finite elements, finite volume}
\thanks{${}^*$Corresponding author}
\email{\{fritzm, koepplto, wagneran, wohlmuth\}@ma.tum.de}
\email{pjha@utexas.edu}
\email{oden@oden.utexas.edu}

\begin{document}

\maketitle
\vspace*{-2mm}
\begin{center} \footnotesize
	$^1$Department of Mathematics, Technical University of Munich, Germany \\
	$^2$Oden Institute for Computational Engineering and Sciences, The University of Texas at Austin, USA \\
	$^3$Department of Mathematics, University of Bergen, Allegaten 41, 5020 Bergen, Norway
\end{center}

\vspace{6mm}
\begin{abstract}
 In this work, we present a coupled 3D-1D model of {solid} tumor growth within a dynamically changing vascular network to facilitate realistic simulations of angiogenesis. Additionally, the model includes erosion of the extracellular matrix, interstitial flow, and coupled flow in {blood} vessels and tissue. We employ continuum mixture theory with stochastic Cahn–Hilliard type phase-field models of tumor growth. The interstitial flow is governed by a mesoscale version of Darcy’s law. The flow in the blood vessels is controlled by Poiseuille flow, and Starling’s law is applied to model the mass transfer in and out of blood vessels. The evolution of the network of blood vessels is orchestrated by the concentration of the tumor angiogenesis factors {(TAFs); blood vessels grow towards the increasing TAFs concentrations}. {This} process is not deterministic, allowing random growth of blood vessels and, therefore, due to the coupling of nutrients in tissue and vessels, {makes the growth of tumors }stochastic. We demonstrate the performance of the model by applying it to a variety of scenarios. Numerical experiments illustrate the flexibility of the model and its ability to generate satellite tumors. Simulations of the effects of angiogenesis on tumor growth are presented as well as sample-independent features of cancer.
\end{abstract}

\vspace{0.2cm}
{\bf{Keywords: }} tumor growth, 3D-1D coupled blood flow models, angiogenesis, finite elements, finite volume
\vspace{0.1cm}

\section{Introduction}
In this work, we present new computational models and algorithms for simulating and predicting a broad range of biological and physical phenomena related to cancer at the tissue scale. We consider the growth of {solid} vascular tumors inside living tissue containing a dynamically evolving vasculature. One of the main goals of this work is to provide realistic simulations of the vascular growth characterizing angiogenesis, whereby blood vessels sprout and invade the domain of the {solid} tumor when prompted by concentrations of proteins collectively referred to as tumor angiogenesis factors (TAFs); these proteins are produced by nutrients-starved cancerous cells. The tumor growth is necessarily depicted by a multispecies model in which tumor cell concentrations are categorized as proliferative, hypoxic, or necrotic. To capture the complex interaction of cell species and the evolving interfaces between species, continuum mixture theory is used as a framework for constructing mesoscale stochastic phase-field models of the Cahn--Hilliard type. 
Other critical phenomena are also addressed by this class of models, including the erosion of the extracellular matrix (ECM) due to concentrations of matrix-{degrading} enzymes (MDEs{, such as matrix metalloproteeinase and urokinase plasminogen activators}) {that erode the ECM and permit the invasion of tumor cells as a prelude to metastasis \cite{gerisch2008mathematical, nargis2016effects}}.

The volume of an isolated colony of tumor cells will not generally grow beyond approximately $1.0$ \si{mm}$^3$  \cite{holmgren1995dormancy, parangi1996antiangiogenic, nishida2006angiogenesis} unless sufficient nutrients and oxygen are supplied for proliferation. To acquire such nutrients, cancerous cells promote angiogenesis \cite{carmeliet2011molecular, patsch2015generation}. Low levels of oxygen and nutrient result in tumor cells entering the hypoxia phase during which they remain dormant and release various proteins such as TAFs that promote the proliferation {of} endothelial cells and new vessel formation. {Similarly, low oxygen levels can generate irregular invasive tumors governed by haptotaxis \cite{gerisch2008mathematical, nargis2016effects}.}
Because angiogenesis is one of the major processes through which tumors grow, anti-angiogenic drugs that inhibit the formation of the new vascular structure are often identified as one of the approaches to delay or arrest the growth of cancer. Thus, a realistic model of angiogenesis is of critical importance for studying the effectiveness of anti-angiogenic drugs. 

Typically, the vasculature near the tumor core in the early stages of tumor growth may not effectively supply nutrients to the tumor. The vasculature evolves rapidly and, therefore, the vessel walls are not fully developed and may be destroyed due to pressure (proliferation of tumor cells result in higher pressure nearby), the pruning of vessels due to insufficient flow for a sustained period and, vasculature adaptation and remodeling \cite{owen2009angiogenesis, pries1998structural, pries2001structural, pries2001structural2, stephanou2005mathematical}. Highly interconnected and irregular vasculature with inefficient blood vessels causes low blood flow rates to the tumor, making it possible that therapeutic drugs miss the tumor mass altogether \cite{stephanou2005mathematical}. Shear and circumferential stresses due to blood flow result in vascular adaptation effects such as vessel radii adaptation, see \cite{stephanou2006mathematical, mcdougall2006mathematical, pries2001structural2}. All of these phenomena are represented by the models described herein. 

Earlier models taking into account angiogenesis include lattice-probabilistic network models, see \cite{anderson1998continuous, zheng2005nonlinear, mcdougall2002mathematical, stephanou2005mathematical, stephanou2006mathematical, mcdougall2006mathematical, owen2009angiogenesis}. An overview of this class of models is given in \cite{dorraki2020angiogenic}. Another class of models referred to as agent-based models has been proposed and extensively studied. There, the idea is to introduce a phase-field for the tip endothelial cells that takes a value $1$ inside the vessel and $0$ outside and through the agents, which can move anywhere in the simulation domain following certain rules, which can be designed to trigger the sprouting of new vessels; see \cite{lima2014hybrid, travasso2011tumor, vilanova2018computational, phillips2020hybrid}. These models do not capture blood circulation in the vessel and, therefore, are unable to be truly coupled to the tumor growth. In \cite{wu2020patient}, a dimensionally coupled model for drug delivery based on MRI data and a study of dosing protocols is considered with drug flow in the vessels governed by algebraic rules instead of PDEs. More recently, vasculature models involving a network of straight cylindrical vessels supporting the 1D flow of nutrient, oxygen, and therapeutic drugs and coupled to the 3D tissue domain by versions of the Starling or Kedem--Katchalsky law have been presented; see \cite{xu2016mathematical, xu2017full,  koppl20203d}.

We consider a class of 3D-1D vascular tumor models \cite{fritz2020analysis} that approximates the flow within the blood vessels by one-dimensional flow based on the Poiseuille law effectively reducing the flow in the three-dimensional vessels to the flow in a network of one-dimensional vessel segments. While coupling the flow in the vessels and tissue, the blood vessels' three-dimensional nature is retained by approximating the vessels as a network of straight-cylinders and applying the fluid exchange at the walls of cylindrical segments. From a mathematical and computational point of view, a complicating factor is the use of one-dimensional characterizations of vessel segments in the vascular network embedded in three-dimensional domains of the tissue and the tumor within the tissue and the assignment of mechanical models to this 3D-1D system to depict interstitial flow and pressure fluctuations. Mathematical analysis showing well-posedness and existence of weak solutions for the class of 3D-1D model considered in this work is performed in a recent paper \cite{fritz2020analysis}. 

In our model, flow in vessels is governed by 1D Poiseuille law, whereas the flow in tissue is derived by treating the tissue domain as a porous medium and applying a version of Darcy’s law. The model consists of nutrients in the tissue and vessels; nutrients in the vessels are governed by the 1D advection-diffusion equation and advection-diffusion-reaction equation in the tissue. Flow and nutrients in the tissue and vessel are coupled; we assume that vessel walls are porous, resulting in the advection and diffusion-driven exchange of nutrients and coupling of the extravascular and intravascular pressures. Some aspects of the 1D model architecture and coupling of 3D and 1D models are based on previous works, see \cite{hodneland2020well, koppl20203d, koch2020modeling}. The 3D tissue domain includes, in addition to the nutrients, ECM, tumor species such as proliferative, hypoxic, necrotic, and diffusive molecules such as TAF and MDE. 

As noted earlier, the 3D tumor model is derived from the balance laws of continuum mixture theory as in \cite{byrne2003modelling, cristini2009nonlinear, oden2016toward, lima2014hybrid, garcke2018multiphase, cristini2010multiscale}, and representations of the principal mechanisms governing the development and evolution of cancer, see, e.g., \cite{lima2014hybrid, hanahan2011hallmarks}.
Especially, we note the comprehensive developments of diffuse-interface multispecies models presented in \cite{wise2008three, frieboes2010three}, the ten species models derived in \cite{lima2014hybrid}, and the multispecies nonlocal models of adhesion and promotes a tumor invasion due to ECM degradation described in \cite{fritz2019local}. Angiogenesis models embedded in models of hypoxic and cell growth are presented in \cite{lima2014hybrid, xu2016mathematical, xu2017full, fritz2020analysis}. Related models of extracellular matrix (ECM) degradation due to matrix-{degrading} enzymes (MDEs) and subsequent tumor invasion and metastasis are discussed in \cite{fritz2019local, chaplain2005mathematical, chaplain2011mathematical, engwer2017structured}. Several of the earlier mechanistic models of tumor growth focused on modeling the effects of mechanical deformation and blood flow, and fluid pressure on tumor growth, e.g., \cite{ambrosi2005review, ambrosi2009cell, preziosi2003cancer, bellomo2008foundations, bellomo2000modelling, koumoutsakos2013fluid}. 

A key new feature of the models proposed here is the dynamic growth/deletion of the vascular network and full coupling between the dynamic network and tumor system in the tissue microenvironment. In response to TAF generated by nutrient-starved hypoxic cells, new vessels are formed. Due to the formation of new vessels, the local conditions such as nutrient concentration changes near the tumor, affecting TAF production and promoting a higher proliferation of tumor cells. The rules by which the network grows, or existing vessels are deleted due to insufficient flow and dormancy, are based on the experimentally known causes of angiogenesis and are parameterized so that various aspects of the network growth algorithm can be adjusted based on available experimental data. By including the time-evolution of the larger vascular tissue domain and the sprouting, growth, bifurcation, and pruning of the vascular network orchestrated by a combination of blood supply and tumor-generated growth factors, a more realistic depiction of tumor growth than the more common isolated-tumor (avascular) models is obtained. 

This article is organized as follows: In \cref{Sec:Derivation}, we introduce various components of the model, such as the tissue and 1D network domain and the equations governing various fields. The details associated with the vessel network growth are presented in \cref{Sec:NetworkGrowth}. Spatial and temporal discretization and solver schemes for the highly nonlinear coupled system of equations are discussed in \cref{Sec:Solver}. We apply mixed finite volumes, finite element approximations to the model equations. The systems of equations arising in each time step are solved using a semi-implicit fixed point iteration scheme. In \cref{Sec:Simulation}, the model is applied to various situations, and several simulation experiments are presented.  {For further details on our implementation of the solver, we refer to the open-source code at \url{https://github.com/CancerModeling/Angiogenesis3D1D}. } Concluding comments are given in \cref{Sec:Conclusion}.

\section{Mathematical Modeling} 
\label{Sec:Derivation}

\begin{figure}[!htb] 
	\centering
	\includegraphics[width=.8\textwidth]{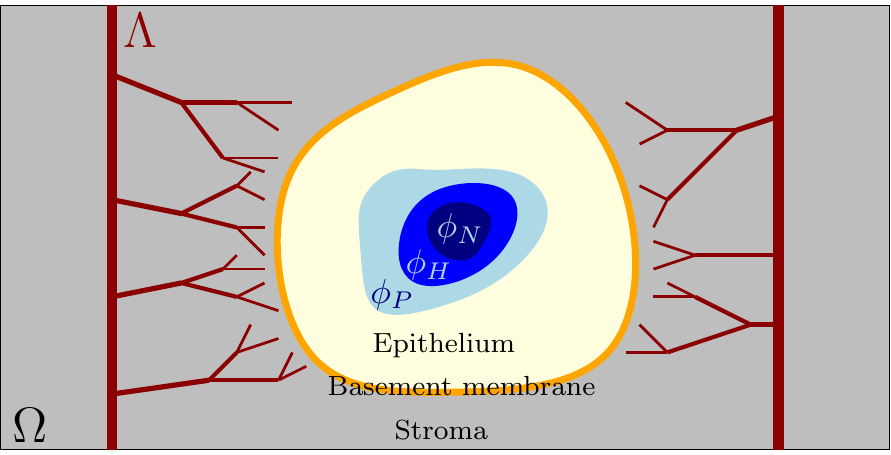}
	\caption{Setup of the domain $\Omega$ with the 1D microvascular network $\Lambda$ and a tumor mass, which is composed of its proliferative ($\phi_P$), hypoxic ($\phi_H$) and necrotic ($\phi_N$) phases. }
	\label{Figure:Sketch}
\end{figure}

In this work, a colony of tumor cells in an open bounded domain $\Omega \subset \R^3$, e.g., representing an organ, is considered. It is supported by a system of macromolecules  consisting of collagen, enzymes, and various proteins, that constitute the extracellular matrix. We focus on phenomenological characterizations to capture mesoscale and macroscale events.  Additionally, we consider a one-dimensional graph-like structure $\Lambda$ inside of $\Omega$ forming a microvascular network, see \cref{Figure:Sketch}.

\begin{figure}[!htb]
	\centering
	\includegraphics[width=.48\textwidth]{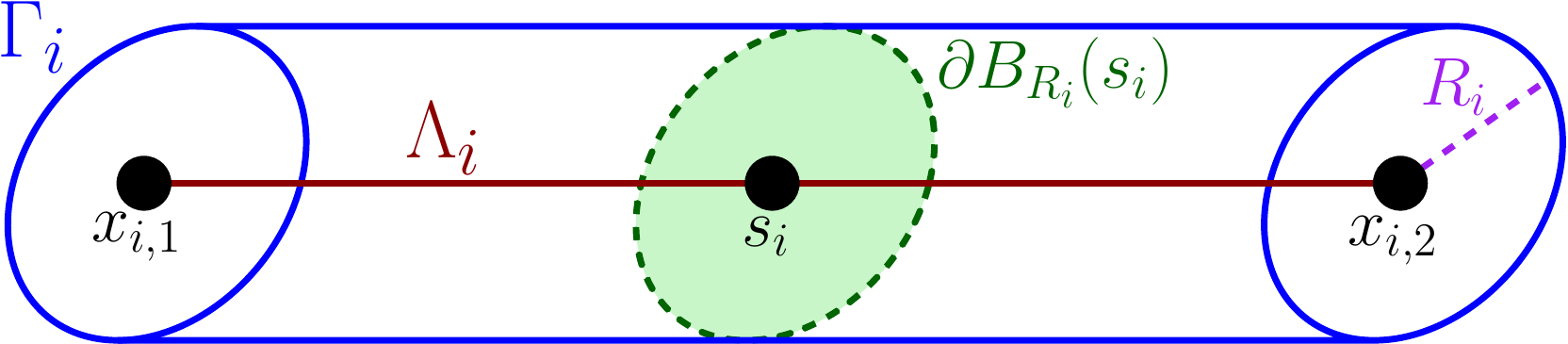}
	\caption{Modeling of a single edge $\Lambda_i$ contained in the 1D graph-like structure with boundary nodes $\bx_{i,1}$ and $\bx_{i,2}$. The cylinder $\Gamma_i$ has a constant radius $R_i$.}
	\label{Figure:Gamma}
\end{figure}

The single edges of $\Lambda$ of vessel components are denoted by $\Lambda_i$ such that $\Lambda$ is given by
$
\Lambda = \bigcup_{i=1}^N \Lambda_i
$
and each edge $\Lambda_i$, $i \in \{1,\dots,N\}$, is parameterized by a corresponding curve parameter $s_i$ such that
\begin{equation*}
\Lambda_i = \left\{ \bx \in \Omega \left|\; \bx = \Lambda_i ( s_i ) = \bx_{i,1} + s_i \cdot ( \bx_{i,2}-\bx_{i,1} ),\;s_i \in (0,1 )  \right. \right\},
\end{equation*}
where $\bx_{i,1} \in \Omega$ and $\bx_{i,2} \in \Omega$ mark the boundary nodes of $\Lambda_i$, see \cref{Figure:Gamma}.
For the total 1D network $\Lambda$, we introduce a global curve parameter $s$, which is interpreted in the following way: $s=s_i$, if $\bx = \Lambda ( s ) = \Lambda_i (s_i )$. At each value of the curve parameter $s$, various 1D constituents exist, which interact with their respective 3D counterpart in $\Omega$. 

We introduce the surface $\Gamma$ of the microvascular network $\Lambda$ to formulate the coupling between the 3D and 1D constituents in \cref{sec:3Dmodel} and \cref{sec:1Dmodel}. For simplicity, it is assumed that the surface for a single vessel is approximated by a cylinder with a constant radius, see \cref{Figure:Gamma}. The radius of a vessel that is associated with $\Lambda_i$ is given by $R_i$ and the corresponding surface is denoted by $\Gamma_i$; i.e., we have as the total surface 
$
\Gamma = \bigcup_{i=1}^N \Gamma_i.
$

\subsection{Governing constituents}
The principal dependent variables characterizing the growth and decline of the tumor mass are taken to be a set of scalar-valued fields $\phi_\alpha$ with values $\phi_\alpha(\bx,t)$ at a time $t \in [0,T]$ and point $\bx \in \Omega \subset \R^3$, representing the volume fractions of constituents in the space-time {domain} $\Omega \times [0,T]  $. The primary feature of our model of tumor growth is the application of the framework of continuum mixture theory in which multiple mechanical and chemical species can exist at a point $\bx \in \Omega$ at time $t>0$. Therefore, for a medium with ${N_\alpha} \in \N$ interacting constituents, the volume fraction of each species $\phi_\alpha$, $\alpha \in \{1,\dots,{N_\alpha}\}$, is represented by a field $\phi_\alpha$ with the value $\phi_\alpha(\bx,t)$ at $(\bx,t)$ and the property $\sum_{\alpha} \phi_\alpha(\bx,t) = 1$.

We separate the tumor volume fraction $\phi_T=\phi_T(\bx,t)$ into the sum of three phases $\phi_T = \phi_P + \phi_H + \phi_N$, where $\phi_P=\phi_P(\bx,t)$ is the volume fraction of proliferative cells, $\phi_H=\phi_H(\bx,t)$ that of hypoxic cells, and $\phi_N=\phi_N(\bx,t)$ is the volume fraction of necrotic cells, see \cref{Figure:Sketch}. Proliferative cells have a high probability of mitosis, i.e., division into twin cells, and to produce growth of the tumor mass. Hypoxic cells are those tumor cells which are deprived of sufficient nutrient to become or remain proliferative. Lastly, necrotic cells have died due to the lack of nutrients. 

The nutrient concentration and the tumor angiogenesis factor (TAF) over $\Omega \times [0,T]  $ are represented by scalar fields $\phi_\sigma = \phi_\sigma(\bx,t)$ and $\phi_\TAF = \phi_\TAF(\bx,t)$, respectively. The tumor cells response to hypoxia, i.e., when $\phi_\sigma$ is below a certain threshold, is the production of an enzyme that increases cell mobility and activates the secretion of angiogenesis promoting factors characterized by $\taf$. As a particular case of TAFs, we consider the vascular endothelial growth factor (VEGF), which promotes sprouting of endothelial cells forming the tubular structure of blood vessels, which grow into new vessels and supply nutrients to the hypoxic volume fraction $\phi_H$. 

Another consequence of hypoxia is the release of matrix-{degrading} enzymes (MDEs), e.g., urokinase plasminogen and matrix metalloproteinases, by the hypoxic cells. We denote the volume fraction of the MDEs by $\mde=\mde(\bx,t)$. The primary feature of the MDEs is the erosion of the extracellular matrix, whose volume fraction is denoted by $\ecm=\ecm(\bx,t)$. Consequently, the erosion of the ECM produces room for the invasion of tumor cells, which increases $\phi_T$ in the ECM domain and therefore, raises the likelihood of metastasis. Below a certain level necrosis occurs and cells die, entering the necrotic phase $\phi_N$. Tumor cells may also die naturally, in a process which is called apoptosis.

Within the one-dimensional network $\Lambda$, we introduce the constituents $\phi_v=\phi_v(s,t)$ and $v_v=v_v(s,t)$, which represent the one-dimensional counterparts of the local nutrient concentration $\phi_\sigma$ and the volume-averaged velocity $v$. Additionally, we consider the pressures $p_v=p_v(s,t)$ and $p=p(\bx,t)$ in the network and tissue domain, respectively.  { In summary, we refer to the table below for the primary variables and constituents of the model.}

\begin{center}
	\centering {
	\begin{tabular}{ |p{0.05\textwidth}|p{0.7\textwidth}|}
		\hline
		\multicolumn{2}{|c|}{Constituents} \\
		\hline
		$\bphi$ & Vector of all 3D species volume fractions \\
		$\phi_\alpha$ & Volume fraction of 3D species $\alpha \in \A=\{P,H,N,\sigma,MDE,TAF,ECM\}$ \\
		$\mu_\beta$ & Chemical potential, $\beta \in \{P,H\}$  \\
		$\phi_{v}$ & Volume fraction of nutrients in 1D network $\Lambda$  \\
		\hline
		\multicolumn{2}{|c|}{Flow model} \\
		\hline
		$v$ & Convective velocity in tissue domain $\Omega$ \\
		$p$ & Pressure in tissue domain $\Omega$ \\
		$p_v$ & Pressure in 1D network domain $\Lambda$ \\
		$v_v$ & Velocity of interstitial flow in tissue domain $\Lambda$ \\
				\hline
		\multicolumn{2}{|c|}{Functions} \\
		\hline
		$\Psi$ & Double-well potential, see \cref{Eq:Psi} \\
		$m_\alpha$ & Mobility function, see \cref{Eq:Mob} \\
		$S_\alpha$ & Mass source, see \cref{Eq:DerivationSourceCH} \\
		$W_\alpha$ & Wiener process, see \cref{eq:WeinerProc} \\
		$J_{\sigma v}$ & Mass source density of nutrient due to 1D network \cref{eq:KedemKatalchsky} \\
		\hline
	\end{tabular}}
\end{center}

\subsection{Three-dimensional model governing the tumor constituents}
\label{sec:3Dmodel}
The evolution of the constituents $\phi_\alpha$ must obey the balance laws of continuum mixture theory (e.g., see \cite{lima2014hybrid,fritz2019unsteady}). Assuming constant and equal mass densities of the constituents, the mass balance equations for the mixture read as follows:
\begin{equation*} \label{Eq:MassBalance}
\partial_t \phi_\alpha+\text{div}(\phi_\alpha v_\alpha)=- \text{div} J_\alpha(\phib) +S_\alpha(\phib),
\end{equation*}
where $v_\alpha$ is the cell velocity of the $\alpha$-th constituent, and $S_\alpha$ describes a mass source term that may depend on all species $\phib=(\phi_P,\phi_H,\phi_N,\phi_\sigma,\mde,\taf,\ecm)$. Moreover, $J_\alpha$ represents the mass flux of the $\alpha$-th constituent and is given by:
\begin{equation*}J_\alpha(\phib) = - m_\alpha(\phib) \nabla \mu_\alpha,\end{equation*}
 where $\mu_\alpha$ denotes the chemical potential of the $\alpha$-th species, and $m_\alpha$ is its corresponding mobility function. Generally, the mobilities may depend on many species, but in this work we consider the following cases,
\begin{equation} \label{Eq:Mob} \begin{aligned} m_\alpha(\phib) &=M_\alpha \phi_\alpha^2 (1-\phi_T)^2 I_d, && \alpha \in \{P,H\}, \\ m_\beta(\phib) &= M_\beta I_d, && \beta \in \{\sigma,\MDE,\TAF\}, 
\end{aligned}\end{equation} where $M_\alpha$ are mobility constants, and $I_d$ is the $(d\times d)$-dimensional identity matrix.  For the remaining species $\phi_N$ and $\ecm$, we choose $m_N=m_{\ECM}=0$ in accordance to the non-diffusivity of the necrotic cells and the ECM; see \cite{nargis2016effects}.  Following \cite{hawkins2012numerical,lima2014hybrid,wise2008three}, we define the chemical potential $\mu_\alpha$ as the first variation (G\^{a}teaux derivative) with respect to $\phi_\alpha$ of the Ginzburg--Landau--Helmholtz free energy functional $\mathcal{E}(\phib)$. {{The free energy in this work is designed to capture the following key effects:
\begin{itemize}[leftmargin=.18in]
\item {\bf Phase change in tumor species} $\phi_T, \phi_P, \phi_H$. For example, $\phi_T$ can change (conditions permitting) from a healthy phase $\phi_T = 0$ to a tumor phase $\phi_T = 1$. This is typically achieved by introducing a double-well potential 
\begin{equation} \label{Eq:Psi} \Psi = \Psi(\phi_T, \phi_P, \phi_H) = \sum_{\alpha \in \{T, P, H\}} C_{\Psi_\alpha} \phi_\alpha^2 (1 - \phi_\alpha)^2
\end{equation}
to the free energy, where $C_{\Psi_\alpha}$, $\alpha \in \{T, P, H\}$, are constants. In addition to phase separation between healthy and cancer phases (using the energy term $C_{\Psi_T}\phi_T^2 (1- \phi_T)^2$), we have also introduced energy terms that promote phase separation between proliferative and non-proliferative and hypoxic and non-hypoxic phases. It is possible to consider different forms of the double-well potential \cite{fritz2020analysis}, however, in this work we will consider $\Psi$ in \cref{Eq:Psi} with $C_{\Psi_P} = C_{\Psi_H} = 0$, see \cref{Sec:Simulation}.

\item {\bf Promote phase separation} between two phases of species $\phi_T, \phi_P, \phi_H$. For example, a model could exhibit phase values at $\bx$ between, say, $\phi_\alpha = 0$ and $\phi_\alpha = 1$, with a change in gradient, $\nabla \phi_\alpha$, at the interface of these phases. Such changes are manifested as surface energy terms in the form of penalties on the magnitude of $\nabla \phi_\alpha$ of the form $$\frac{\eps_\alpha^2}{2} |\nabla \phi_\alpha|^2 ,$$ where $\eps_\alpha$ controls the thickness of the phase interface.

\item {\bf Diffusion driven mobilities} of species $\phi_\sigma, \taf, \mde$. These effects are captured by adding the diffusive energies $$\frac{D_\beta}{2} \phi_\beta^2 ,$$
where $D_\beta$, $\beta \in \{\sigma, \TAF, \MDE\}$, are diffusion coefficients.

\item {\bf Chemotaxis and haptotaxis effects}. Chemotaxis represents a movement of cells towards a gradient of nutrients (i.e., along the direction of increasing nutrients). Similar to chemotaxis, the tumor cells show a tendency to move along the ECM gradient, and this phenomenon is referred to as haptotaxis. These effects are incorporated via the terms \cite{hillen2013convergence,tao2011chemotaxis} $$-(\chi_c \phi_\sigma + \chi_h \ecm) \sum_{\alpha\in \{P,H\}} \phi_\alpha ,$$
where $\chi_c,\chi_h$ are chemotaxis and haptotaxis coefficients, respectively. In the above energy terms, we exclude necrotic cells to be consistent with our assumption that necrotic cells are immobile.
\end{itemize}
Combining these effects, the free energy takes the form
\begin{equation*}
\mathcal{E}(\phib)= \int_\Omega \Big\{ \Psi(\phi_P, \phi_H, \phi_N) + \sum_{\alpha \in \{P,H\}}  \frac{\eps_\alpha^2}{2} |\nabla \phi_\alpha|^2  + \sum_{\beta \in \RD} \frac{D_\beta}{2} \phi_\beta^2 -  (\chi_c \phi_\sigma+\chi_h \ecm) \sum_{\alpha \in \{P,H\}} \phi_\alpha \Big\} \text{ d}\bx,
\label{Eq:GinzburgLandau}
\end{equation*}
where $\RD = \{\sigma,\MDE,\TAF, \ECM\}$ is the set of species driven by reaction-diffusion type equations.
} 
}
We assume a volume-averaged velocity $v$ for the proliferative cells, hypoxic cells, and the nutrients concentration. This assumption is regarded as reasonable whenever cells are tightly packed. 

{In thin subdomains at the interfaces of the phase fields, stochastic variations of the phase concentrations are possible. The variations in these regions of random behavior are bounded by noise parameters $\phi_\alpha^\omega$ and noise intensity $\omega_\alpha$; the variations (along with the noise intensity) in $\phi_\alpha$, $\alpha \in \{P,H\}$, are restricted to interface regions using function $G_\alpha$ given by
\begin{equation}\label{eq:WeinerProc}
	G_\alpha(\phi_P,\phi_H,\phi_N)=\omega_\alpha  \mathcal{H}((\phi_\alpha-\phi_\alpha^\omega)(1-\phi_\alpha-\phi_\alpha^\omega)) \mathcal{H}((\phi_T-\phi_T^\omega)(1-\phi_T-\phi_T^\omega)).
\end{equation}
Here, $\mathcal{H}$ denotes the Heaviside step function. Typically, the randomness in the evolution of species near the interface is incorporated in the model in the form of cylindrical Wiener process on $L^2(\Omega)$, see \cite{da1996stochastic, orrierioptimal, antonopoulou2019numerical}; we add $G_P \dot{W}_P$ and $G_H \dot{W}_H$ to the mass balance equation for $\phi_P$ and $\phi_H$. To keep the mass balance equations in standard form, we slightly abuse the standard notation and use notation $\dot{W}_\alpha$ such that $\dot{W}_\alpha \dd t = \dd W_\alpha$. Further details on Wiener processes $W_\alpha$ and numerical discretization are provided in \cref{Sec:Solver}. 
}

Following these assumptions and conventions, we arrive at the following system of equations governing the model:
\begin{equation} \label{Eq:Model3D} \begin{aligned}
\pt \phi_P+ \div(\phi_P v)            & = \div (m_P( \phib) \nabla \mu_P)+ S_P( \phib) + G_P(\phi_P,\phi_H,\phi_N) \dot W_P,                                                               \\
\mu_P                                 & =   \p_{\phi_P} \Psi(\phi_P,\phi_H,\phi_N) - \eps^2_P \Delta \phi_P - \chi_c \phi_\sigma-\chi_h \ecm ,                                                           \\
\pt \phi_H+ \div(\phi_H v)            & = \div (m_H( \phib) \nabla \mu_H)+S_H( \phib) + G_H(\phi_P,\phi_H,\phi_N) \dot W_H ,                                                              \\
\mu_H                                 & =   \p_{\phi_H}\Psi(\phi_P,\phi_H,\phi_N) - \eps^2_H \Delta \phi_H - \chi_c \phi_\sigma-\chi_h \ecm ,                                                           \\
\pt \phi_N                            & = S_N( \phib)   ,                                                                                                                                   \\
\pt \phi_\sigma + \div(\phi_\sigma v) & = \div (m_\sigma( \phib) (D_\sigma \nabla \phi_\sigma \!-\! \chi_c  \nabla (\phi_P +\phi_H))+S_\sigma( \phib) + J_{\sigma v} (\phi_\sigma, p, \Pi_\Gamma \phi_v,  \Pi_\Gamma p_v ) \delta_\Gamma ,\\
\pt \mde + \div(\mde v)    & = \div (m_{\MDE}( \phib) D_{\MDE} \nabla \mde)+S_{\MDE}( \phib), \\
\pt \taf + \div(\taf v)  & = \div (m_{\TAF}( \phib) D_{\TAF} \nabla \taf)+S_{\TAF}( \phib), \\
\pt \ecm                        & = S_{\ECM}(\phib),  \\
-\div ( K \nabla p ) &=    J_{pv}( p, \Pi_\Gamma p_v ) \delta_\Gamma - \div (KS_p( \phib,\mu_P,\mu_H)),\\
v  & = - K(\nabla p -S_p(\phib,\mu_P,\mu_H)) , 
\end{aligned} \end{equation}
in the space-time domain $\Omega \times (0,T)$ and we supplement the system with homogeneous Neumann boundary conditions. {{In the above set of governing equations, the velocity $v$ is given by modified Darcy's law, where $K$ denotes the hydraulic conductivity.  The source term $S_p$ (defined below) represents a form of the elastic Korteweg force, e.g., see \cite{frigeri2018on},
and includes a correction of the chemical potential by the haptotaxis and chemotaxis adhesion terms following \cite{garcke2018multiphase}.}} Here $J_{pv}$ and $J_{\sigma v}$ are the fluid flux and nutrient flux as described in \cref{Sec:Interaction}.  We consider  the following choices of the coupling source functions; see \cite{fritz2020analysis}, 
\begin{equation} \label{Eq:DerivationSourceCH}
\begin{aligned}
S_P(\phib) &= \lambdap_{P} \phi_\sigma \phi_P(1- \phi_T) - \lambdad_{P} \phi_P - \lambda_{P\!H}  \H(\sigma_{P\!H} - \phi_\sigma)\phi_P + \lambda_{H\!P} \H(\phi_\sigma - \sigma_{H\!P}) \phi_H,  \\
S_H(\phib) &=  \lambdap_{H} \phi_\sigma \phi_H(1-\phi_T) -\lambdad_{H} \phi_H + \lambda_{P\!H}  \H(\sigma_{P\!H} - \phi_\sigma) \phi_P - \lambda_{H\!P} \H(\phi_\sigma - \sigma_{H\!P}) \phi_H \\[-.1cm]&\qquad - \lambda_{H\!N} \H(\sigma_{H\!N} - \phi_\sigma)\phi_H,  \\
S_N(\phib) &= \lambda_{H\!N} \H(\sigma_{H\!N} - \phi_\sigma) \phi_H, \\
S_{\ECM}( \phib) 
&=-\lambdad_{\ECM}\ecm\mde + \lambdap_{\ECM} \phi_\sigma (1- \ecm) \H(\ecm - \phi^\text{pro}_{\ECM}),\\
S_{\sigma}(\phib) &= -\lambdap_{P} \phi_\sigma\phi_P -\lambdap_{H} \phi_\sigma\phi_H + \lambdad_{P} \phi_P + \lambdad_{H} \phi_H  - \lambdap_{\ECM} \phi_\sigma (1- \ecm) \H(\ecm - \phi^\text{pro}_{\ECM}) \\[-.1cm] &\qquad + \lambdad_{\ECM}\ecm\mde,  \\
S_{\MDE}(\phib) &= -\lambdad_{\MDE}\mde + \lambdap_{\MDE}(\phi_P + \phi_H) \ecm\frac{\sigma_{H\!P}}{\sigma_{H\!P} + \phi_\sigma} (1-\mde) - \lambdad_{\ECM} \ecm \mde,  \\[-.05cm]
S_{\TAF}(\phib) &= \lambdap_{\TAF} (1- \taf) \phi_H \H(\phi_H-\phi_{H_P}) -\lambda_\TAF^{\deg} \phi_\TAF, \\
S_p(\phib,\mu_P,\mu_H) &=(\mu_P+\chi_c\phi_\sigma + \chi_h \ecm) \nabla \phi_P +(\mu_H+\chi_c \phi_\sigma+ \chi_h \ecm) \nabla \phi_H.
\end{aligned}
\end{equation}
Here, $\lambdap_{\alpha}$ and $\lambdad_{\alpha}$ denote the proliferation and degradation rate of the $\alpha$-th species, respectively, $\lambda_{\alpha \beta}$ the transition rate from the $\alpha$-th to the $\beta$-th volume fraction, $\sigma_{\alpha \beta}$ the corresponding nutrient threshold for the transition, and $\mathcal{H}$ is the Heaviside step function. Further, $\phi^\text{pro}_{\ECM}$ denotes the threshold level for the ECM density in order to activate the production of ECM fibers. 
Moreover, we introduce the projection $\Pi_\Gamma$ of the 1D quantities onto the cylinder $\Gamma$ via extending its function values $\Pi_\Gamma \phi_v(s) = \phi_v(s_i)$ for all $s \in \partial B_{R_i}(s_i)$.

\subsection{Interaction between the 3D and 1D model} \label{Sec:Interaction}
We apply the Kedem--Katchalsky law  \cite{ginzburg1963frictional} to quantify the flux of nutrients across the vessel surface; i.e., $J_{\sigma v}$ in \cref{Eq:Model3D} is given by
\begin{equation}
\label{eq:KedemKatalchsky}
J_{\sigma v} (\ov \phi_\sigma, \ov p, \phi_v, p_v ) = ( 1-r_{\sigma} ) J_{pv} ( \ov p,p_v )  \phi_{\sigma}^v + L_\sigma ( \phi_v - \ov \phi_\sigma ),
\end{equation}
where  $J_{pv}$ denotes the flux, which is caused by the flux of blood plasma from the vessels into the tissue or vice versa. Further, $J_{pv}$ is governed by Starling's law \cite{salathe1976mathematical}, i.e.,
$
J_{pv}( \ov p, p_v ) = L_p ( p_v - \overline{p} )
$
where $\overline{p}$ denotes an averaged pressure over the circumference of cylinder cross-sections and is computed in the following way: For each point $s_i$ on the curve $\Lambda_i$, we consider the circle  $\partial B_{R_i}( s_i )$ of radius $R_i$, which is perpendicular to $\Lambda_i$; see \cref{Figure:Gamma}. Thus, the tissue pressure $p$ is averaged with respect to $\partial B_{R_i}( s_i )$,
$$
\overline{p} ( s_i ) = \frac{1}{2 \pi R_i} \int_{\partial B_{R_i}( s_i )} p|_\Gamma \,\dd S.
$$

The part $J_{pv} \phi_\sigma^v$ in the Kedem--Katchalsky law \cref{eq:KedemKatalchsky} is weighted by a factor $1-r_{\sigma}$, $r_{\sigma}$ being a reflection parameter, introduced to account for the permeability of the vessel wall with respect to the nutrients. The value of $\phi_\sigma^v$ is set to $\phi_v$ for $p_v \geq \ov p$ and to $\ov\phi_\sigma$ otherwise. 
The second term on the right hand side of \cref{eq:KedemKatalchsky} is a Fickian type law, accounting for the tendency of the nutrients to balance their concentration levels, where the permeability of the vessel wall is represented by the parameter $L_\sigma$.

The interaction between the vascular network and the tissue occur at the vessel surface $\Gamma$, and thus, we concentrate the flux $J_{\sigma v}$ by means of the Dirac measure $\delta_\Gamma$; i.e., we define
$$\delta_\Gamma(\varphi) = \int_\Gamma \varphi|_{\Gamma} \, \dd S,$$ for a sufficiently smooth test function $\varphi$ with compact support. 

\subsection{One-dimensional model for transport in the vascular network}
\label{sec:1Dmodel}
The one-dimensional vessel variables $\phi_v$ and $p_v$ represent averages across cross-section of the blood vessels.
Thus, the one-dimensional variables $\phi_v$ and $p_v$ on a 1D vessel $\Lambda_i$, $i \in \{1,\dots,N\}$, depend only on $s_i$. See also \cite{koppl20203d} for more details related to the derivation of the 1D pipe flow and transport models. With these conventions, the 1D model equations for flow and transport on $\Lambda_i$ are given by
\begin{equation} \label{Eq:Model1D}
		\begin{aligned}
			\pt \phi_v + \partial_{s_i} (v_v \phi_v) & =  \partial_{s_i} (m_v(\phi_v)D_v \partial_{s_i} \phi_v) -2\pi R_i J_{\sigma v} (\ov \phi_\sigma, \ov p, \phi_v,  p_v),\\
			R_i^2 \pi  \;\partial_{s_i} ( K_{v,i} \; \partial_{s_i} p_v )   & = 2 \pi R_i J_{pv}( \overline{p}, p_v ).
		\end{aligned} 
\end{equation}
Here, we have introduced the permeability
$
K_{v,i} = \tfrac18 R_i^2/\mu_{bl} 
$
of the $i$-th vessel with $\mu_{bl}$ being the viscosity of blood. We assign $\mu_{bl}$ a constant value, i.e.,  non-Newtonian behavior of blood is not considered. The diffusivity parameter $D_v$ is set to the same value as $D_\sigma$. The blood velocity $v_v$ is given by the Darcy equation
$
v_v = - K_{v,i} \partial_{s_i} p_v. 
$

In order to interconnect $p_v$ and $\phi_v$ on $\Lambda_i$ at the inner networks nodes on the intersections
$\bx \in \partial \Lambda_i \setminus \partial \Lambda,$ we require continuity conditions on $p_v$ and $\phi_v$. Moreover, we enforce conservation of mass to obtain a physically relevant solution. To formulate these coupling conditions in a mathematical way, we define for each bifurcation point $\bx$ an index set
$$
N(\bx ) = \left\{\left. i \in \left\{ 1,\ldots,N \right\} \; \right| \; \bx \in \partial \Lambda_i  \right\}.
$$
We  state the following continuity and mass conservation conditions at an inner node $\bx \in \partial \Lambda_i$:
$$\begin{aligned}
	p_v |_{\Lambda_i}(\bx) - p_v |_{\Lambda_j}(\bx) &=0, \quad\text{ for all }  j \in N(\bx) \backslash\{i\},\\ 
	\phi_v |_{\Lambda_i}(\bx) - \phi_v |_{\Lambda_j}(\bx) &=0, \quad\text{ for all }  j \in N(\bx) \backslash\{i\}, \\
	\sum_{j \in N(\bx)}  -\frac{  R_j^4 \pi}{8 \mu_{\mathrm{bl}} } \frac{\partial p_v}{\partial s_j} \Big|_{\Lambda_j}(\bx) &=0, \\
	\sum_{j \in N(\bx)}  \Big( v_v \phi_v - m_v(\phi_v)D_v \frac{\partial \phi_v}{\partial s_j} \Big) \Big|_{\Lambda_j}(\bx)  &= 0.
\end{aligned}$$

\section{Angiogenesis: Network Growth Algorithm} 
\label{Sec:NetworkGrowth}

As noted earlier, angiogenesis is triggered by an increased TAF concentration $\phi_{\TAF}$ around the pre-existing blood vessels. After the TAF molecules are emitted by the dying hypoxic tumor cells, they move through the tissue matrix and may encounter sensor ligands on the vessel surfaces. If the TAF concentration  is large enough at the vessel surfaces, an increased number of sensors in the vessel wall are activated and a reproduction of endothelial cells forming the vessel walls is initiated. As a result, the affected vessels can elongate, resulting in two different kinds of vessel elongations or growth. In medical literature, see, e.g., \cite{ribatti2012sprouting,eilken2010dynamics}, this process is referred to as apical growth and sprouting of vessels. The term apical growth is derived from the term apex denoting the tip of a blood vessel, i.e., apical growth is the type of growth occurring at the tip of a vessel. On the other hand, the sprouting of new vessels results in the formation of new vessels at other places on the vessel surface. In order to increase or decrease the flow of blood and nutrients through the vessels, it is observed that the newly formed blood vessels can adapt their vessel radii which is caused, e.g., by an increased wall shear stress at the inner side of the vessel walls. Combining these mechanisms, an increased supply of nutrients for both the healthy and cancerous tissue can be achieved such that the tumor can continue to grow. 

In the following, we describe how an angiogenesis step can be realized within our mathematical model. It is assumed that in such a step the apical growth is considered first and then the sprouting of new vessels is simulated. At the end of an angiogenesis step, the radii of the vessels are adapted to regulate the blood flow. The 1D network that is updated during an angiogenesis step is denoted by $\Lambda_{\old}$.    

\subsection{Apical growth} Since the apical growth occurs only at the tips of the blood vessels, we consider all the boundary nodes $\bx$ of the network $\Lambda_{\old}$ contained in the inner part of $\Omega$, i.e., $\bx \in \partial \Lambda_{\old}$ and $\bx \notin \partial \Omega$. Moreover, we assume that $\bx$ is contained in the segment $\Lambda_i \subset \Lambda_{\old}$. At $\bx$, the value of the TAF concentration is denoted by $\phi_{\TAF} (\bx )$. If this value exceeds a certain threshold $Th_{\TAF}$: $\phi_{\TAF} (\bx ) \geq Th_{\TAF}$, the tip of the corresponding vessel is considered as a candidate for growth. 

\begin{figure}[!htb]
	\includegraphics[width=.9\textwidth]{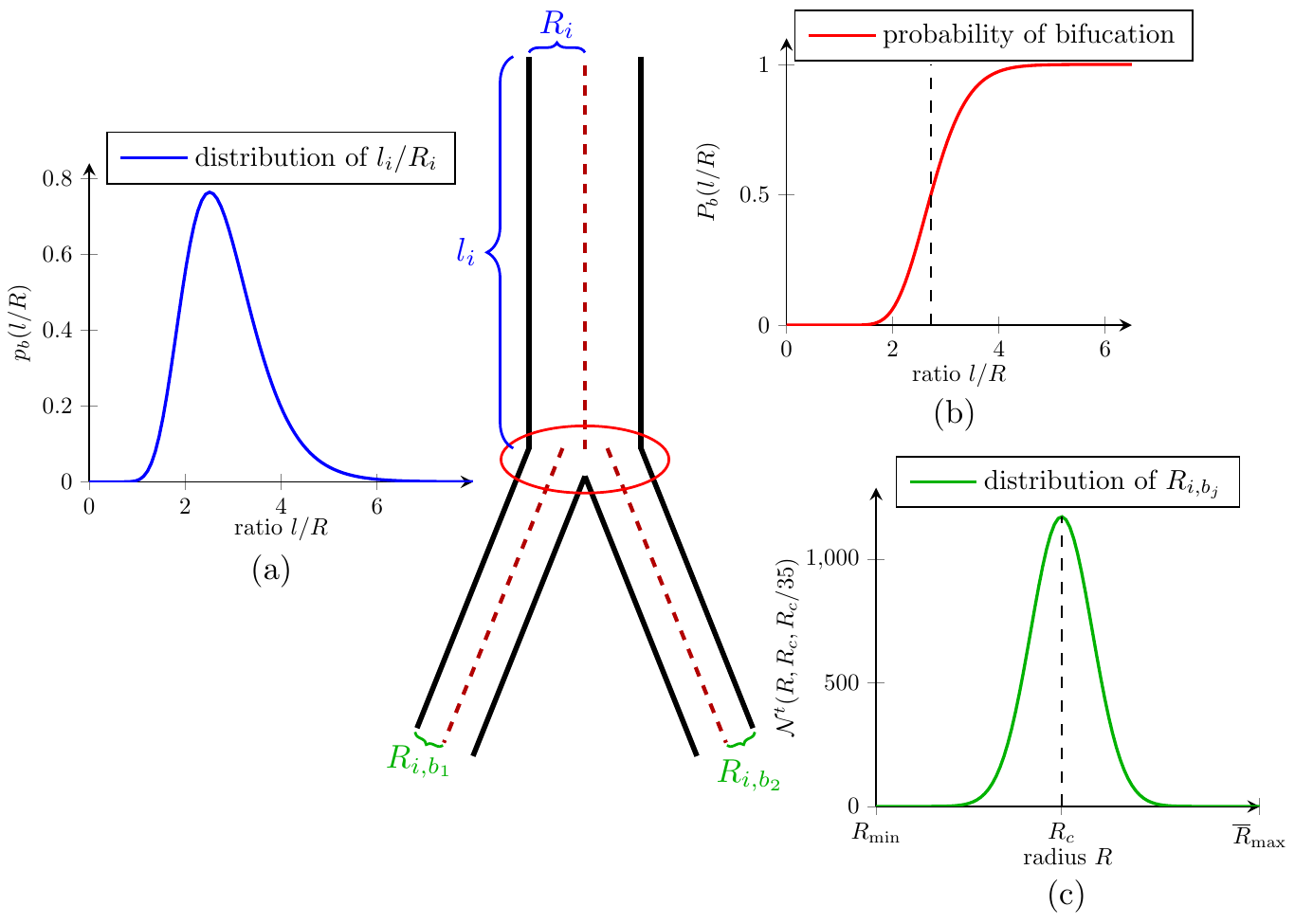}
	\caption{Given a vessel with length $l_i$ and radius $R_i$, we plot the probability of the occurrence of a  bifurcation (red curve in figure (b)), the ratio of its length over the radius (blue curve in figure (a)), and the distribution of the radii of the sproutings (green curve in figure (c)); we choose $R_i=1.5\cdot 10^{-2}$, $R_c = 2^{-\frac{1}{3}} R_i$ according to \cref{eq:radiusbranch1},  $\mu_r=1$, $\sigma_r=0.2$, $R_{\min}=9 \cdot 10^{-3}$, $R_{\max}=3.5 \cdot 10^{-2}$ according to \cref{tab:growth}, $\ov{R}_{\max}=\max\{R_{\max},R_i\}=R_i$. }
	\label{Fig_Distributions}	
\end{figure}

There are two types of growth that are allowed to occur at the apex of a vessel: either the vessel can further elongate or it can bifurcate. In order to decide which event occurs, a probabilistic method is used. According to \cite{schneider2012tissue} and the references therein, the ratio $r_i = l_i/R_i$ of the vessel $\Lambda_i$ follows a log-normal distribution:
\begin{equation}
\label{eq:Pb}
p_b( r ) \sim \mathcal{L}\mathcal{N} ( r, \mu_r, \sigma_r ) = \frac{1}{r \sqrt{2 \pi \sigma_r^2}} \exp \bigg( -\frac{( {\ln r} - \mu_r )^2}{2 \sigma_r^2} \bigg).
\end{equation}
The parameters $\mu_r$ and $\sigma_r$ represent the mean value and standard deviation of the probability distribution $p_b$, respectively. Using the cumulative distribution function of $p_b$, we decide whether a bifurcation is considered or not. This means that a bifurcation at $\bx \in \partial \Lambda_i \cup \partial \Lambda$ is formed with a probability of:
\begin{equation}
\label{eq:Pbif}
P_{b}(r) = \Phi \bigg( \frac{\ln r - \mu_r}{\sigma_r} \bigg) = \frac 12 + \frac 12 \text{erf}\bigg( \frac{\ln r-\mu_r}{\sqrt{2\sigma_r^2}}\bigg),
\end{equation}
where $\Phi$ denotes the standard normal cumulative distribution function and $x\mapsto \text{erf}(x)$ the Gaussian error function. We refer to \cref{Fig_Distributions} for the illustration of an exemplary vessel, which bifurcates. Moreover, we depict the distribution of the ratio $l_i/R_i$ according to \cref{eq:Pb}, the radii of its bifurcations, see \cref{eq:radiusbranch1} below, and the probability of the occurrence of a bifurcation, see \cref{eq:Pbif}.

If a single vessel is formed at $\bx$, the direction of growth $\mathbf{d}_g$ is based on the TAF concentration:
\begin{equation}\label{eq:growthDirection}
\mathbf{d}_g(\bx ) = \frac{\nabla \phi_{\TAF} (\bx )}{\left\| \nabla \phi_{\TAF} (\bx ) \right\|} + \lambda_g \frac{\mathbf{d}_i}{\left\| \mathbf{d}_i \right\|},
\end{equation}
where $\|\cdot\|$ denotes the Euclidean norm. The vector $\mathbf{d}_i=\bx_{i,2} -\bx_{i,1}$ is the orientation of the vessel $\Lambda_i$, and
the value $\lambda_g \in \left(0,1 \right]$ represents a regularization parameter that can be used to circumvent the formation of sharp bendings and corners. This is necessary if the TAF gradient at $\bx$ encloses an acute angle with $\mathbf{d}_i$. The radius $R_{i^\prime}$ of the new vessel $\Lambda_{i^\prime}$ is taken over from $\Lambda_{i}$ i.e. $R_{i^\prime} = R_i$. Having the radius $R_{i^\prime}$ at hand, we use \eqref{eq:Pbif} to determine the length $l_{i^\prime}$ of $\Lambda_{i^\prime}$. Before $\Lambda_{i^\prime}$ is incorporated into the network $\Lambda_{\old}$, we check, whether it intersects another vessel in the network. If this is the case, $\Lambda_{i^\prime}$ is not added to $\Lambda_{\old}$. {{In order to test whether a new vessel intersects an existing vessel that is not directly connected, we compute the distance between the centerlines of the new vessel and the existing vessel. 
If this distance is smaller than the sum of the radii for any of the existing vessels, the new vessel is considered too close to existing vessels, and, therefore, the new vessel is not inserted into the network.}}

In the case of bifurcations, we have to choose the radii, orientations and lengths of the new
branches $b_1$ and $b_2$. The radii of the new branches are computed based on a Murray-type law. It relates the radius $R_i$ of the father vessel to the radius $R_{i,b_1}$ of branch $b_1$ and the radius $R_{i,b_2}$ of branch $b_2$ as follows \cite{murray1926physiological}:
\begin{equation}
\label{eq:Murray}
R_i^{\gamma} = R_{i,b_1}^{\gamma} + R_{i,b_2}^{\gamma},
\end{equation}
where $\gamma$ denotes the bifurcation exponent. According to  \cite{schneider2012tissue}, $\gamma$ can vary
between $2.0$ and $3.5$. In addition to \eqref{eq:Murray}, we require an additional equation to determine the radii of the branches. Towards this end, it is assumed that $R_{b_1}$ follows a truncated Gaussian normal distribution:
\begin{equation}
\label{eq:radiusbranch1}
R_c = 2^{ -\frac{1}{\gamma} }R_i,\;\qquad R_{b_k}  \sim \mathcal{N}^t( R, \mu = R_c, \sigma = R_c/35 ),\;\qquad k \in \left\{1,2\right\},
\end{equation}
which is set to zero outside of the interval $[R_{\min}, \overline{R}_{\max}]$ with $\overline{R}_{\max}=\max\{R_{\max},R_i\}$; we refer to \cref{tab:growth} for a choice of parameters for $R_{\min}$ and $R_{\max}$. Additionally, the radius of the parent vessel acts as a natural bound for the radius of its bifurcations.

The selection of $R_{b_k}$ is motivated as follows: Using the radius $R_i$ of $\Lambda_{i}$, we compute the expected radius $R_c$ resulting
from Murray's law for a symmetric bifurcation $( R_{b_1} = R_{b_2} )$. Here, $R_c$ is used as a mean value for a Gaussian
normal distribution, with a small standard {deviation}. This yields bifurcations that are slightly deviating
from a symmetric bifurcation which is in accordance with Murray's law. Having $R_{b_1}$ and $R_{b_2}$ at hand, we
compute the corresponding lengths $l_{b_1}$ and $l_{b_2}$ as in the case of a single vessel. 

\begin{figure}[!htb]
	\includegraphics[width=.85\textwidth]{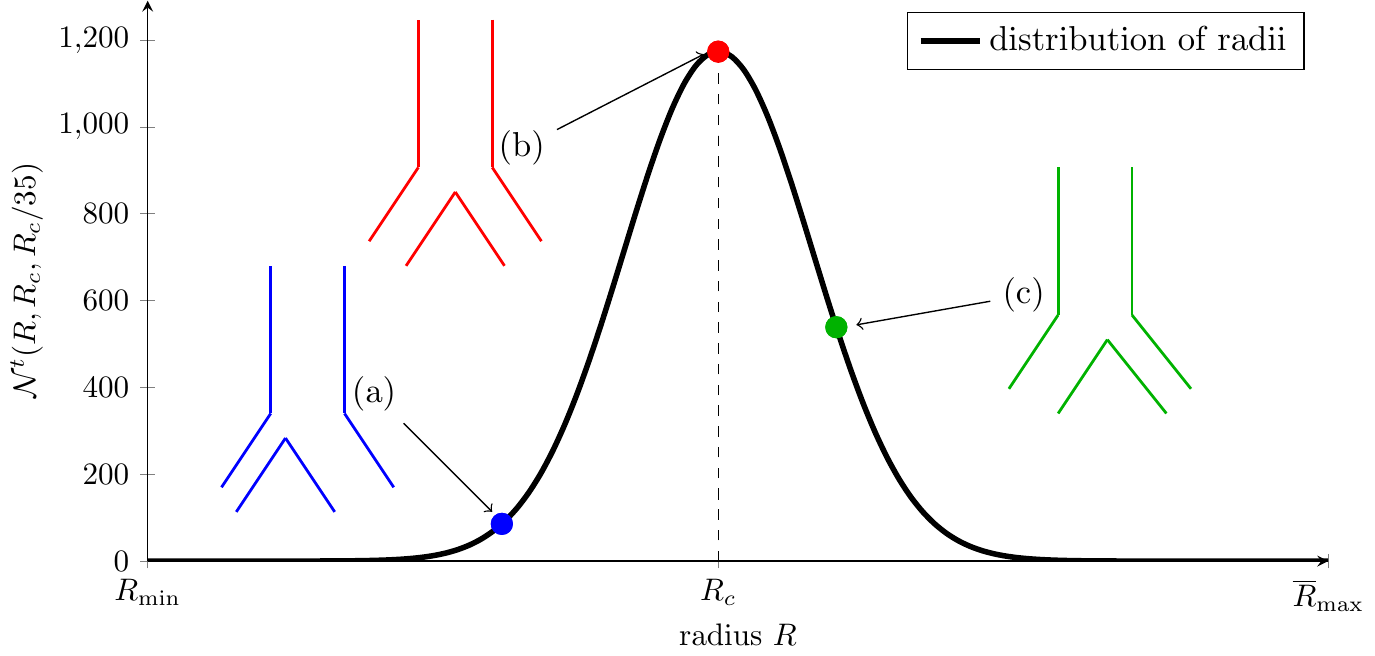}
	\caption{Distribution of the radii of the bifurcating vessels, choosing $R_i=0.015$, $R_c=2^{-\frac{1}{3}} R_i$. Examples of bifurcations with different radii are given, $R=1.08 \cdot 10^{-2}$ (case (a)), $R=R_c$ (case (b)), $R=1.25\cdot 10^{-2}$ (case (c)).}
	\label{Figure_Radii}
\end{figure}

We refer to \cref{Figure_Radii} for the distribution of the radii of the bifurcating vessels. We note that the ideal case is a symmetric bifurcation, that means both radii which correspond to the mean. Further, we also depict two asymmetric cases where the radii deviate from the mean.

The creation of a bifurcation
is accomplished by specifying the orientations of the two branches. At first, we define the plane in which
the bifurcation is contained. The normal vector $\mathbf{n}_p$ of this plane is given by the cross product of the vessel
orientation $\mathbf{d}_i$ and the growth direction $\mathbf{d}_g$ from the non-bifurcating case:
\begin{equation}\label{eq:normalNp}
\mathbf{n}_p (\bx ) = \frac{\mathbf{d}_i \times \mathbf{d}_g}{\left\| \mathbf{d}_i \times \mathbf{d}_g \right\|}.
\end{equation}
The exact location of the plane is determined such that the vessel $\Lambda_{i}$ is contained in this plane. Further
constraints for the bifurcation configuration are related to the bifurcation angles. In \cite{murray1926physiological,murray1926physiological2}, it is shown how optimality principles like minimum work and minimum energy dissipation can be utilized to derive formulas relating the radii of the branches to the branching angles $\alpha_i^{(1)}$ and $\alpha_i^{(2)}$:
\begin{equation}
\label{eq:bifurcation_angles}
\cos\big( \alpha_i^{(1)} \big) = \frac{R_i^4+R_{b_1}^4-R_{b_2}^4}{2 \cdot R_i^2 R_{b_1}^2} \;\text{ and }\;
\cos\big( \alpha_i^{(2)} \big) = \frac{R_i^4+R_{b_2}^4-R_{b_1}^4}{2 \cdot R_i^2 R_{b_2}^2}.
\end{equation}
The value $\alpha_i^{(k)}$ denotes the bifurcation angle between branch $k \in \left\{1,2\right\}$ and the father vessel. Rotating the vector $\mathbf{d}_g$ at $\mathbf{x}$ around the axis defined by $\mathbf{n}_p (\bx )$ counterclockwise by
$\alpha_i^{(1)}+\alpha_i^{(2)}$, we obtain two new growth directions $\mathbf{d}_{b_1} = \mathbf{d}_g$ and $\mathbf{d}_{b_2}$. These vectors are used to define the main axes of the two cylinders representing the two branches. This choice of the growth directions can be considered as a compromise between the optimality principles provided by \cite{murray1926physiological,murray1926physiological2} and the tendency of the network to adapt its growth direction to the nutrient demand of the surrounding tissue. At the end of the apical growth phase, we obtain a 1D network denoted by $\Lambda_{\text{ap}}$.
{{
\begin{algorithm}[h!] 
	\SetAlgoLined
	\caption{Apical growth algorithm} \label{Alg_Apical_Growth} 
	\textbf{Input}:  Network $\Lambda_{old}$, \textbf{Output}: New network $\Lambda_{ap}$ \ \\
	\For{ each $\mathbf{x} \in \partial\Lambda \cap \Omega$}{ 
		Compute the TAF concentration at $\mathbf{x}$: $\phi_{\TAF} (\bx )$; \ \\
		Consider the TAF threshold $Th_{\TAF}$;\ \\
		\If{$\phi_{\TAF} (\bx ) \geq Th_{\TAF}$}{
			Consider the edge $\Lambda_i$ containing 
			$\mathbf{x}$ i.e. $\mathbf{x} \in \partial \Lambda_i \cap \partial \Lambda$; \ \\
			$\Lambda_i$ has the orientation $\mathbf{d}_i$, the radius $R_i$  \ \\
			Compute the gradient $\nabla\phi_{\TAF} (\bx )$; \ \\	
			Compute the new growth direction $\mathbf{d}_g$ using \eqref{eq:growthDirection}; \ \\
			Compute the probability $P_b \left( \mathbf{x} \right)$ given by \eqref{eq:Pbif}; \ \\
			Form a bifurcation with probability $P_b \left( \mathbf{x} \right)$;\ \\
			\If{ a bifurcation is formed }{
			   Determine the radii of the new branches $R_{b_1}$ and $R_{b_2}$ according to \eqref{eq:Murray} and \eqref{eq:radiusbranch1};\ \\
			   Compute the bifurcation angels $\alpha_i^{(1)}$ and $\alpha_i^{(2)}$ according to \eqref{eq:bifurcation_angles}; \ \\
			   Rotate $\mathbf{d}_g(\bx )$ by the angle $\alpha_i^{(1)}+\alpha_i^{(2)}$ counterclockwise around the rotation axis defined by the vector $\mathbf{n}_p(\bx)$ (computed using \eqref{eq:normalNp}) to obtain a second growth direction $\mathbf{d}_{b_2}(\bx )$; \ \\
	           Determine the ratios $r_{b_1}$ and $r_{b_2}$ according to the probability distribution \eqref{eq:Pb}; \ \\	
               Construct new edges $\Lambda_{b_1}$ and $\Lambda_{b_2}$ having the radii $R_{b_1}$ and $R_{b_2}$, \ \\
               the lengths $l_{b_1}=r_{b_1}\cdot R_{b_1}$ and $l_{b_2}=r_{b_2}\cdot R_{b_2}$ as well as the orientations \ \\
               $\mathbf{d}_{b_1} = \mathbf{d}_g\left( \mathbf{x} \right)$ and $\mathbf{d}_{b_2}$;\ \\			   
			   \If{$\Lambda_{b_1}$ and $\Lambda_{b_2}$ are not intersecting and $R_{b_1},R_{b_2} \in \left[R_{\min},\overline{R}_{\max} \right]$}{
			   	Add $\Lambda_{b_1}$ and $\Lambda_{b_2}$ to $\Lambda_{i}$ at the node $\mathbf{x}$;\ \\	           	
			   }
			   \Else{
			   	Continue;\ \\
			   }		   
		    }
	        \Else{
	           The radius for the new edge is set to $R_i$; \ \\
	           Determine the ratio $r_i$ according to the probability distribution \eqref{eq:Pb}; \ \\	
	           Construct a new edge $\Lambda_{i^\prime}$ having the radius $R_i$, \ \\
	           the length $l_{i^\prime}=r_i\cdot R_i$ and the orientation $\mathbf{d}_g\left( \mathbf{x} \right)$;   \ \\
	           Check whether $\Lambda_{i^\prime}$ intersects; \ \\
	           \If{$\Lambda_{i^\prime}$ is not intersecting and $R_i \in \left[R_{\min},\overline{R}_{\max} \right]$}{
	             Add $\Lambda_{i^\prime}$ to $\Lambda_{i}$ at the node $\mathbf{x}$;\ \\	           	
	           }
               \Else{
	             Continue;\ \\
	           }
	        }		
		}
        \Else{
          Continue;	
        }
	}
\end{algorithm}
}}
\subsection{Sprouting of new vessels} In the second phase of the angiogenesis process, we examine each vessel or segment $\Lambda_i \subset \Lambda_{\text{ap}}$. As ligands has been already mentioned, the sprouting of inner vessels is triggered by TAF molecules touching some sensor ligands in the vessel walls. Therefore, we determine for the middle region of each segment, i.e., $\Lambda_i ( s_i ) \subset \Lambda_i,\; s_i \in ( 0.25,0.75 )$ at which place an averaged TAF concentration $\overline{\phi}_{\TAF}$ attains its maximum $\overline{\phi}_{\TAF}^{(\max)}$. As in the previous section  $\overline{\phi}_{\TAF}$ is determined by means of an integral expression:
$$
\overline{\phi}_{\TAF}( s_i ) = \frac{1}{2 \pi R_i} \int_{\partial B_{R_i}( s_i )} \phi_{\TAF}(\bx ) \,\dd S,\; s_i \in ( 0.25,0.75 ).
$$
We consider only the parameters $s_i \in ( 0.25,0.75 )$, since we want to avoid a sprouting of new vessels at the boundaries of $\Lambda_i$. Furthermore, boundary edges are not considered, and we demand that the edges should have a minimal length $l_{\min}$ to avoid the formation of tiny vessels. If $\overline{\phi}_{\TAF}^{(\max)}$ is larger than $Th_{\TAF}$, we attach a new vessel $\Lambda_{i^\prime}$ at $\bx$. As in the case of apical growth, the local TAF gradient is considered as the preferred growth direction of the new vessel:
$$
\mathbf{d}_g(\bx ) = \frac{\nabla \phi_{\TAF} (\bx )}{\left\| \nabla \phi_{\TAF} (\bx ) \right\|}.
$$
In order to prevent $\Lambda_{i^\prime}$ from growing in the direction of $\Lambda_{i}$, we demand that $\mathbf{d}_g$ draws an angle of at least $\frac{10}{180} \pi$. The new radius $R_{i^\prime}$ is computed as follows:
$$
\tilde{R}_i = {\zeta} R_i,\;\tilde{R}_{i^\prime} = ( \tilde{R}_i - R_i )^{\frac{1}{\gamma}} = ( {\zeta} - 1 )^{\frac{1}{\gamma}} R_i,\;R_{i^\prime} =
\begin{cases}
R_{i^\prime} \sim \mathcal{U}(1.25 \cdot R_{\min},\tilde{R}_{i^\prime}) \text{ if } 1.25 \cdot R_{\min}<\tilde{R}_{i^\prime} \\
R_{\min} \text{ otherwise.}
\end{cases}
$$
${\zeta}>1$ is a fixed parameter, $R_{\min}$ denotes the minimal radius of a blood vessel, and $\mathcal{U}$ stands for the uniform distribution,
i.e., new segment radius $R_{i^\prime}$ is chosen from the interval $[R_{\min}, \tilde{R}_{i^\prime}]$ based on a uniform distribution. For a given radius $\tilde{R}_{i^\prime}$, the new length $l_{i^\prime}$ of $\Lambda_{i^\prime}$ is determined by means of \eqref{eq:Pb}.

Finally, three new vessels $\Lambda_{i_1}$, $\Lambda_{i_2}$ and $\Lambda_{i^\prime}$ are added to the network $\Lambda_{\text{ap}}$. As in the case of apical growth, we test whether a new vessel intersects an existing vessel, before we incorporate $\Lambda_{i^\prime}$ into $\Lambda_{\text{ap}}$. In addition, we check whether a terminal vessel, i.e., a vessel that is part of $\partial \Lambda_{\text{ap}}$ can be linked to another vessel. For this purpose, the distance of the point $\bx_b \in \partial \Lambda_{\text{ap}} \cup \partial \Lambda_{i}$ to its neighboring network nodes that are not directly linked to $\bx_b$ is computed. If the distance is below a certain threshold $\text{dist}_{\text{link}}$, the corresponding network node is considered as a candidate to be linked with $\bx_b$. If $\bx_b$ is part of an artery or the high pressure region of $\Lambda_{\text{ap}}$, we link it preferably with a candidate at minimal distance and whose pressure is in the low pressure region (venous part). If $\bx_b$ is part of a vein, the roles are switched.

\subsection{Adaption of the vessel radii} In the final phase of the angiogenesis step, we iterate over all the vessels $\Lambda_i \subset \Lambda_{\text{sp}}$ and compute for each vessel the wall shear stress $\mathbf{\tau}_w$ by:
$$
\mathbf{\tau}_{w,i} = \frac{4.0 \;\mu_{\text{bl}}}{\pi R_i^3}\left| Q_i \right|,\;Q_i = - K_{v,i} \frac{R_i^2 \pi \Delta p_{v,i}}{l_i},
$$
where $\Delta p_{v,i}$ is the pressure drop along $\Lambda_i$. By means of the 1D wall shear stress, the wall shear stress stimulus for the vessel adaption is given by \cite{stephanou2006mathematical}:
\begin{equation}\label{eq:SWSS}
S_{\WSS,i} = {\ln}( \mathbf{\tau}_{w,i} + \tau_{\text{ref}} ).
\end{equation}
Here, $\tau_{\text{ref}}$ is a constant that is included to avoid a singular behavior at lower wall shear stresses \cite{pries2001structural}. 
{{Following the model for radius adaptation in \cite{secomb2013angiogenesis}, the change in radius $\Delta R_i$ over a time step $\Delta t$ is assumed to be proportional to the stimulus $S_{\WSS,i}$ and current radius $R_i$:
\begin{equation}\label{eq:DeltaRi}
\Delta R_i = \left( k_{\WSS} \cdot S_{\WSS,i} -k_s \right) \cdot \Delta t \cdot R_i ,
\end{equation}
where $k_{s}$ is a constant that controls the natural shrinking tendency of the blood vessel and $k_{\WSS}$ a proportionality constant that controls the effect of stimulus $S_{\WSS, i}$. Once we have $\Delta R_i$, we can compute the updated radius of vessels using $R_{\new, i} = R_i + \Delta R_i$.}} If 
$$
R_{\new,i} \in \left[ R_{\min}, 1.25 \cdot R_i \right],
$$
{where $R_{\min}$ is some fixed constant,} we update the vessel radius of $\Lambda_i$. Otherwise, if $R_i<R_{\min}$ and $\partial \Lambda_i \cup \partial \Lambda_{\text{sp}} \neq \emptyset$, the vessel is removed from the network. Finally, {after following the procedure discussed in this section,} we obtain a new network $\Lambda_{\new}$.

\section{Numerical Discretization}
\label{Sec:Solver}

\begin{figure}[!htb]
	\begin{center}		
		\includegraphics[width=.82\textwidth]{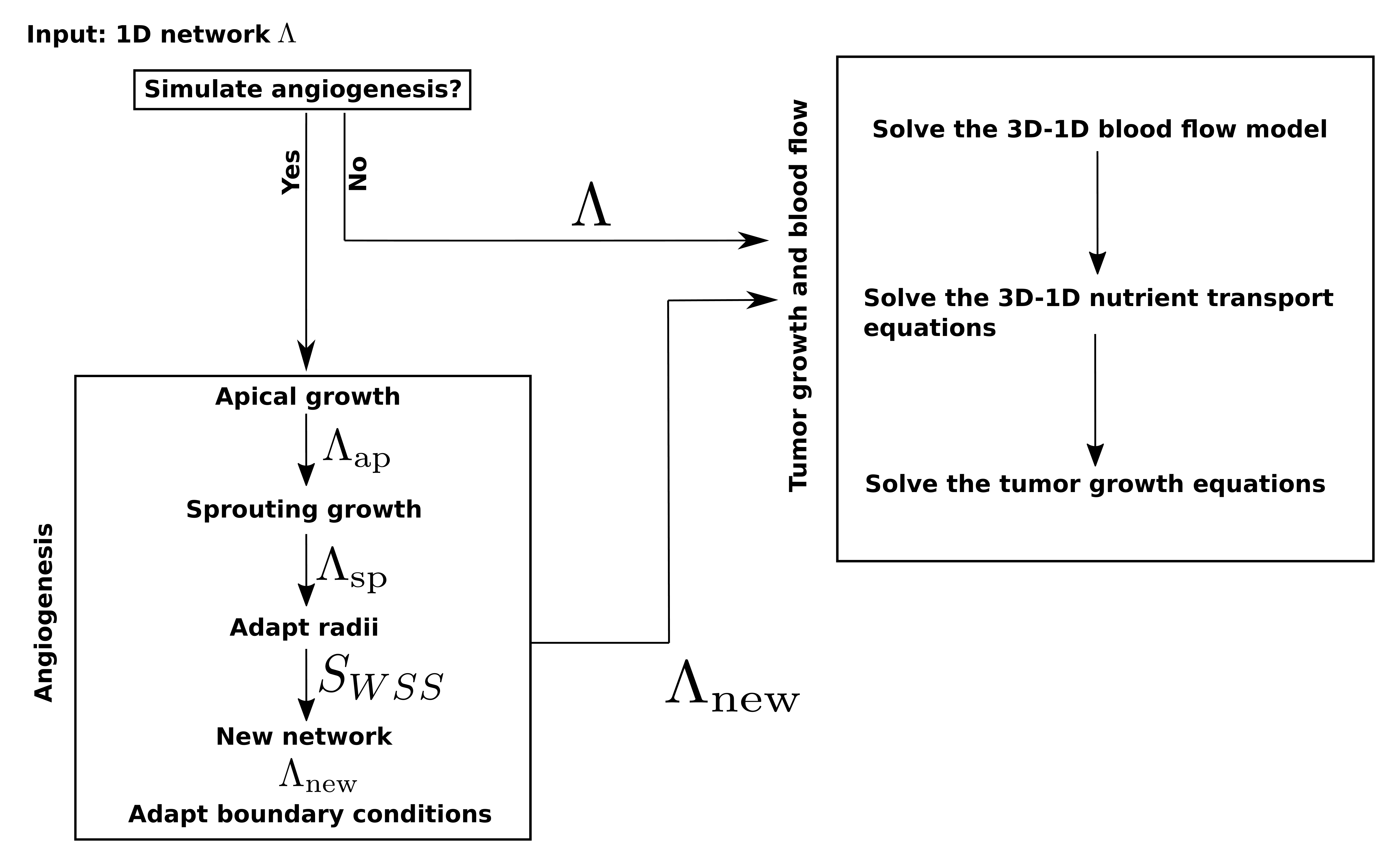}
	\end{center}
	\caption{\label{fig:SchemeSimulation} Simulation steps within a single time step.}
\end{figure}

With our mathematical models for tumor growth, blood flow and nutrient transport as well as angiogenesis processes presented in previous sections, we now turn our attention to numerical solution strategies. Toward this end, let us consider a time step $n$ given by the interval $\left[ t_n, t_{n+1} \right]$, with $\Delta t = t_{n+1}-t_n$. At the beginning of a time step $n$, we decide whether an angiogenesis process has to be simulated or not. As examples of relevant simulations, we consider an angiogenesis process after each third time step. If angiogenesis has to be taken into account, we follow the steps described in \cref{Sec:NetworkGrowth}. Given the 1D network $\Lambda$ at the time point $t_n$, we first apply the algorithm for the apical growth. Afterwards, the sprouting of new vessels and the adaption of the vessel radii is simulated. Finally, we obtain a new network $\Lambda_{\new}$ for the new time point $t_{n+1}$.  If the simulation of angiogenesis is omitted in the respective time step $n$, $\Lambda$ is directly used for the simulation of the tumor growth as well as blood flow and nutrient transport, see \cref{fig:SchemeSimulation}.

For the time discretization of the 3D model equations in \cref{Sec:Derivation}, the semi-implicit Euler method is used i.e. we keep the linear terms implicit and the nonlinear terms explicit with respect to time. Discretizing the model equations in space, standard conforming trilinear $Q1$ finite elements are employed for the partial differential equations governing the tumor growth \eqref{Eq:Model3D}, whereas the PDEs for pressure $( p )$ and nutrient transport $( \phi_\sigma )$ are solved by means of cell centered finite volume methods. The computational mesh is given by a union of cubes having an edge length of $h_{3D}$. 

We use finite elements to approximate the higher-order Cahn--Hilliard type equations as well advection-reaction-diffusion equations corresponding to the species $\phi_{TAF}, \phi_{ECM}, \phi_{MDE}$ in \cref{Eq:Model3D}. In order to ensure mass conservation for both flow and nutrient transport in the interstitial domain, finite volume schemes are taken into account, since they are locally mass conservative. 

In order to solve the 3D-1D coupled system, such as pressure $(p_{v}, p)$, the iterative Block-Gauss-Seidel method is used, i.e., in each iteration, we first solve the equation system for the 1D system. Then the updated 1D solution is used to solve the equation system derived from the 3D problem. We stop the iteration when the change in the current and previous iteration solution is within a small tolerance. At each time step, we first solve the $(p_v, p)$ coupled system. Afterwards the $(\phi_v, \phi_\sigma)$ coupled system is solved. Next, we solve the remaining equations in the 3D system. This is summarized in Algorithm \ref{Alg_AlgImpExp}. In the remainder of this section, the discretizations of the 1D and 3D systems are outlined.

\subsection{VGM discretization of the 1D PDEs}\label{ss:Discret1D}

It remains to specify the numerical solution techniques for the 1D network equations. The time integration is based on the implicit Euler method. For the spatial discretization of the 1D equations, the Vascular Graph Method (VGM) is considered. This method corresponds to a node centered finite volume method \cite{reichold2009vascular,vidotto2019hybrid}. We then briefly describe this numerical method as well as the discretization of the terms arising in the context of the 3D-1D coupling. We restrict ourselves to the pressure equations. 

\begin{figure}[!htb]
	\centering	
		\includegraphics[width=.8\textwidth]{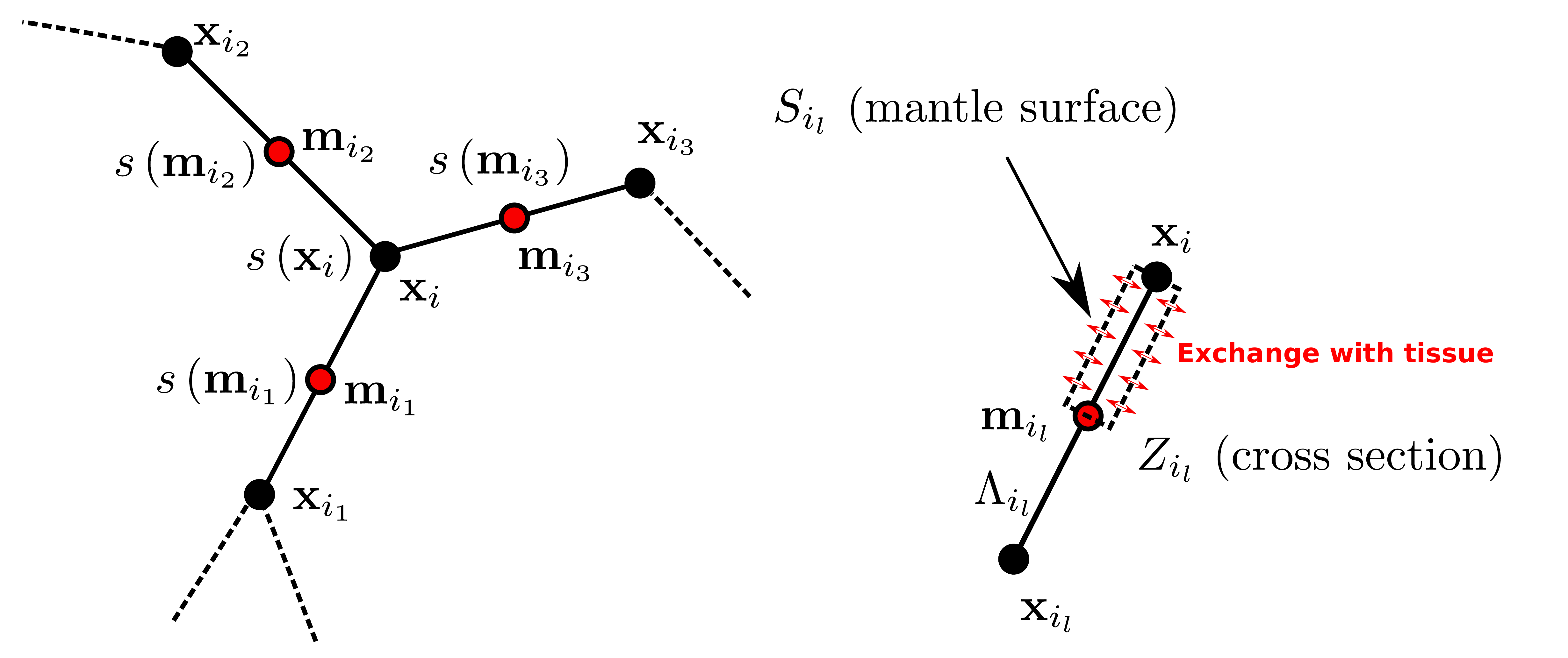}
	\caption{\label{fig:VGM} Notation for the Vascular Graph Method.}
\end{figure}

As mentioned in \cref{Sec:Derivation}, the 1D network is given by a graph-like structure, consisting of edges $\Lambda_{i} \subset \Lambda$ and network nodes $\bx_i \in \Lambda$. In a first step, we assign to each network node $\bx_i$ an unknown for the pressure that is denoted by $p_{v,i}$. Let us assume that the edges containing $\bx_i$ are given by $\Lambda_{i_1},\ldots,\Lambda_{i_N}$ and its midpoints by $\mathbf{m}_{i_1},\ldots,\mathbf{m}_{i_N}$, see \cref{fig:VGM}. 

On each edge, $\Lambda_{k} \in \left\{ \Lambda_{i_1},\ldots,\Lambda_{i_N} \right\}$, we consider the following PDE for the pressure; see also \eqref{Eq:Model1D}. For convenience, the curve parameter is simply denoted by $s$.
$$
-R_k^2 \pi \;\partial_{s} ( K_{v,k} \; \partial_{s} p_v ) = -2 \pi R_k L_p  ( p_v - \overline{p} ).
$$

Next, we establish for the node $\bx_i$ a mass balance equation taking the fluxes across the cylinders $Z_{i_l}$ into account. $Z_{i_l}$ is a cylinder having the edge $\Lambda_{i_l}$ as a rotation axis and the radius $R_{i_l}$. Furthermore its top and bottom facets are located at $\mathbf{m}_{i_l}$ and $\bx_i$, respectively (see Figure \ref{fig:VGM}). The corresponding curve parameters are denoted by $s ( \bx_i  )$ and $s ( \bx_{i_l} ),\;l \in \left\{ 1,\ldots,N\right\}$. Accordingly, the mass balance equation reads as follows:
$$
-\sum_{l=1}^N \int_{s ( \bx_i  ) }^{s ( \mathbf{m}_{i_l} )} R_{i_l}^2 \pi \;\partial_{s} ( K_{v,{i_l}} \; \partial_{s} p_v )\, \dd s =
-2 \pi L_p \sum_{l=1}^N \int_{s ( \bx_i  ) }^{s ( \mathbf{m}_{i_l} ) } R_{i_l} ( p_v - \overline{p} ) \, \dd s.
$$
Integration yields:
$$
-\sum_{l=1}^N R_{i_l}^2 \pi  K_{v,{i_l}} \left. \partial_{s} p_v \right|_{s ( \mathbf{m}_{i_l} )} + \sum_{l=1}^N R_{i_l}^2 \pi  K_{v,{i_l}} \left. \partial_{s} p_v \right|_{s ( \bx_{i} )}= -2 \pi L_p \sum_{l=1}^n  \int_{s ( \bx_i  ) }^{s ( \mathbf{m}_{i_l} )}   {R_{i_l} (p_v - \overline{p})} \, \dd s.
$$

Approximating the derivatives by central finite differences and using the mass conservation equation (see \cref{sec:1Dmodel}):
$$
\sum_{l=1}^N R_{i_l}^2 \pi  K_{v,{i_l}} \left. \partial_{s} p_v \right|_{s ( \bx_i  )} = 0,
$$
at an inner node $\bx_i$, it follows that
$$
\sum_{l=1}^N R_{i_l}^2 \pi  K_{v,{i_l}} \; \frac{p_{v,i}-p_{v,i_l}}{l_{i_l}}= -2 \pi L_p \sum_{l=1}^N \int_{s ( \bx_i  ) }^{s ( \mathbf{m}_{i_l} )}  {R_{i_l} (p_v - \overline{p})} \, \dd s,
$$
where $l_{i_l}$ denotes the length of the edge $\Lambda_{i_l}$. Denoting the mantle surface of $Z_{i_l}$ by $S_{i_l}$, we have:
$$
\sum_{l=1}^N \frac{R_{i_l}^2 \pi  K_{v,{i_l}}}{l_{i_l}} \; ( p_{v,i}-p_{v,i_l} )= -L_p \sum_{l=1}^N \left|  S_{i_l} \right| p_{v,i} + L_p \sum_{l=1}^N \int_{S_{i_l}} p \, \dd S.
$$
Computing the integrals $\int_{S_{i_l}} p \, \dd S$, we introduce the decomposition of $\Omega$ into $M$ finite volume cells $\CV_k$: \linebreak $\Omega = \bigcup_{k=1}^M \CV_k$. The pressure unknown assigned to $\CV_k$ is given by $p_{k}$. Using this notation, one obtains:
$$
\int_{S_{i_l}} p \, \dd S = \sum_{ \CV_k \cap S_{i_l} \neq \emptyset} \int_{ \CV_k \cap S_{i_l} } p \, \dd S 
\approx \left|S_{i_l} \right| \sum_{ \CV_k \cap S_{i_l} \neq \emptyset} \underbrace{\frac{\left| \CV_k \cap S_{i_l} \right|}{\left|S_{i_l} \right|}}_{=: w_{ki_l}} p_{k} = \left|S_{i_l} \right| \sum_{ \CV_k \cap S_{i_l} \neq \emptyset} w_{ki_l} \; p_{k}.
$$
{{
In order to estimate the weights $w_{ki_l}$	we discretize the mantle surface $S_{i_l}$ by $N_s$ nodes. For our simulations, we used $N_s = 400 $ nodes. $S_{i_l}$ intersects some finite volume cells $CV_k$. The number of nodes contained in $CV_k$ is denoted by $N_{ki_l}$. Using these definitions, the weights $w_{ki_l}$ are computed as follows: $w_{ki_l} = N_{ki_l}/N_s$. As an example, consider Figure \ref{fig:discCyl} below, where we show the discretization of the surface of a cylinder.}}
We note that one has to guarantee that the relation $\sum_{ \CV_k \cap S_{i_l} \neq \emptyset}  w_{ki_l} = 1$ holds. Otherwise, a consistent mass exchange between the vascular system and the tissue {could} not be enforced. All in all, we obtain a linear system of equations for computing the pressure values. 
\begin{figure}[h!]
	\centering
	\includegraphics[width=0.4\linewidth]{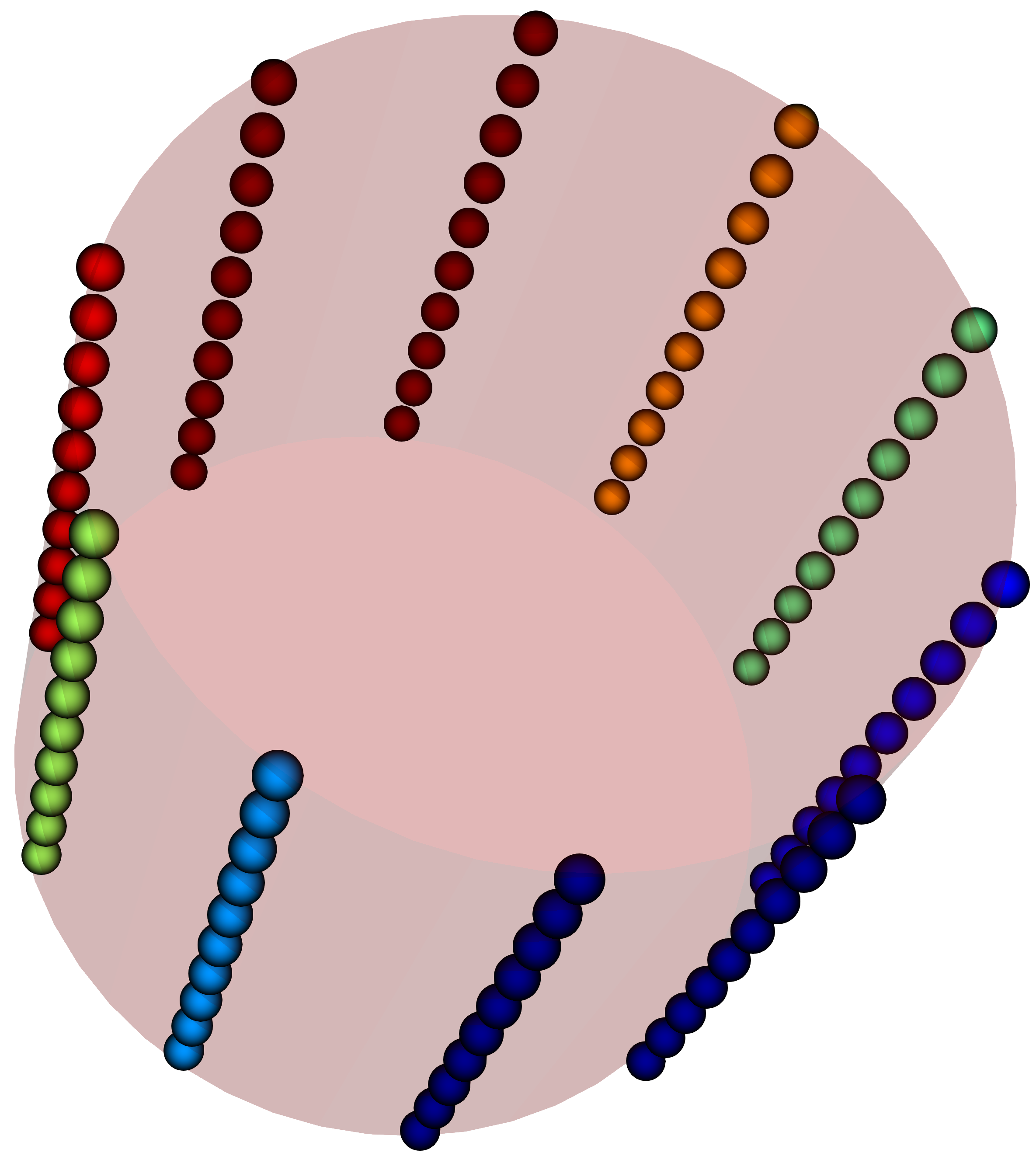}
	\hspace{0.6cm}
	\includegraphics[width=0.4\linewidth]{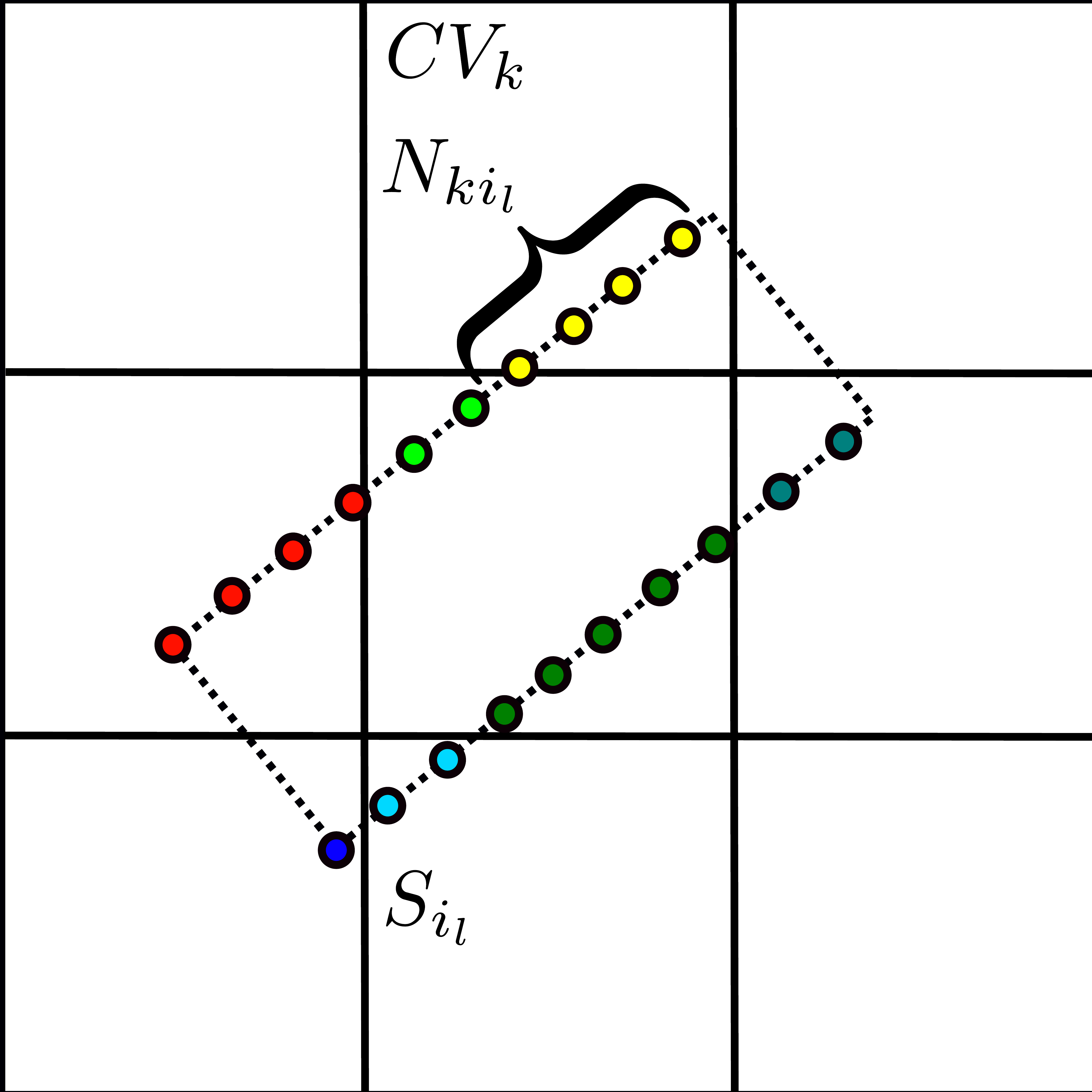}
	\caption{ Typical discretization of the cylinder surface (left). Cross section through a mesh composed of finite volume cells $CV_k$ and a cylinder with the mantle surface $S_{i_l}$. $S_{i_l}$ is discretized by $N_s$ nodes, which are contained in different finite volume cells. Nodes belonging to different cells are colored differently. The number of nodes contained in $CV_k$ is denoted by $N_{ki_l}$.}\label{fig:discCyl}
\end{figure}
On closer examination, it can be noted that {the corresponding} matrix is composed of four blocks, i.e., two coupling blocks as wells as a block for the 1D diffusion term and the 3D diffusion term. As said earlier, at each time step, we decouple the 1D and 3D pressure equations and use a Block-Gauss-Seidel iteration to solve the two systems until the 3D pressure is converged. The discretization of the nutrient equation is exerted in a similar manner, where the main difference consists in adding an upwinding procedure for the convective term. At each time step, the nutrient equations are also solved using a Block-Gauss-Seidel iteration.

\subsubsection{Initial and boundary conditions for the 1D PDEs}
Since we use a transient PDE to simulate the transport of nutrients, we require an initial condition for the variable $\phi_v$. In doing so, a threshold $R_T$ for the radii is introduced in order to distinguish between arteries and veins. If the radius of a certain vessel is below $R_T$, the vessel is considered as an artery and otherwise as a vein. In case of an artery, we set $\phi_v(t = 0 ) = 1$ and in case of a vein $\phi_v(t = 0 ) = 0$ is used. When the network starts to grow, initial values for the newly created vessels have to be provided. If a new vessel is created due to sprouting growth, we consider the vessel or edge to which the new vessel is attached. At the point of the given vessel, where the new vessel or edge is added, $\phi_v$ is interpolated linearly. For this purpose the two values of $\phi_v$ located at the nodes of the existing vessel are used. The interpolated value is assigned to both nodes of the newly created vessel. 

When apical growth takes {place, a} new vessel is added to $\bx \in \partial\Lambda$. In this case, we consider $\phi_v(\bx,t )$ for a time point $t$ and assign it to the newly created node, since we assume that no flow boundary conditions are enforced. With respect to the boundary conditions for the 1D pressure PDE, the following distinction of cases is made:
\ \\
\begin{itemize}[leftmargin=.18in]
	\item \textbf{Dirichlet boundary for $p_v$ if $\bx \in \partial\Lambda \cap \partial \Omega$.} In this case, we set: $p_v(\bx) = p_{v,D}(\bx)$, where $p_{v,D}$ is a given Dirichlet value at $\bx$. Numerically, we enforce this boundary condition by setting in the corresponding line of the matrix all entries to zero except for the entry on the diagonal which is fixed to the value $1$. Additionally, the corresponding component of the right hand side vector contains the Dirichlet value $p_D$. Let us assume that $\by$ is the neighbor of $\bx$ on the edge $\Lambda_1$. The other edges adjacent to $\by$ are denoted by $\Lambda_2,\ldots,\Lambda_N$. Then the balance equation for $\by$ has to be adapted as follows to account for the Dirichlet boundary condition $p_{v,D}$:
    \begin{align*}    	
     &\frac{R_{1}^2 \pi  K_{v,{1}}}{l_{1}} \; ( p_v\left( \by \right)-p_{v,D} ) + \sum_{j=2}^N \frac{R_{j}^2 \pi  K_{v,{j}}}{l_{j}} \; ( p_v\left( \by \right)-p_{v,j} )  \\
      &\quad =- L_p \left|  \tilde{S}_1 \right| p_v\left( \by \right) + L_p \sum_{ \CV_k \cap \tilde{S}_1 \neq \emptyset} w_{k1} \; p_{k} -L_p \sum_{j=2}^N \left|  S_j \right| p_v\left( \by \right) + L_p \left|S_{j} \right| \sum_{ \CV_k \cap S_{j} \neq \emptyset, j>1} w_{kj} \; p_{k},
    \end{align*} 
    where $\tilde{S}_1$ is the mantle surface of the cylinder covering the whole edge $\Lambda_1$.
	\ \\
	\item \textbf{Homogeneous Neumann boundary for $p_v$ if $\bx \in \partial\Lambda \cap \Omega$.} Let $\bx \in \partial \Lambda_i \cap \partial\Lambda \cap \Omega$, then we set $\left. -R_i^2 \pi \;K_{v,i} \partial_{s} p_v \right|_{\bx}=0$, resulting in the following discretization:
	$$
	R_i^2 \pi \;K_{v,i} \frac{p_v( \bx ) - p_v( \by )}{l_i} = -L_p   \left| S_{i_1} \right| p_{v,i} +  L_p \left|S_{i_1} \right| \sum_{ \CV_k \cap S_{i_1} \neq \emptyset} w_{ki_l} \; p_{k},
	$$ 
	where $\by \in \partial \Lambda_i \cap \Lambda$ and $l_i$ is the length of the edge $\Lambda_i$.
\end{itemize}
\ \\
Summarizing, we consider for the pressure in the network Dirichlet boundaries at the boundary of the 3D domain $\Omega$ and homogeneous Neumann boundary conditions in the inner part of $\Omega$. For the nutrients, the implementation of boundary conditions is  more challenging, since an upwinding procedure has to be taken into account.
\ \\
\begin{itemize}[leftmargin=.18in]
	\item \textbf{Dirichlet boundary for $\phi_v$ if $\bx \in \partial \Lambda_i \cap \partial\Lambda \cap \partial \Omega$ and }
	$$
	v_v( \bx )\approx - K_{v,i} \frac{p_v\left( \by \right) - p_{v,D}(\bx)}{l_i} > 0.
	$$
	In this case, we set: $\phi_v(\bx,t) =  \phi_{v,D}(\bx)$, where $\phi_{v,D}$ is a given Dirichlet value at $\bx$. The numerical implementation can be exerted analogously to the case of the pressure $p_v$.
	\ \\
	\item \textbf{Homogeneous Neumann boundary for $\phi_v$ if $\bx \in \partial\Lambda \cap \Omega$.} Let $\bx \in \partial \Lambda_i \cap \partial\Lambda \cap \Omega$ and we set 
	$$ 
	( \left.v_v \phi_v - D_v \partial_{s_i} \phi_v ) \right|_{\bx} =0,
	$$
	resulting in the following discretization:
	\begin{align*}
	&\frac{l_i}{2} \frac{\phi_v( \bx,t+\Delta t ) - \phi_v( \bx,t )}{\Delta t} + \left.  v_v \phi_v\right|_{\mathbf{m}_i} - D_v \frac{\phi_v( \by,t+\Delta t )-\phi_v( \bx,t+\Delta t )}{l_i} \\
	&\quad =-2\pi R_i \left[ (1-r_\sigma ) \int_{s( \bx )}^{s( \mathbf{m}_i )} J_{pv} ( \ov p,p_v )\cdot \phi_\sigma^v( s,t+\Delta t )  \;\dd s + L_\sigma \int_{s( \bx )}^{s( \mathbf{m}_i )} \phi_v( s,t+\Delta t )-\overline{\phi}_\sigma( s,t+\Delta t ) \, \dd s\right],
	\end{align*}
	where $\by \in \partial \Lambda_i \cap \Lambda$. The integrals modeling the exchange terms are discretized as in the case of the pressure equations.
	\ \\
    \item \textbf{Upwinding boundary for $\phi_v$ if $\bx \in \partial\Lambda \cap \partial \Lambda_i \cap \partial \Omega$ and }
    $$
    v_v( \mathbf{m}_i ) \approx - K_{v,i} \frac{p_v\left( \by \right) - p_v\left( \mathbf{x} \right)}{l_i} \leq 0.
    $$ 
    Here, we obtain the following semi-discrete equation:
    \begin{align*}
    &\frac{l_i}{2} \frac{\phi_v( \bx,t+\Delta t ) - \phi_v( \bx,t )}{\Delta t} + \left.  v_v \right|_{\mathbf{m}_i} \phi_v( \by,t+\Delta t ) -\left.  v_v \right|_{\mathbf{m}_i}\phi_v( \bx,t+\Delta t ) \\
     &\quad =-2\pi R_i \left[ (1-r_\sigma ) \int_{s( \bx )}^{s( \mathbf{m}_i )} J_{pv} ( \ov p,p_v ) \cdot \phi_\sigma^v( s,t+\Delta t ) \;\dd s + L_\sigma \int_{s( \bx )}^{s( \mathbf{m}_i )} \phi_v( s,t+\Delta t )-\overline{\phi}_\sigma( s,t+\Delta t ) \, \dd s\right],
    \end{align*}
    where $\by \in \partial \Lambda_i \cap \Lambda$.
\end{itemize}
\subsection{Discretization of the 3D PDEs}\label{ss:Discret3D}
Suppose that $\phi_{\alpha_{n}}, \mu_{\alpha_{n}}, p_{n}, p_{v_{n}}, \bv_n$ denote the various fields at time $t_n$. Let $V_h$ be a subspace of $H^1(\Omega, \bbR)$ consisting of continuous piecewise trilinear functions on a uniform mesh $\Omega_h$. We consider $\phi_{\alpha_{n}} \in V_h$ for $\alpha \in \{P, H, N, TAF, ECM, MDE\}$, $\mu_{P_n}, \mu_{H_n} \in V_h$, and $\bv_h \in [V_h]^3$. The test functions are denoted by $\phit\in V_h$ for species in $V_h$,  $\mut\in V_h$ for chemical potentials $\mu_P, \mu_H$, and $\tilde{\bv}\in [V_h]^3$ for velocity. 

Given a time step $n$ and solutions $\phi_{\alpha_{n}}, \mu_{\alpha_{n}}, p_{n}, p_{v_{n}}, \bv_n$, we are interested in the solution at the next time step. For the 3D-1D coupled pressure $(p_{v_{n+1}}, p_{n+1})$, as mentioned earlier, we utilize a block Gauss-Seidel iteration, where the discretization of the 1D equation is discussed in \cref{ss:Discret1D} and discretization of the 3D equation using finite-volume scheme is provided in  \cref{eq:formPresExp}. Similarly, $(\phi_{v_{n+1}}, \phi_{\sigma_{n+1}})$ is solved using a block Gauss-Seidel iteration with the discretization of the 1D  equation along the lines of the discretization of the 1D pressure equation and discretization of the 3D equation provided in \cref{eq:formNutExp}. We then solve the proliferative, hypoxic, necrotic, MDE, ECM, and TAF systems sequentially. Once we have pressure $p_{n+1}$, we compute the velocity $\bv_{n+1} \in [V_h]^3$ using the weak form:
\begin{equation}\label{eq:velDiscret}
	(\bv_{n+1},\tilde{\bv}) = (- K (\nabla p_{n+1} - S_{p_n}),\tilde{\bv}), \qquad \forall \tilde{\bv} \in [V_h]^3, 
\end{equation}
where $S_{p_{n}} = S_{p}(\bphi_{n}, \mu_{P_n}, \mu_{H_n})$, see \cref{Eq:DerivationSourceCH}. For the advection terms, using the fact that $\nabla p \cdot \bn = 0$ on $\partial \Omega$ and so $\bv_{n+1} \cdot\bn = 0$ on $\partial \Omega$, we can write
\begin{align}\label{eq:advect}
\weakDot{\nabla \cdot (\phi_\alpha \bv_{n+1})}{\phit} = -\weakDot{\phi_\alpha \bv_{n+1}}{\nabla \phit}, \qquad \forall \phit \in V_h, \; \forall \alpha \in \{P,H,TAF,MDE\}.
\end{align}
In what follows, we consider the expression on the right hand side in above equation for the advection terms. 

For fields $\phi_a$, $a\in \{P,H,N,\sigma, TAF, ECM, MDE\}$, and chemical potentials $\mu_P, \mu_H$, we assume homogeneous Neumann boundary condition on $\partial \Omega$. Next, we describe the discretization of the {scalar fields in the 3D model}.

\begin{itemize}[leftmargin=.18in]
	
\item \textbf{Pressure.} Let $\CV \in \Omega_h$ denote the typical finite volume cell and $\sigma \in \partial \CV$  face of a cell $\CV$. Let $(p_{v_{n+1}}^k, p_{{n+1}}^k)$ denote the pressures at $k^{\text{th}}$ iteration and time $t_{n+1}$. Suppose we have solved for $p_{v_{n+1}}^\knew$ following \cref{ss:Discret1D}.  To solve $p_{n+1}^\knew$, we consider, for all $\CV \in \Omega_h$,
\begin{align}\label{eq:formPresExp}
	-\surfIntSum{K\nabla p_{n+1}^\knew} &= -\surfIntSum{KS_{p_{n}}} + \int_{\Gamma \cap \CV} J_{pv}(p_{n+1}^\knew, \Pi_{\Gamma}p_{v_{n+1}}^\knewO) \,\dd S \notag \\ 
	&= -\surfIntSum{KS_{p_{n}}} + \sum_{i=1}^N \int_{\Gamma_i \cap \CV} J_{pv}(p_{n+1}^\knew, \Pi_{\Gamma_i}p_{v_{n+1}}^\knewO)  \,\dd S.
\end{align}
Above follows by the integration of the pressure equation in \cref{Eq:Model3D} over $\CV$ and using the divergence theorem. Here $J_{pv}(p, p_v) = L_p(p_v - p)$, $\Gamma = \cup_{i=1}^N \Gamma_i$ is the total vascular surface, and $\Pi_{\Gamma_i}(p_v)$ is the projection of the 1D pressure defined on the centerline $\Lambda_i$ onto the surface of the cylinder, $\Gamma_i$. 
\\
\item \textbf{Nutrients.} Suppose we have solved for $\phi_{v_{n+1}}^\knew$. To solve $\phi_{{\sigma}_{n+1}}^\knew$, we consider, for all $\CV$,
\begin{align}\label{eq:formNutExp}
&\elemInt{\frac{\phi_{\sigma_{n+1}}^\knew - \phi_{\sigma_n}}{\Delta t}} + 
\surfIntSum{\phi_{\sigma_{n+1}}^\knew \hat{\bv}_{n+1}}  - \surfIntSum{m_\sigma(\phib_{n}) \left( D_\sigma \nabla \phi_{\sigma_{n+1}}^\knew \right)} \notag \\
&\quad + \elemInt{\lambdap_P \phi_{P_{n}} \phi_{\sigma_{n+1}}^\knew}  + \elemInt{\lambdap_{H} \phi_{H_{n}} \phi_{\sigma_{n+1}}^\knew} \notag \\
&\quad + \elemInt{\lambdap_{ECM} (1- {\ecm}_{n}) \H({\ecm}_{n} - \phi^{\textmd{pro}}_{\ECM})  \phi_{\sigma_{n+1}}^\knew} \notag \\
&= \elemInt{\lambdad_{P}\phi_{P_{n}} + \lambdad_{H}\phi_{H_{n}}} 
+ \elemInt{\lambdad_{\ECM} {\ecm}_{n} {\mde}_{n}} \notag \\
&  -\surfIntSum{m_\sigma(\phib_{n}) \chi_c \nabla \left(\phi_{P_{n}} + \phi_{H_{n}} \right)} 
 + \sum_{i=1}^N \int_{\Gamma_i \cap \CV} J_{\sigma v}(\phi_{\sigma_{n+1}}^\knew, p_{n+1}^\knewO, \Pi_{\Gamma_i} \phi_{{v}_{n+1}}^\knewO, \Pi_{\Gamma_i}p_{v_{n+1}}^\knewO) \, \dd S,
\end{align}
where $J_{\sigma v}$ is given by \cref{eq:KedemKatalchsky}. Noting that the velocity is $$\hat{\bv}_{n+1} = -K \nabla p_{n+1} + K S_{p_n},$$
we divide the advection term, for $\sigma \in \partial \CV$, into two parts:
\begin{align}\label{eq:nutAdvect}
\surfInt{\phi_{\sigma_{n+1}}^\knew \hat{\bv}_{n+1}} &= -K\surfInt{\phi_{\sigma_{n+1}}^\knew \nabla p_{n+1}} + K\surfInt{\phi_{\sigma_{n+1}}^\knew S_{p_n}}.
\end{align}
For the first term we apply the upwinding scheme. For the second term, we perform quadrature approximation to compute the integral over the face $\sigma$. 
\\
\item \textbf{Proliferative.}  For a general double-well potential $\Psi(\phi_P,\phi_H,\phi_N)=\sum_{a\in \{T,P,H\}}C_{\Psi_a} \phi_a^2(1-\phi_a)^2$, we consider the convex-concave splitting, see \cite{eyre1998unconditionally}, as follows
\begin{align}\label{eq:splitDWExp}
\Psi(\phi_P,\phi_H,\phi_N) = \sum_{a\in \{T,P,H\}}\frac{3}{2}C_{\Psi_a} \phi_a^2 + \sum_{a\in \{T,P,H\}}C_{\Psi_a}(\phi_a^4 - 2\phi_a^3 - \frac{1}{2}\phi_a^2).
\end{align}
This results in
\begin{align}\label{eq:derSplitDWExp}
\p_{\phi_P} \Psi(\phi_P,\phi_H,\phi_N) = \sum_{a\in \{T,P\}} 3C_{\Psi_a} \phi_a +  \sum_{a\in \{T,P\}} C_{\Psi_a} \phi_a (4\phi_a^2 - 6\phi_a - 1).
\end{align}
The expression for $\p_{\phi_H} \Psi$ can be derived analogously. In our implementation, $\phi_P, \phi_H, \phi_N$ are  the main state variables and $\phi_T$ is computed using $\phi_T = \phi_P + \phi_H + \phi_N$. Let the mobility $\bar{m}_{P_n}$ at the current step be given by
\begin{align}
\bar{m}_{P_n} = M_P \left[ (\phi_{P_{n}})^+ (1 - \phi_{T_{n}})^+ \right]^2,
\end{align}
where for a field $f$, $\left( f \right)^+$ is the projection onto $[0,1]$ given by
\begin{align}\label{eq:ProjectFn}
\left( f \right)^+ &= \begin{cases}
f \qquad \text{if } f \in [0,1], \\
0 \qquad \text{if } f\leq 0, \\
1 \qquad \text{if } f \geq 1.
\end{cases}
\end{align}

We solve for $\phi_{P_{n+1}}, \mu_{P_{n+1}}$ using the weak forms below
\begin{align}\label{eq:formProlificExp}
&\weakDot{\frac{\phi_{P_{n+1}} - \phi_{P_n}}{\Delta t}}{\phit} 
- \weakDot{\phi_{P_{n+1}} \bv_{n+1}}{\nabla\phit} 
+ \weakDot{\bar{m}_{P_n}\nabla \mu_{P_{n+1}}}{\nabla \phit} \notag \\
&\quad - \weakDot{\lambdap_P \phi_{\sigma_{n+1}} (1 - \phi_{T_{n}})^+ \phi_{P_{n+1}}}{\phit} + \weakDot{\lambdad_P\phi_{P_{n+1}}}{\phit} \notag \\
&= \weakDot{\lambda_{HP}\H(\phi_{\sigma_{n+1}} - \sigma_{HP}) \plusOp{\phi_{H_{n}}}}{\phit} - \weakDot{\lambda_{P\!H}  \H(\sigma_{P\!H} - \phi_{\sigma_{n+1}})\plusOp{\phi_{P_{n}}}}{\phit}\notag \\
&\quad + \frac{1}{\Delta t} \weakDot{ G_{P_n}\int_{t_n}^{t_{n+1}}  \dd {W}_P}{\phit}
\end{align}
and 
\begin{align}\label{eq:formProlificMuExp}
&\weakDot{\mu_{P_{n+1}}}{\mut} - \weakDot{3(C_{\Psi_T} + C_{\Psi_P})\phi_{P_{n+1}}}{\mut} -\weakDot{\epsilon_P^2 \nabla \phi_{P_{n+1}}}{\nabla \mut} \notag \\
&= \weakDot{C_{\Psi_T} \phi_{T_n}(4\phi_{T_n}^2 - 6\phi_{T_n} - 1)}{\mut} + \weakDot{C_{\Psi_P} \phi_{P_n}(4\phi_{P_n}^2 - 6\phi_{P_n} - 1)}{\mut} 
\notag \\
&\quad + \weakDot{3C_{\Psi_T}(\phi_{H_{n}} + \phi_{N_{n}})}{\mut} - \weakDot{\chi_c \phi_{\sigma_{n+1}} + \chi_h {\ecm}_{n}}{\mut},
\end{align}
where $(\cdot)^+$ is the projection to $[0,1]$ defined in \cref{eq:ProjectFn}, $ G_{P_n} = G_P(\phi_{P_n}, \phi_{H_n}, \phi_{N_n})$ is given by \cref{eq:WeinerProc}, and $W_P$ is the cylindrical Wiener process. We discuss the computation of stochastic term in more detail in \cref{ss:StochDisc}. 
\\
\item \textbf{Hypoxic.} Let the mobility $\bar{m}_{H_n}$ be given by
\begin{align}
\bar{m}_{H_n} = M_H \left[ (\phi_{H_{n}})^+ (1 - \phi_{T_{n}})^+ \right]^2.
\end{align}
To solve for $\phi_{H_{n+1}}, \mu_{H_{n+1}}$, we consider
\begin{align}\label{eq:formHypoxicExp}
&\weakDot{\frac{\phi_{H_{n+1}} - \phi_{H_n}}{\Delta t}}{\phit} 
- \weakDot{\phi_{H_{n+1}} \bv_{n+1}}{\nabla\phit} 
+ \weakDot{\bar{m}_{H_n}\nabla \mu_{H_{n+1}}}{\nabla \phit} \notag \\
&\quad - \weakDot{\lambdap_{H} \phi_{\sigma_{n+1}} (1 - \phi_{T_{n}})^+ \phi_{H_{n+1}}}{\phit} + \weakDot{\lambdad_H\phi_{H_{n+1}}}{\phit} \notag \\
&= \weakDot{\lambda_{PH}\H(\sigma_{PH} - \phi_{\sigma_{n+1}}) \plusOp{\phi_{P_{n}}}}{\phit} - \weakDot{\lambda_{H\!P}  \H(\phi_{\sigma_{n+1}} - \sigma_{H\!P})\plusOp{\phi_{H_{n}}}}{\phit} \notag \\
&\quad -\weakDot{\lambda_{H\!N} \H(\sigma_{H\!N} - \phi_{\sigma_{n+1}}) \plusOp{\phi_{H_{n}}}}{\phit} + \frac{1}{\Delta t} \weakDot{ G_{H_n}\int_{t_n}^{t_{n+1}}  \dd {W}_H}{\phit}
\end{align}
and 
\begin{align}\label{eq:formHypoxicMuExp}
&\weakDot{\mu_{H_{n+1}}}{\mut} - \weakDot{3(C_{\Psi_T} + C_{\Psi_H})\phi_{H_{n+1}}}{\mut} -\weakDot{\epsilon_H^2 \nabla \phi_{H_{n+1}}}{\nabla \mut} \notag \\
&= \weakDot{C_{\Psi_T} \phi_{T_n}(4\phi_{T_n}^2 - 6\phi_{T_n} - 1)}{\mut} + \weakDot{C_{\Psi_H} \phi_{H_n}(4\phi_{H_n}^2 - 6\phi_{H_n} - 1)}{\mut} \notag \\
&\quad + \weakDot{3C_{\Psi_T}(\phi_{P_{n}} + \phi_{N_{n}})}{\mut} - \weakDot{\chi_c \phi_{\sigma_{n+1}} + \chi_h {\ecm}_{n}}{\mut},
\end{align}
where $ G_{H_n} = G_H(\phi_{P_n}, \phi_{H_n}, \phi_{N_n})$ is given by \cref{eq:WeinerProc} and $W_H$ is the cylindrical Wiener process.
\\
\item \textbf{Necrotic.} 
\begin{align}\label{eq:formNecroticExp}
\weakDot{\frac{\phi_{N_{n+1}} - \phi_{N_n}}{\Delta t}}{\phit} = \weakDot{\lambda_{HN}\H(\sigma_{HN} - \phi_{\sigma_{n+1}}) \plusOp{\phi_{H_{n}}}}{\phit}.
\end{align}
\item \textbf{MDE.} 
\begin{align}\label{eq:formMDEExp}
&\weakDot{\frac{{\mde}_{n+1} - {\mde}_{n}}{\Delta t}}{\phit} 
- \weakDot{{\mde}_{n+1} \bv_{n+1}}{\nabla\phit}  
+ \weakDot{m_{\MDE}(\phib_{n}) D_{\MDE}\nabla {\mde}_{n+1}}{\nabla \phit}  \notag \\
&\quad + \weakDot{\lambdad_{\MDE}{\mde}_{n+1}}{\phit} \notag %
+ \weakDot{\lambdap_{\MDE}(\phi_{P_{n}} + \phi_{H_{n}}) {\ecm}_{n} \frac{\sigma_{H\!P}}{\sigma_{H\!P} + \phi_{\sigma_{n+1}}} {\mde}_{n+1}}{\phit}  \notag \\
&= \weakDot{\lambdap_{\MDE}(\phi_{P_{n}} + \phi_{H_{n}}) {\ecm}_{n} \frac{\sigma_{H\!P}}{\sigma_{H\!P} + \phi_{\sigma_{n+1}}}}{\phit}
 \notag \\ &\quad 
 - \weakDot{\lambdad_{ECM} {\ecm}_{n} {\mde}_{n}}{\phit} .
\end{align}
\item \textbf{ECM.}
\begin{align}\label{eq:formECMExp}
\weakDot{\frac{{\ecm}_{n+1} - {\ecm}_{n}}{\Delta t}}{\phit} &
= \weakDot{\lambdap_{ECM} \phi_{\sigma_{n+1}} \H({\ecm}_{n} 
  - \phi^\text{pro}_{\ECM}) (1 - {\ecm}_{n})}{\phit}  
- \weakDot{\lambdad_{ECM} {\mde}_{n} . {\ecm}_{n}}{\phit} 
\end{align}
\item \textbf{TAF.} 
\begin{align}\label{eq:formTAFExp}
&\weakDot{\frac{{\taf}_{n+1} - {\taf}_{n}}{\Delta t}}{\phit} 
- \weakDot{{\taf}_{n+1} \bv_{n+1}}{\nabla\phit}  
+ \weakDot{D_{TAF}\nabla {\taf}_{n+1}}{\nabla \phit} \notag \\
&\quad + \weakDot{\lambdap_{\TAF} \phi_{H_{n+1}} {\taf}_{n+1} \H(\phi_{H_{n+1}} - \phi_{H_{P}})}{\phit} 
\notag \\&
= \weakDot{\lambdap_{\TAF} \phi_{H_{n+1}}\H(\phi_{H_{n+1}} - \phi_{H_{P}})}{\phit}
- \weakDot{\lambdad_{\TAF} \phi_{\TAF_n}}{\phit}.
\end{align}
\end{itemize}
\ \\ \\
The steps followed in solving the coupled system of equations including the angiogenesis step are summarized in Algorithm \ref{Alg_AlgImpExp}. 
\ \\ \\
\begin{algorithm}[H] 
	\SetAlgoLined
	\caption{The 3D-1D tumor growth model solver with angiogenesis step} \label{Alg_AlgImpExp} 
	\textbf{Input}:Model parameters, $\phi_{\alpha_0}, v_0, \Delta t, T, \text{TOL}$ for $\alpha \in \A:=\{P,H,N,\sigma, \TAF,\MDE,\ECM\}$\\
	\textbf{Output}: $\phi_{\alpha_n}, \mu_{P_n}, \mu_{H_n}, p_n, \bv_n, p_{v_n}, \phi_{{\sigma v}_b}$ for all $n$  \\
	$n=0$, $t=0$\\
	\While{$t\leq T$ }{ 
		$\phi_{\alpha_{n}} = \phi_{\alpha_{n+1}}$ $\forall \alpha \in \A$, 
		$\mu_{P_n} = \mu_{P_{n+1}}$, $\mu_{H_n} = \mu_{H_{n+1}}$, 
		$p_{v_n} = p_{v_{n+1}}$, $\phi_{{v}_n} = \phi_{{v}_{n+1}}$\\		
		\If{$\text{apply\_angiogenesis(n)} == \text{True}$} {
			\textbf{apply} angiogensis model described in \cref{Sec:NetworkGrowth}\\
			\textbf{update} 1D systems if the network is changed
		}
		
		\textbf{solve} coupled linear system $(p_{v_{n+1}}, p_{n+1})$ using block Gauss-Seidel iteration and \cref{ss:Discret1D} \\
		\textbf{solve} coupled linear system $(\phi_{v_{n+1}}, \phi_{\sigma_{n+1}})$  using block Gauss-Seidel iteration and \cref{ss:Discret1D} \\
		\textbf{solve} velocity $\bv_{n+1}$ using \cref{eq:velDiscret} \\		
		\textbf{solve} $\left( \phi_{P_{n+1}}, \mu_{P_{n+1}}\right)$ using the semi-implicit scheme in \cref{eq:formProlificExp} and \cref{eq:formProlificMuExp}  \\
		\textbf{solve} $\left( \phi_{H_{n+1}}, \mu_{H_{n+1}}\right)$  using the semi-implicit scheme in  \cref{eq:formHypoxicExp}  and \cref{eq:formHypoxicMuExp}\\
		\textbf{solve} $\phi_{N_{n+1}}$  using the semi-implicit scheme in  \cref{eq:formNecroticExp}  \\
		\textbf{solve} ${\mde}_{n+1}$  using the semi-implicit scheme in  \cref{eq:formMDEExp}  \\
		\textbf{solve} ${\ecm}_{n+1}$  using the semi-implicit scheme in  \cref{eq:formECMExp}  \\
		\textbf{solve} ${\taf}_{n+1}$  using the semi-implicit scheme in  \cref{eq:formTAFExp}  \\
		$n \mapsto n+1$, $t \mapsto t+\Delta t$} 
\end{algorithm}

{{
\begin{remark}
	If we ignore the advection and reaction terms of the given system and set $\chi_c = 0$ we can show that our algorithm is unconditionally gradient stable.

	This is due to the fact that if we freeze the field $\phi_{H_{n}}$, 
	in both the convex and concave part of our double-well potential and solve Eq.~\ref{eq:formProlificExp} for $\phi_{P_{n+1}}$, then due to the given convex-concave splitting we get 
	\[ 
		\mathcal E(\phi_{\sigma_{n+1}}, \phi_{P_{n+1}}, \phi_{H_{n}}, \phi_{N_{n}}, \phi_{MDE_{n}}, \phi_{ECM_{n}}, \phi_{TAF_{n}} ) - \mathcal E(\phi_{\sigma_{n+1}}, \phi_{P_{n}}, \phi_{H_{n}}, \phi_{N_{n}}, \phi_{MDE_{n}}, \phi_{ECM_{n}}, \phi_{TAF_{n}} ) \leq 0 .
    \]
	Similarly, if we now freeze $\phi_{P_{n+1}}$ in both parts of the potential and solve \ref{eq:formHypoxicExp} for $\phi_{H_{n+1}}$ we get from the unconditional gradient stability of the sub scheme that
	\[
		\mathcal E(\phi_{\sigma_{n+1}}, \phi_{P_{n+1}}, \phi_{H_{n+1}}, \phi_{N_{n}}, \phi_{MDE_{n}}, \phi_{ECM_{n}}, \phi_{TAF_{n}} ) - \mathcal E(\phi_{\sigma_{n+1}}, \phi_{P_{n+1}}, \phi_{H_{n}}, \phi_{N_{n}}, \phi_{MDE_{n}}, \phi_{ECM_{n}}, \phi_{TAF_{n}} ) \leq 0 .
	\]
	This observation extends to arbitrarily large systems of Cahn-Hilliard equations 
	and since Eqs.~\eqref{eq:formNutExp}, \eqref{eq:formMDEExp}, \eqref{eq:formTAFExp} can be considered as very simple Cahn-Hilliard equations it also extends to them.
	Finally we note that $\phi_{N_{n+1}} = \phi_{N_{n}}$ and $\phi_{ECM_{n+1}} = \phi_{ECM_{n}}$ holds trivially without source terms.
	Hence, using a telescope sum over all the energy decrements due to solving Eqs.~\eqref{eq:formNutExp}, \eqref{eq:formHypoxicExp}, \eqref{eq:formProlificExp}, \eqref{eq:formNecroticExp}, \eqref{eq:formMDEExp}, \eqref{eq:formECMExp} and \eqref{eq:formTAFExp} we get
	\[
		\mathcal E(\mathbf \phib_{n+1})
		- 
		\mathcal E(\mathbf \phib_{n})
		\leq 0
	\]
	independent of our time-step size, which provides a strong motivation for the stability of our algorithm.

	With the stochastic terms we can generate tumor mass and hence $\mathcal E$ does not have to decrease.
	And even though the reaction terms all add up to zero, this does not necessarily mean that $\mathcal E$ has to decrease, since they are not part of our gradient-flow.
	For arbitrary initial states we therefore cannot expect that $\frac{d}{dt}\mathcal E \leq 0$ holds even for our continuous system.
\end{remark}
}}

\subsubsection{Stochastic component of the system}\label{ss:StochDisc}
Generally, the cylindrical Wiener processes $W_\alpha$, $\alpha \in \{P,H\}$, on $L^2(\Omega)$ with $\Omega=(0,2)^3$ can be written as
$$W_\alpha(t)(\bx) = \sum_{i,j,k=1}^\infty \eta^\alpha_{ijk}(t) \underbrace{\cos(i\pi x_1/L) \cos(j\pi x_2/L) \cos(k\pi x_3/L)}_{=: e_{ijk}},$$
where  $\bx = (x_1,x_2,x_3)$, {{$L$} is the edge length of the cubed domain $\Omega$}, $\{e_{ijk}\}$ form the orthonormal basis of $L^2(\Omega)$, and $\{\eta^\alpha_{ijk}\}_{i,j,k \in \mathbb{N}}$ is a family of real-valued, independent, and identically distributed Brownian motions. Following \cite{chai2018conforming,antonopoulou2019numerical}, we approximate the term involving the Wiener process in the fully discretized system as follows
\begin{equation}\label{eq:WeinerProcApprox}
\frac{1}{\Delta t} \left( \int_{t_n}^{t_{n+1}} \dd W_\alpha(t) , \xi \right)_{L^2} \approx \frac{1}{\Delta t} \sum_{\substack{i,j,k, \\ i+j+k < I_\alpha}} \eta^\alpha_{ijk} (e_{ijk},\xi)_{L^2},
\end{equation}
where $\xi \in V_h$ is a test function, $\eta^\alpha_{ijk} \sim \mathcal{N}(0,\Delta t)$ are independent Gaussians, and $I_\alpha$ controls the number of basis functions.

\section{Numerical simulations}
\label{Sec:Simulation}

In this section, we apply the models described in Sections \ref{Sec:Derivation} and \ref{Sec:NetworkGrowth} and use the numerical discretization steps discussed in Section \ref{Sec:Solver}. We consider examples that showcase the effects of angiogenesis on the  tumor growth. For this purpose, the model parameters and the basic setting for our simulations are introduced in \cref{sec:setup}. In the base setting, we consider two vessels, one representing an artery and the other a vein, and introduce an initially spherical tumor core. Based on this setting, tumor growth is simulated first without considering angiogenesis, i.e., the growth algorithm from \cref{Sec:NetworkGrowth} is not applied. Afterwards, we repeat the same simulation including the angiogenesis effects and study the differences between the corresponding simulations results. 

\begin{figure}[!htb]
	\centering
	\includegraphics[clip,width=0.4\textwidth]{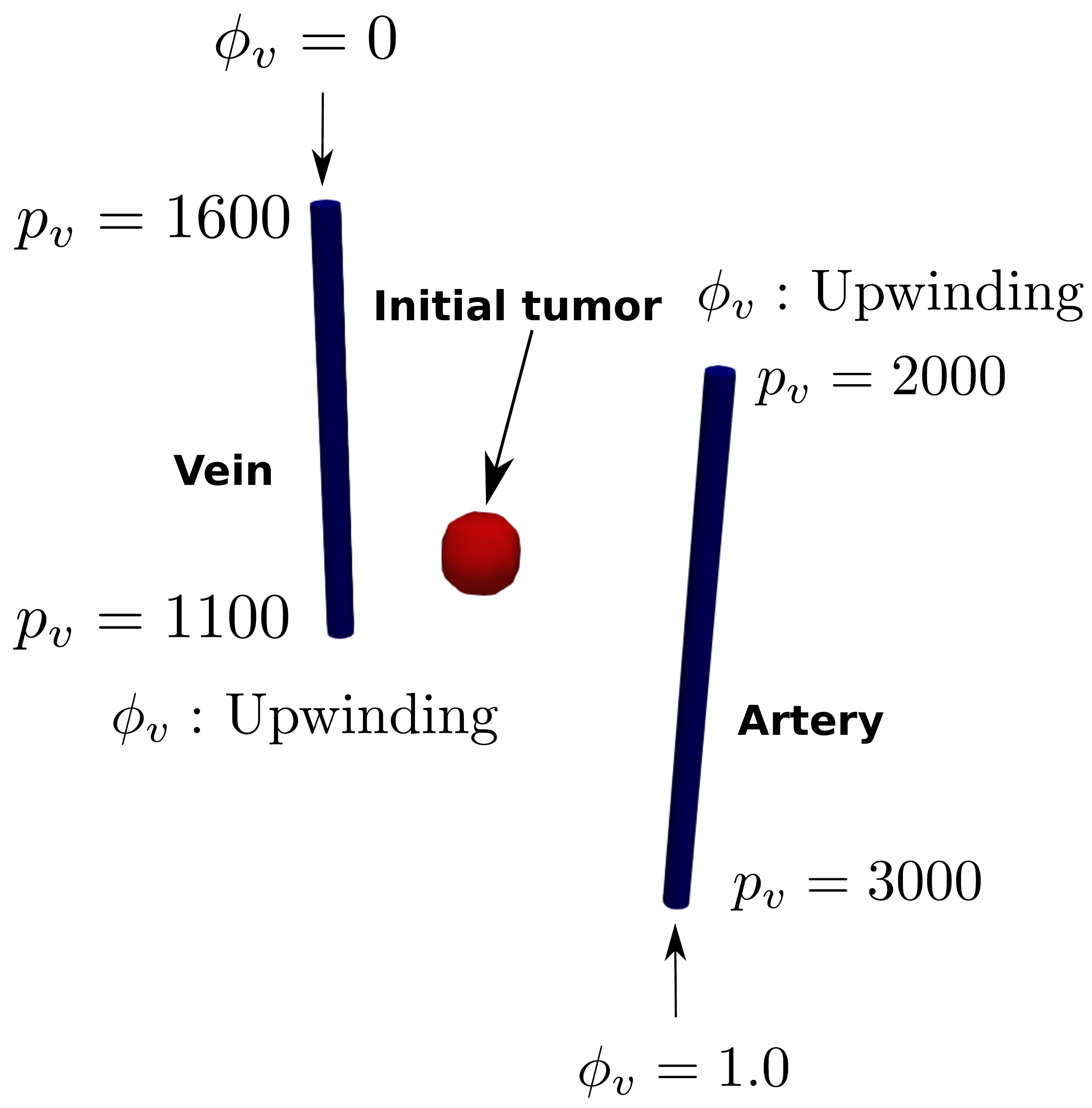}
	\caption{\label{fig:Setting} Initial setting with boundary conditions for a first numerical experiment. Pressure at top plane ($z=0$) and bottom ($z=2$) ends of the artery are $3000$ and $2000$ respectively. Similarly, the pressure at top and bottom ends of the vein are fixed at $1100$ and $1600$, respectively. The high pressure end of the artery is the inlet end, and there we assign the nutrients value to $1$. The high pressure end of the vein is also the inlet, and here we assign the nutrients value to $0$. At the remaining ends, we apply the upwinding scheme for solving the nutrient equation.}
\end{figure} 

We then consider a scenario consisting of a tumor core surrounded by a small capillary network. We obtain the network from  source\footnote{https://physiology.arizona.edu/sites/default/files/brain99.txt}. First, we rescale the network so that it fits into the domain $\Omega = (0,2)^3$. The vessel radii are rescaled such that the maximal and minimal vessel radius is given non-dimensionally by $0.05606$ and $0.025$, respectively. In all of the simulations, we consider the double-well potential of the form:
\begin{equation*}
	\Psi = C_{\Psi_{T}} \phi_T^2 (1- \phi_T)^2,
\end{equation*}
where $C_{\Psi_T}$ is a constant. Since the model involves stochastic PDEs as well as stochastic angiogenesis, we employ Monte-Carlo approximation based on samples of the probability distributions characterizing the white noise terms, using 10 samples for the case without angiogenesis and 50 samples for the case with angiogenesis. We use the samples to report statistics of quantity of interests such as total tumor volume, vessel volume density, etc.

\subsection{Setup and model parameters for the two vessel problem}
\begin{table}[!htb]
	\begin{center}
		\caption{List of parameters and their values for the numerical simulations. Unlisted parameters are set to zero. $\phi_\alpha^\omega$, $\omega_\alpha$, and $I_\alpha$ are parameters related to Wiener process, see \cref{eq:WeinerProc} and \cref{eq:WeinerProcApprox}.} \label{tab:params}
		\begin{tabular}{|l | l || l | l || l | l |}
			\hline
			\textbf{Parameter} &
			\textbf{Value} & \textbf{Parameter} &
			\textbf{Value} & \textbf{Parameter} &
			\textbf{Value}  \\ \hline \hline
			$\lambdap_{P}$ & 5 & $\lambdap_H$ & 0.5 & $\lambdad_{P}, \lambdad_H$ & 0.005  \\ 
			$\lambdap_{\ECM}$ & 0.01 & $\lambdap_{\MDE}, \lambdad_{\MDE}$ & 1 & $\lambdad_{\ECM}$ & 5 \\ 
			$\lambda_{P\!H}$ & 1 & $\lambda_{H\!P}$ & 1 & $\lambda_{H\!N}$ & 1 \\
			$\sigma_{P\!H}$ & 0.55 & $\sigma_{H\!P}$ & 0.65 & $\sigma_{H\!N}$ & 0.44 \\
			$M_P$ & 50 & $M_H$ & 25 &  $C_{\Psi_T}$ & 0.03  \\
			$\varepsilon_P$ & 0.005 & $\varepsilon_H$ & 0.005 & $\lambdap_{\TAF}$ & 10  \\
			$D_{\TAF}$ & 0.5 & $M_{\TAF}$ & 1 & $L_p$  & $10^{-7}$  \\
			$D_\sigma$ & 3 & $M_\sigma$ & 1 & $K$ & $10^{-9}$  \\
			$D_{\MDE}$ & 0.5 & $M_{\MDE}$ & 1 &  $L_\sigma $ & $4.5$ \\
			$D_v$  & $0.1$  & $\mu_{\text{bl}}$ & 1 & $r_\sigma$ & $0.95$ \\
			$I_\alpha$, $\alpha \in \{P,H\}$  & $17$  & $\phi_\alpha^{\omega}$, $\alpha \in \{P, H, T\}$  & $0.1$  & $\omega_\alpha$, $\alpha \in \{P,H\}$  & $0.0025$ \\
			\hline
		\end{tabular}
	\end{center}
\end{table}

\begin{table}[!htb]
	\begin{center}
		\caption{List of parameters and their values for the growth algorithm and numerical discretization}\label{tab:growth}
		\begin{tabular}{|l | l | l |}
			\hline
			\textbf{Parameter} & \textbf{Value} & \textbf{Function} \\
			\hline
			\hline
			$Th_{\TAF}$  & $7.5 \cdot 10^{-3}$ & Threshold for the TAF concentration (sprouting) \\
			\hline 
			$\mu_r$  & $1.0$ & Mean value for the log-normal distribution (ratio radius/vessel length) \\
			\hline
			$\sigma_r$  & $0.2$ & Standard dev. for the log-normal distribution (ratio radius/vessel length) \\
			\hline
			$\lambda_g$ & $1.0$ & Regularization parameter to avoid bendings and sharp corners \\
			\hline
			$\gamma$ & $3.0$ & Murray parameter determining the radii at a bifurcation \\
			\hline
			$R_{\min}$ & $9.0 \cdot 10^{-3}$ & Minimal vessel radius \\
			\hline
			$l_{\min}$ & $0.13$ & Minimal vessel length for which sprouting is activated \\
			\hline
			$R_{\max}$ & $0.035$ & Maximal vessel radius \\
			\hline
			$R_T$ & $0.05$ & Threshold for the radius to distinguish between arteries and veins for $t=0$ \\
			\hline
			${\zeta}$ & $1.05$ & Sprouting parameter \\
			\hline
			$\text{dist}_{\text{link}}$ & $0.08$ & Maximal distance at which a terminal vessel is linked to the network \\
			\hline
			$\tau_{\text{ref}}$ & $0.02$ & Lower bound for the wall shear stress \\
			\hline
			$k_{\WSS}$ & $0.4$ & Proportional constant (wall shear stress) \\
			\hline
			{$k_s$} & {$0.14$} & {Shrinking parameter}\\
			\hline
			$\Delta t$ & $0.0025$ & Time step size \\
			\hline	
			$h_{3D}$ & $0.0364$ & Mesh size of the 3D grid \\
			\hline	
			$h_{1D}$ & $0.25$ & Mesh size of the initial 1D grid \\
			\hline	
			$\Delta t_{net}$ & $2\Delta t$ & Angiogenesis (network update) time interval \\
			\hline	
		\end{tabular}
	\end{center}
\end{table}

\label{sec:setup}
As a computational domain $\Omega$, we choose a cube, $\Omega = \left(0,2 \right)^3$. Within $\Omega$ two different vessels are introduced: an artery and a vein; see \cref{fig:Setting}. The radius of the vein $R_v$ is given by $R_v = 0.0625$, and the radius of the artery $R_a$ is set to $R_a = 0.047$. The centerlines of both vessels are given by straight lines. In case of the artery, the centerline starts at $\left( 0.1875, 0.1875, 0 \right)$ and ends at $\left( 0.1875, 0.1875, 2 \right)$, whereas the vein starts at $\left( 1.8125, 1.8125, 0 \right)$ and ends at $\left( 1.8125, 1.8125, 2 \right)$.

At the boundaries of the vessels, we choose Dirichlet boundaries for the pressure, see also \cref{fig:Setting}. These boundary conditions imply that the artery provides nutrients for the tissue block $\Omega$, while the vein will take up nutrients. For the nutrients in the blood vessels mixed boundary conditions are considered, as depicted in \cref{fig:Setting}. 

As initial conditions for $\phi_v$, we choose $\phi_v=1$ in the artery and $\phi_v=0$ in the vein. The initial value for the nutrient variable $\phi_\sigma$ in the tissue matrix is given by $\phi_\sigma = 0.5$. In order to define the initial conditions for the tumor, we consider a ball $B_T$ of radius $r_c = 0.3$ around the center $\bx_c = \left( 1.0, 0.8, 1.0 \right)$.  Within $B_T$, the total tumor volume fraction $\phi_T$ smoothly decays from $1$ in the center to $0$ on the boundary of the ball:
\begin{align}
	\label{eq:ICTumor}
	\phi_T(\bx, t = 0) &= \begin{cases}\begin{aligned}
		&\exp\left( 1 - \frac{1}{1 - (|\bx - \bx_c|/r_c)^4} \right), &&\text{if } |\bx - \bx_c| < r_c, \\
		& 0, &&\text{otherwise}.
	\end{aligned}\end{cases}
\end{align}
Thereby, the necrotic and hypoxic volume fractions, $\phi_N$ and $\phi_H$, are set initially to zero. In the other parts of the domain, all the volume fractions for the tumor species are set to $0$ at $t=0$. 
In \cref{tab:params}, the parameters for the model equations in \cref{Sec:Derivation} are listed and \cref{tab:growth} contains the parameters for the growth algorithm described in \cref{Sec:NetworkGrowth} as well as the discretization parameters. In particular, the parameters for the stochastic distributions are listed, which determine the radii and vessel lengths, the probability of bifurcations, and the sprouting probability of new vessels.

{{
\subsection{Robustness of the 3D-1D solver}
To ascertain the accuracy and robustness of the proposed solver, we performed several studies where we changed mesh size and time steps and found that the solver is robust and the size of time step and mesh size employed in the studies in sections below balance the computational cost and numerical accuracy pretty well. To strengthen the claims, we consider a two-vessel setup described above without the stochasticity and network growth. We run the simulations using four different time steps $\Delta t_i = 0.01/2^{i-1}$, $i=1,2,3,4$, and compute the rate of convergence of quantity of interests such as $L^2$ or $L^1$ norms of tumor species and nutrients. In \cref{fig:convres1}, we plot the $L^2$ norm of tumor species and nutrients for different time steps. We see that the difference between the curves for different $\Delta t$ is very small. Let $Q_i(t)$ denote the quantity of interest ($L^2$ norm of species) at time $t$ for $\Delta t_i$. We can approximately compute the rate of convergence of $Q$ using the formula:
\begin{align*}
r(t) = \frac{\log(|Q_1(t) - Q_4(t)|) - \log(|Q_2(t) - Q_4(t)|)}{\log(\Delta t_1) - \log(\Delta t_2)} .
\end{align*}
For $Q(t) = \left\|\phi_T(t) \right\|_{L^2}$, we found $r(1) = 0.894, r(2) = 1.03, r(3) = 1.025, r(4) = 0.531, r(5) = 1.692$. 

We also remark that the proposed solver, see Algorithm \ref{Alg_AlgImpExp}, does not involve nonlinear iterations to compute the $\phi_P, \phi_H, \phi_N, \mde, \ecm$ solutions at current time step. We compared the results of current solver and the solver involving nonlinear iterations and observed that the solver with nonlinear iterations still required us to consider small time steps. Also, the error in solution from two solvers decreases with mesh refinement and smaller time steps. These observations motivated us to use the proposed solver for all numerical tests in the sections below.

\begin{figure}[!htb]
\centering
\includegraphics[width=0.9\linewidth]{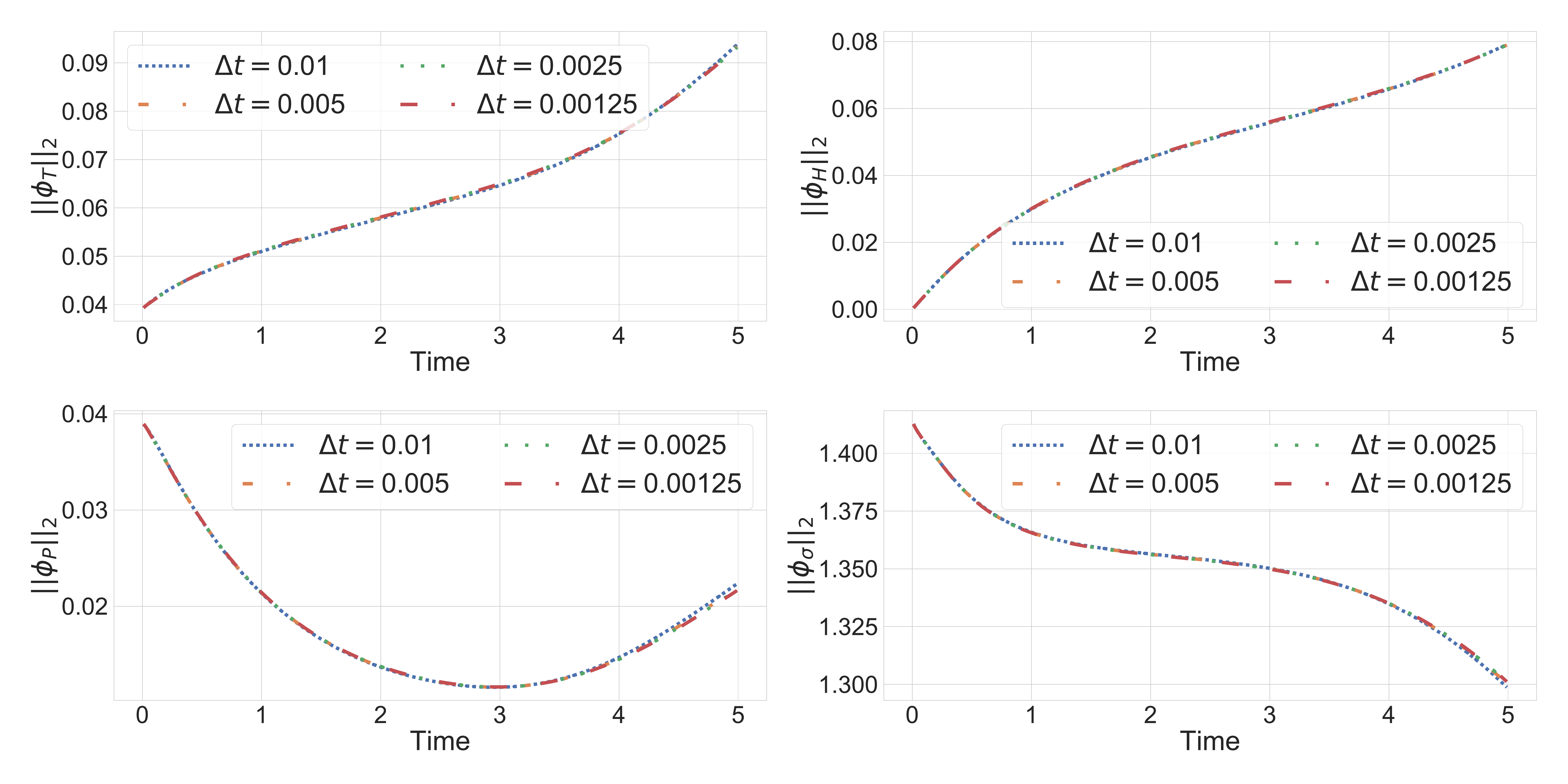}
\caption{{Plot of the $L^2$ norm of various species using four different time steps.}}\label{fig:convres1}
\end{figure}
}}

\subsection{Tumor growth without angiogenesis}

\begin{figure}[!htb]
	\centering
	\includegraphics[clip,width=0.75\textwidth]{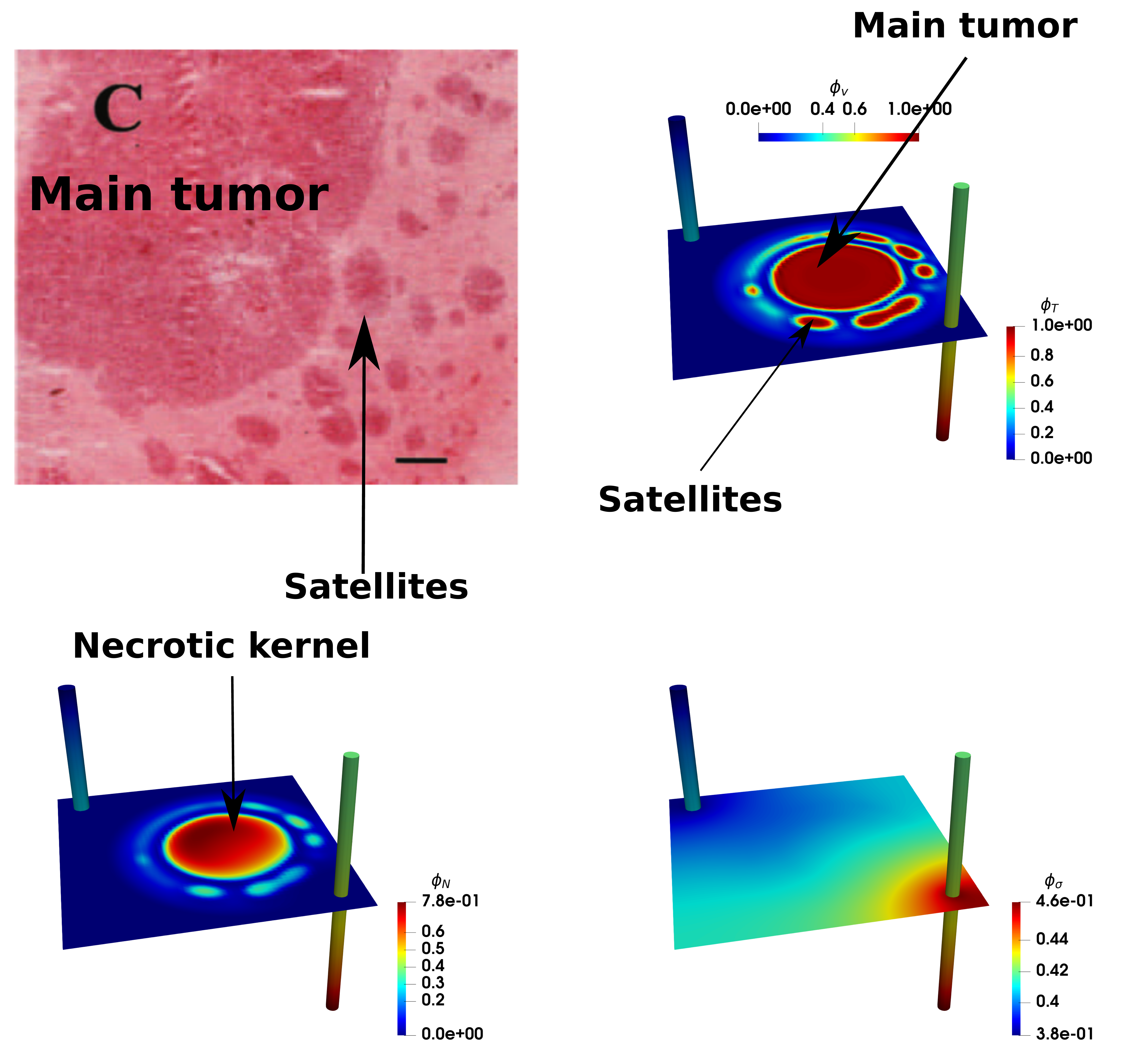}
	\caption[123]{ \textbf{Top left:} Tumor growing in a mouse that is treated with anti-VEGF agents. As a consequence tumor satellites in the vicinity of the main  tumor can be detected. Image taken from \cite{rubenstein2000anti}, with permission from Elsevier. \textbf{Top right:} $\phi_T$ presented in a plane at $z=1$ perpendicular to the $z$-axis. As seen in the medical experiments the formation of satellites and the accumulation of tumor cells at the nutrient-rich artery are reproduced in the simulations.  \textbf{Bottom left:} Distribution of necrotic cells ($\phi_N$). It can be seen that the main tumor consists of a large necrotic kernel.  \textbf{Bottom right:} Distribution of nutrients ($\phi_{\sigma}$). \label{fig:ResWithoutAng}}
\end{figure} 

\label{sec:without_angiogenesis}
The simulation results for tumor growth without angiogenesis are summarized in \cref{fig:ResWithoutAng}.
For $t=8$, 
the tumor cell distribution within the plane perpendicular to the $z$-axis at $z=1.0$ is shown. In three  subfigures, the volume fraction variables $\phi_T = \phi_P + \phi_H + \phi_N$, $\phi_N$, as well as the nutrients $\phi_{\sigma}$ are presented. It can be observed that the primary tumor is enlarged and  small satellites are formed in the vicinity of the main tumor. The distribution of the necrotic cells indicates that the main tumor consists mostly of necrotic cells, while the hypoxic and proliferative cells gather around the nutrient-rich blood vessels. This means that the tumor cells can migrate against the flow from the artery towards the vein. Apparently, the chemical potential caused by the nutrient source dominates the interstitial flow.

These observations are consistent with simulation results and measurements discussed, e.g., in \cite{frieboes2010three,kunkel2001inhibition,rubenstein2000anti}. In \cite{kunkel2001inhibition,rubenstein2000anti} a tumor is introduced into a  mouse. At the same time, anti-VEGF agents were injected into the mouse, such that the sprouting of new vessels growing towards the tumor is prevented. This process leads to the formation of satellites located in the vicinity of the primary tumor as well as the accumulation of tumor cells at nutrient-rich vessels and cells. Furthermore, the primary tumor stops growing and forms a large necrotic core.

\begin{figure}[!htb]
	\centering
	\includegraphics[clip,width=0.8\textwidth]{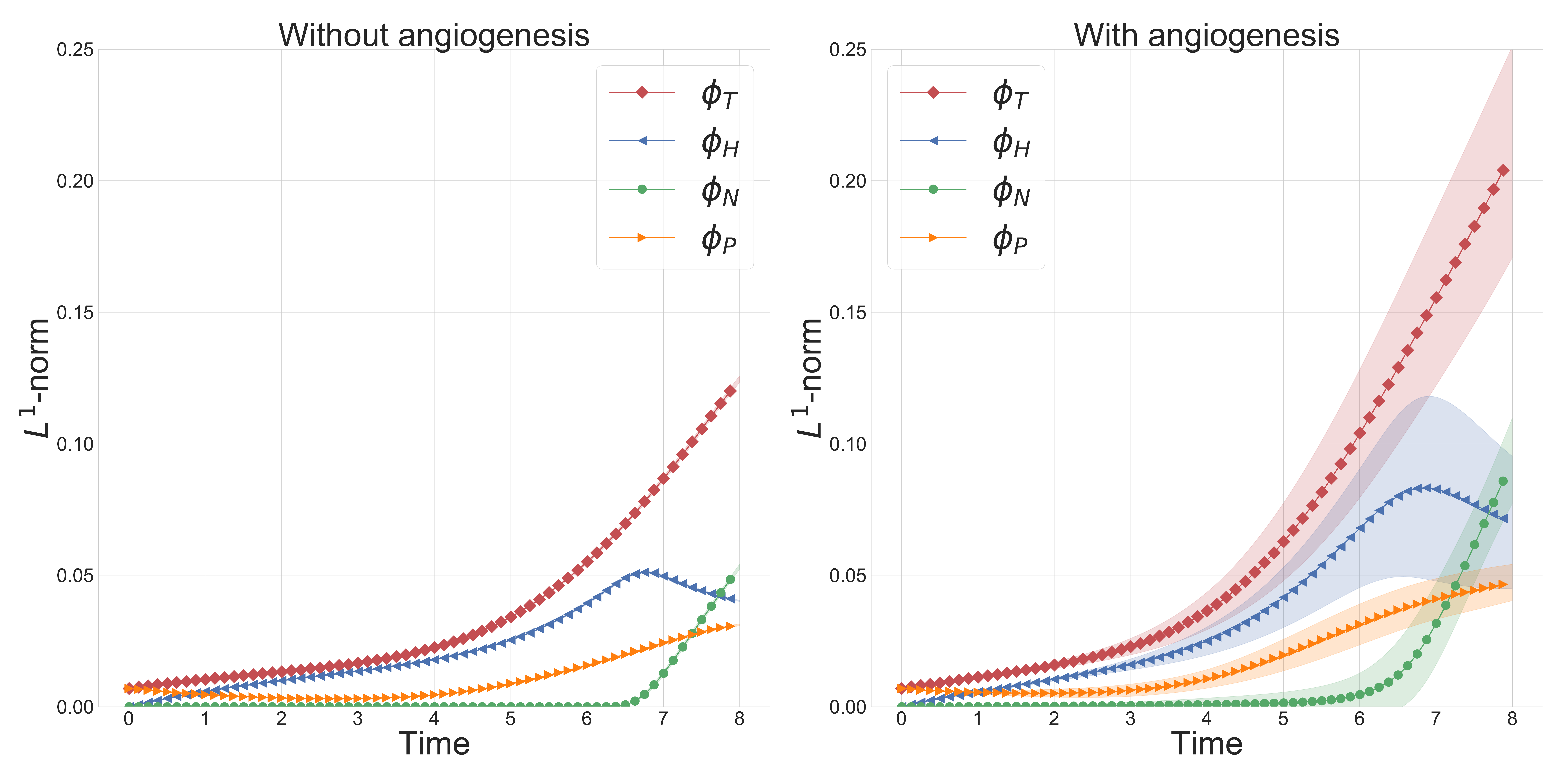}
	\caption{\label{fig:TumorTime} Quantities of interests (QoIs) related to tumor species over a time interval $[0,8]$ for the two-vessels setting. For the case without angiogenesis, the mean QoIs computed using 10 samples is shown. For the angiogenesis case, we compute the mean and standard deviation from 50 samples. The solid line shows the mean QoI as a function of time. The thick layer around the solid line corresponds to interval $(\mu_\alpha(t) - \sigma_\alpha(t), \mu_\alpha(t)+\sigma_\alpha(t))\cap [0,\infty)$, for $t \in [0,T]$, where $\mu_\alpha(t), \sigma_\alpha(t)$ are the mean and standard deviations of QoI $\alpha \in \{\left\|\phi_T \right\|_{L^1}, \left\|\phi_P \right\|_{L^1}, \left\|\phi_H \right\|_{L^1}, \left\|\phi_N \right\|_{L^1}\}$ at time $t$. The variations in the QoIs for the non-angiogenesis case are very small. The mean of the total tumor volume fraction $\left\|\phi_T \right\|_{L^1}$ at the final time for the angiogenesis case is about 1.7 times that of the non-angiogenesis case. }
\end{figure}

In \cref{fig:TumorTime}, the $L^1$-norms of the tumor species over time are presented for the case when angiogenesis was inactive and when it was active. While the profiles for different species in two cases look similar, the total tumor is about 70$\%$ times higher when angiogenesis is active. Diffusivity $D_\sigma = 3$ is large enough, and therefore, the nutrients originating from the nutrient-rich vessels diffuse quickly throughout the domain. 

In summary, one can conclude that without angiogenesis a tumor can grow to a certain extent, before the primary tumor starts to collapse, i.e., a large necrotic core is formed. However, this does not mean that the tumor cells are entirely removed from the healthy tissue. If there is a source of nutrients such as an artery close by, transporting nutrient rich blood, a portion of tumor cells can survive by migrating towards the neighboring nutrient source.

\subsection{Tumor growth with angiogenesis}
\label{sec:with_angiogenesis}

\begin{figure}[!htb] 
	\centering
	\includegraphics[clip,width=0.9\textwidth]{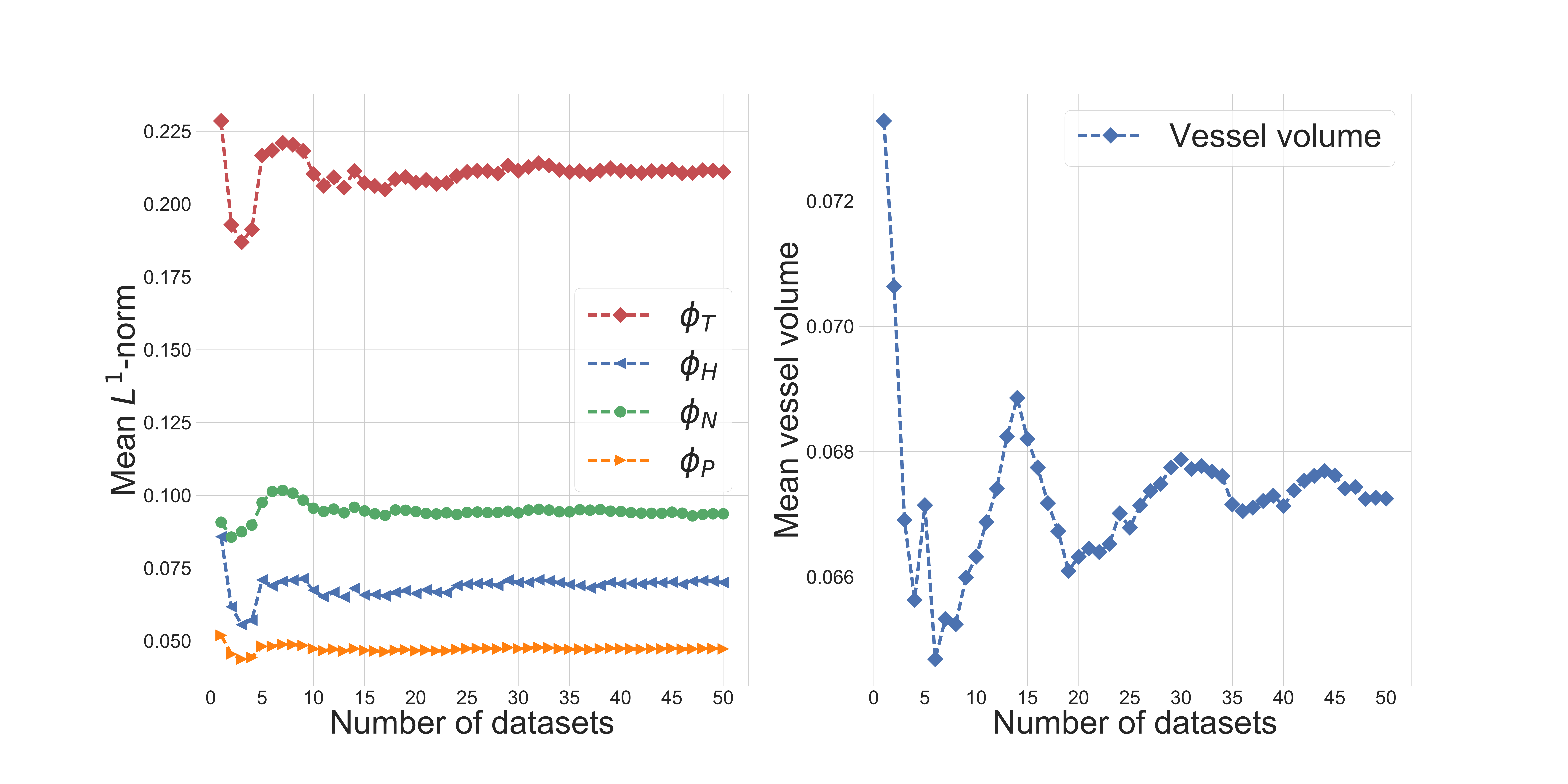}
	\caption{\label{fig:Statistics} The mean values for the $L^1$-norm of the tumor cell volume fractions $\phi_T, \phi_P, \phi_H, \phi_N$ and the volume of the blood vessel network at time $t = 8$ from increasing number of samples. Results correspond to the two-vessel setting. The mean of the total tumor volume fraction appears to be stable after about 28 samples. The mean of the vessel volume shows smaller fluctuations as the number of samples in the data set grows.}
\end{figure}

As in the previous subsection, we compute  the $L^1$-norms of the tumor species $\phi_T$ at time $t=8$. However, since several stochastic processes are involved in the network growth and also Wiener processes appear in the proliferative and hypoxic cell mass balances, several data sets have to be created in order to rule out statistical variations. In this context, the issue arises as to how many data sets are needed to obtain a representative value. In order to investigate this, we compute for every sample $i$, the $L^1$-norm of the tumor species, denoted by $\phi_{\alpha_{L_1,i}},\; \alpha \in \left\{ T, P, H, N \right\}$. Additionally, the volume of the blood vessel network $V_i$ is computed. For each data set with $i$ samples, we compute the mean values:
$$
\mean_i \left( V \right) = \frac{1}{i} \sum_{j=1}^i V_j, \; \qquad \mean \left( \phi_{\alpha_{L_1,i}} \right) = \frac{1}{i} \sum_{j=1}^i \phi_{\alpha_{L_1,j}},\; \qquad \alpha \in \left\{ T, P, H, N \right\}.
$$
In \cref{fig:Statistics}, the mean values $\mean_i \left( V \right)$ and $\mean \left( \phi_{\alpha_{L_1,i}} \right)$, $\alpha \in \{T, P, H, N\}$, are shown. From the plots we see that the mean of $||\phi_T||_{L^1}$ stabilizes after about 25 samples. For the vessel volume, fluctuations in the sample means reduce with increasing sample and get small after 30 samples. While the results in \cref{fig:Statistics} show that mean of the QoIs stabilizes with increasing sample and converge to some number, the trajectory in the figure could change with change in sample values. For example, if we shuffle the samples and recompute the quantities in \cref{fig:Statistics}, various curves may look different.  

As mentioned earlier, \cref{fig:TumorTime} presents the $L^1$-norms of tumor species. For the angiogenesis simulations, we compute the mean and standard deviation using 50 samples. We see that the total tumor volume fraction varies from sample to sample, as expected. 
Both the hypoxic and the tumor cells show an exponential growth after $t \approx 4$; see \cref{fig:TumorTime}. After decreasing until $t \approx 3$, the proliferative cell mass grows from $t \approx 4$ onward.   The result is that in the case of angiogenesis, the overall nutrient concentration is higher compared to the case without angiogenesis, while the spatial variation of the nutrient is the same in the two; and hence the growth of the tumor in the two cases are similar except that the tumor grows more rapidly in the case of angiogenesis. We will see in our second example, where $D_\sigma = 0.05$ is much smaller, that the nutrient concentration is higher near the nutrient rich vessels and tumor growth is more concentrated near these regions, see \cref{fig:TumorTimeComplex}.

\begin{figure}[!htb] 
	\centering
	\includegraphics[clip,width=.8\textwidth]{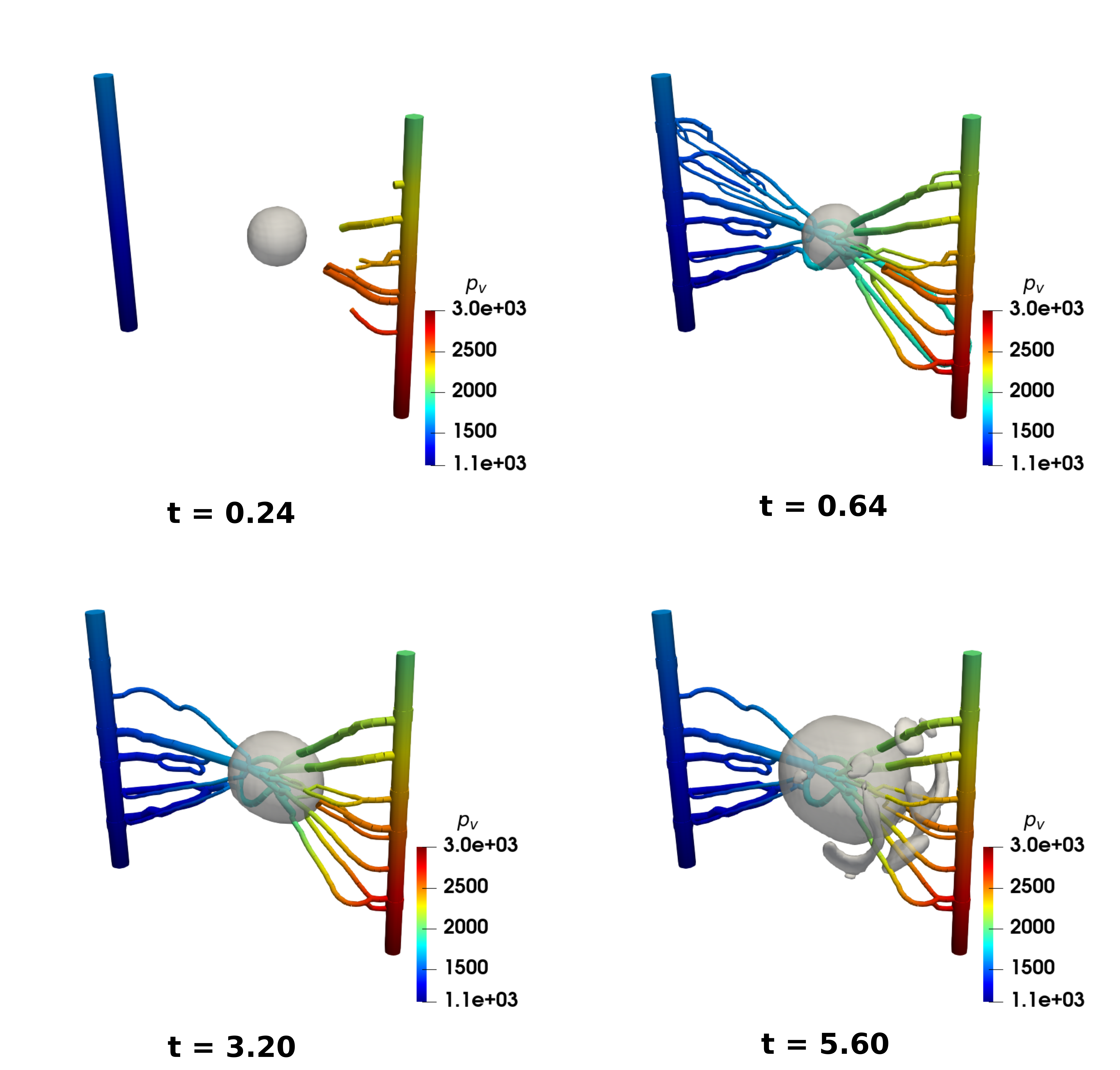}
	\caption{\label{fig:GrowingNetwork} Growth of the tumor and the network at four different time points \\ $t \in \left\{0.24, 0.64, 3.20, 5.60 \right\}$ and one specific sample. We show contour $\phi_T = 0.9$.}
\end{figure} 

\begin{figure}[!htb] 
	\centering
	\includegraphics[clip,width=.8\textwidth]{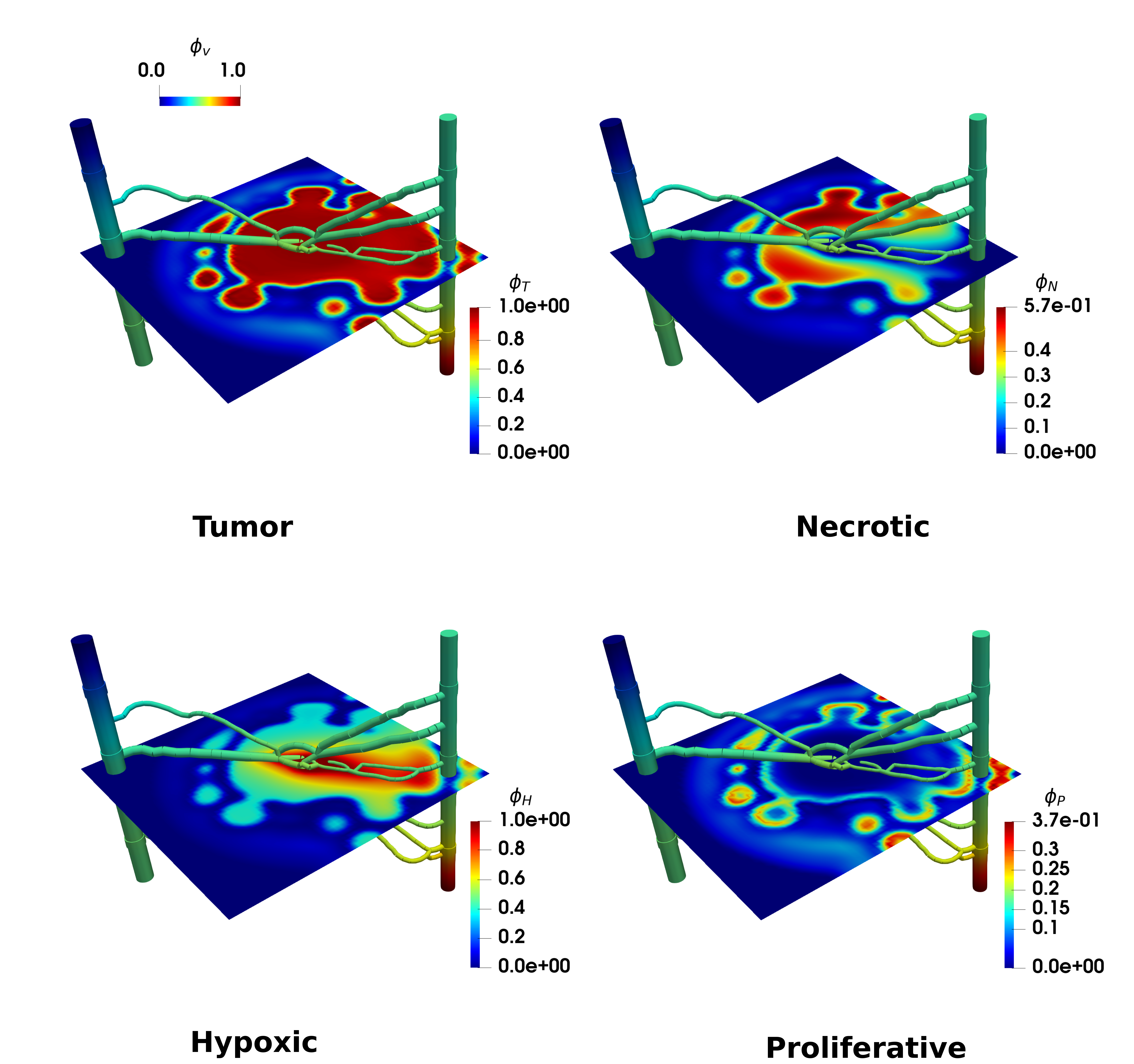}
	\caption{\label{fig:CellsWithAngiogenesis} Tumor cell distributions in the plane $z=1$ with angiogenesis for one sample. \textbf{Top left:} $\phi_T$.  \textbf{Top right:} $\phi_N$. \textbf{Bottom left:} $\phi_H$. \textbf{Bottom right:} $\phi_P$.}
\end{figure}

In \cref{fig:GrowingNetwork}, we show the evolving network together with the contour plot $\phi_T = 0.9$ of the total tumor species. At time $t = 0.24$ (top-left figure), we observe that new vessels originate from the artery and move towards the hypoxic core; the directions of these vessels being based on the gradient of TAF with random perturbations. At $t=0.64$ (top-right), we see a large number of new vessels formed as predicted by the angiogenesis algorithm. However, at time $t = 3.2$ (bottom-left), vessels adapt and due to lower flow rates in some newly created vessels, some vessels are gradually removed, and thus the number of vessels decreases. Comparing $t=3.2$ and $t=5.6$ (bottom-right), we see that the network has stabilized and little has changed in this time window. From \cref{fig:GrowingNetwork}, we can also summarize that the tumor growth is directed towards the nutrient-rich vessels.  

Next, we plot the tumor species at the $z=1$ plane along with the nutrient distribution in the vessels in \cref{fig:CellsWithAngiogenesis}. The plot corresponds to time $t = 8$. The plots corresponding to the necrotic species show that the necrotic concentration is typically higher away from the nutrient-rich vessels. From the hypoxic plot, we see that it is higher near these vessels, and this is explained by the fact that as soon as the proliferation of new tumor cells takes place, due to nutrient concentration below the proliferative-to-hypoxic transition threshold, these newly added proliferative tumor cells convert to hypoxic cells. Further transition to necrotic cells would take place if the nutrients are even below the hypoxic-to-necrotic transition threshold. This is also consistent with the increase concentration of the proliferative cells near the outer tumor-healthy cell phase interface.

{{
\subsection{Sensitivity of the growth parameters}

\begin{figure}[!htb] 
	\centering
	\includegraphics[clip,width=.48\textwidth]{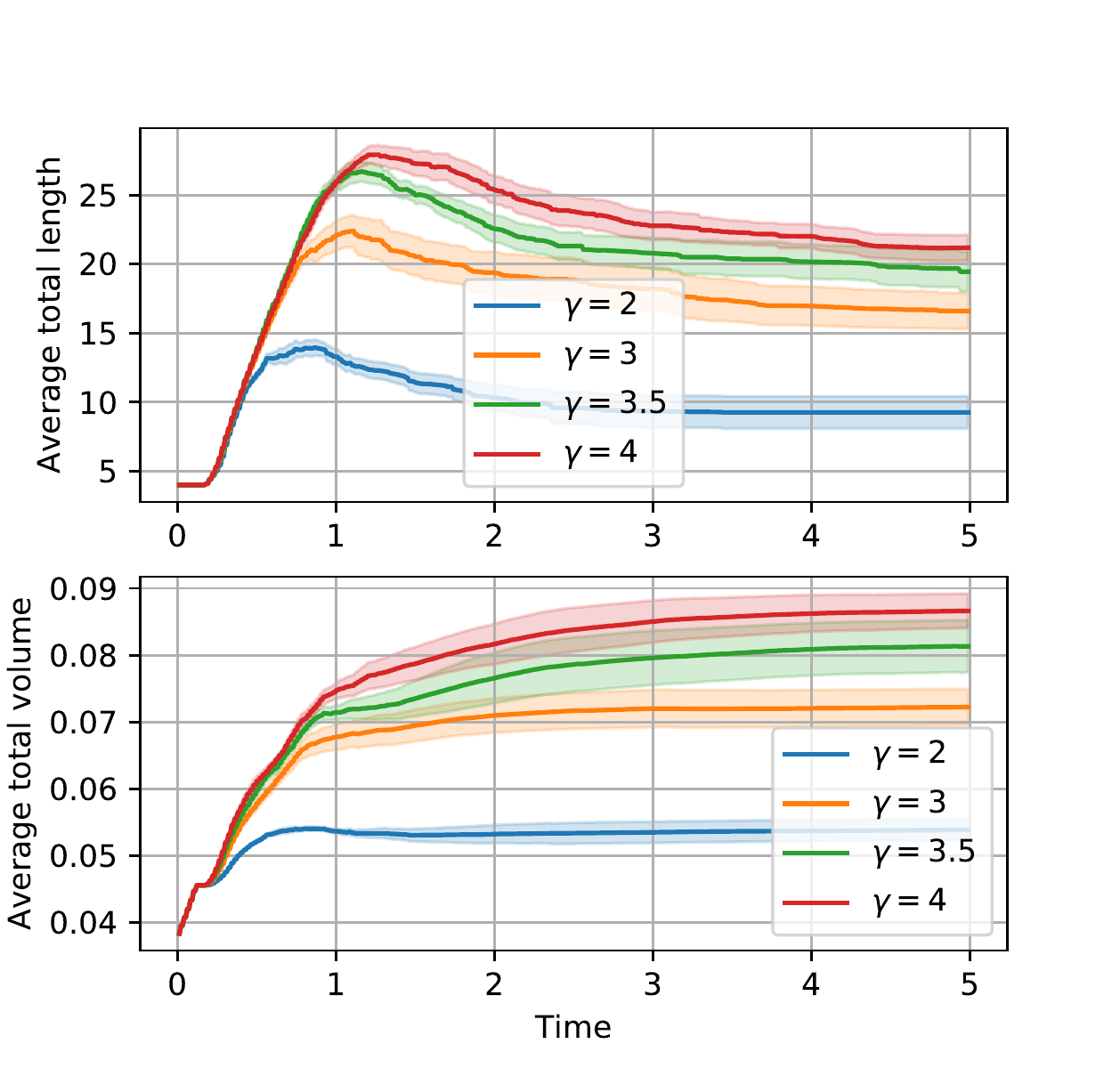}
	\includegraphics[clip,width=.48\textwidth]{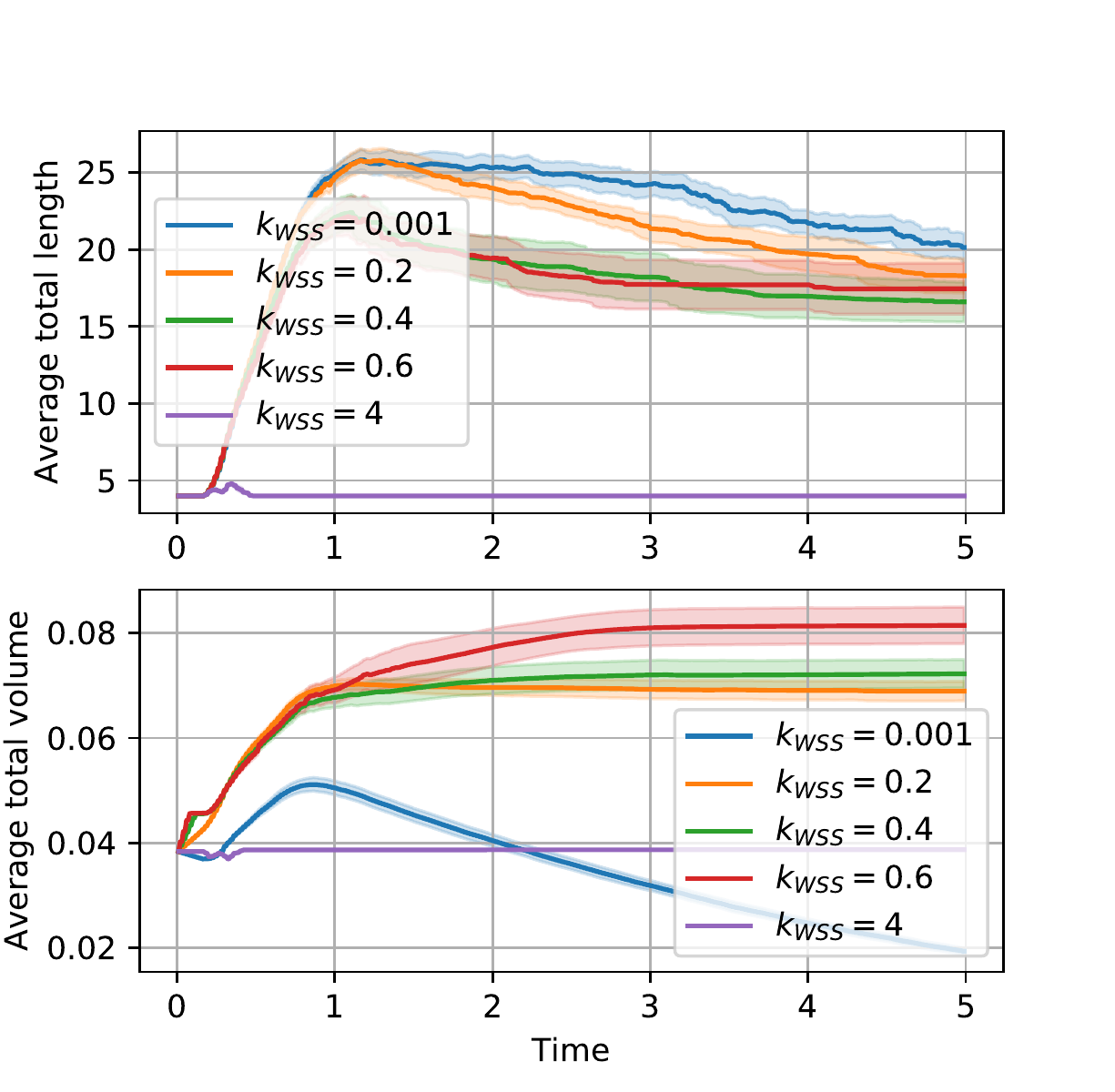}
	\caption{
		\label{fig:network-parameter-study} Results of a parametric study in which certain growth parameters $\gamma$ and $k_{\WSS}$ are varied to measure their impact on the total network length and volume. For these studies, we considered a coarse mesh for the 3D domain with $16^3$ uniform cells and time step $\Delta t = 0.01$. The network update time step was fixed to $\Delta t_{net} = 2\Delta t$.
	}
\end{figure} 

\begin{figure}[!htb] 
	\centering
	\includegraphics[clip,width=1.0\textwidth]{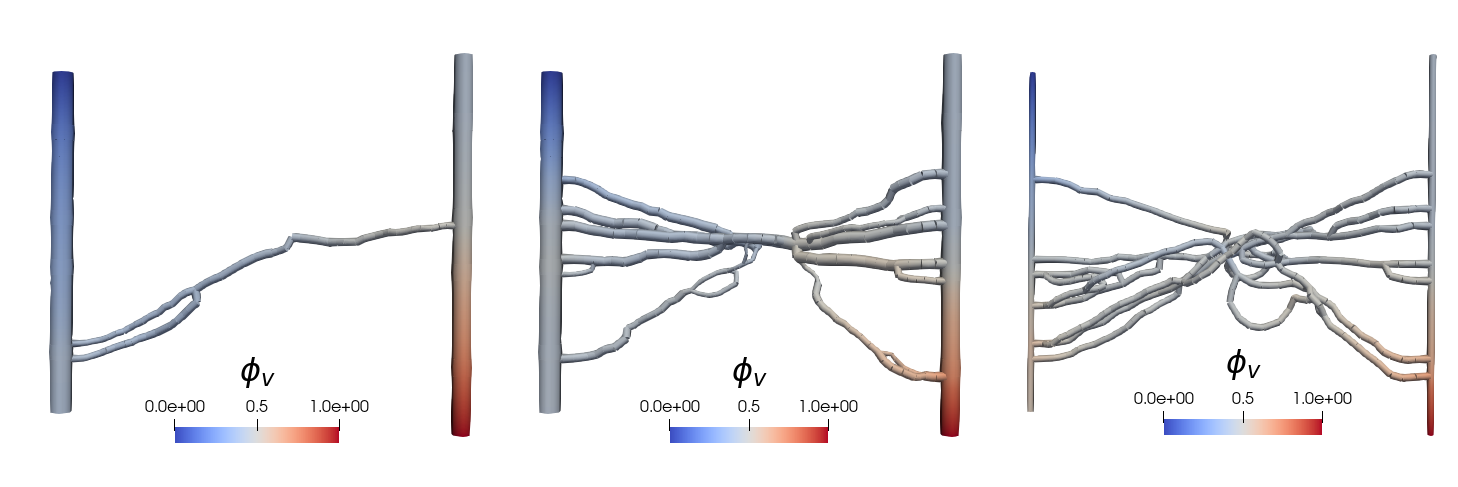}
	\caption{
		\label{fig:network-parameter-study-gamma-pv} 
		Network structure for different values of $\gamma$ at time $t=2.4$. {\bf Left}: $\gamma = 2$. {\bf Middle}: $\gamma = 3$. {\bf Right}: $\gamma = 4$. 
	}
\end{figure} 

\begin{figure}[!htb] 
	\centering
	\includegraphics[clip,width=1.0\textwidth]{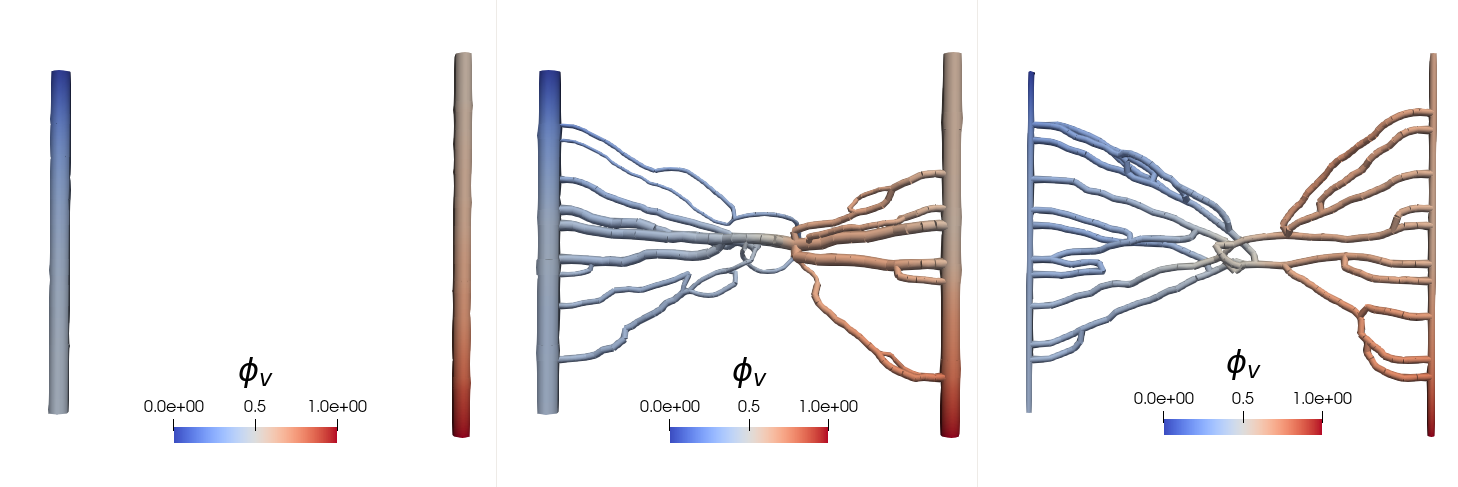}
	\caption{
		\label{fig:network-parameter-study-pv} 
Network structure for different values of $k_{\WSS}$ at time $t=6$. {\bf Left}: $k_{WSS} = 4$. {\bf Middle}: $k_{WSS} = 0.4$. {\bf Right}: $k_{WSS} = 0.001$.
	}
\end{figure} 

In \cref{fig:network-parameter-study}, we present the results of a parametric study designed to test the robustness of the vascular network model to changes in the values of the parameters $\gamma$ and $k_{WSS}$. It is observed that changes in these parameters can produce significant changes in the network structure for given values of the other model parameters. 

The parameter $\gamma$, for example, appears in Murray's law \eqref{eq:Murray}, and controls the radii of network branches, with increase in $\gamma$ leading to larger radii of bifurcating vessels. Such larger radii vessels have a higher probability for connecting with neighboring vessels so as to increase the flow and to continue to evolve; for high $\gamma$, the networks are more dense and the total network length is higher (see \cref{fig:network-parameter-study-gamma-pv}, right). Conversely, small values of $\gamma$ promote thin network segments with lower probability for connecting to neighboring vessels, see \cref{fig:network-parameter-study-gamma-pv}, left.

The change in vessel radius due to the vessel wall shear stress is proportional to the constant $k_{\WSS}$ and stimulus $S_{\WSS, i}$, see \eqref{eq:SWSS} and \eqref{eq:DeltaRi}. Further, the vessels shrink naturally and this effect is controlled by constant $k_s$ (higher $k_s$ means radius decay is higher). 
In our study, we varied the values of parameter $k_{\WSS}$ and found that, for a large $k_{\WSS}$, radii of sprouting vessels decrease, and new sprouts are removed in their early stage of growth before they could join the nearby vessels, see \cref{fig:network-parameter-study-pv} left. As a result of new sprouts getting removed in early stages, the total network length and the vessel volume stay constant with respect to time with constant values very close to the initial values. 
For a very small $k_{WSS}$ with $k_s$ being fixed, the radii during the early phase of simulations do not change much. But in the later phases of the simulation, the radii begin to decay and even with large flow rate (which means large wall shear stress); their decay is unavoidable as the term $k_{\WSS} S_{\WSS,i}$ is small as $k_{\WSS}$ is small and can not counter the effects of $k_s$. In summary, in the long run the radii of vessels decrease with time, see \cref{fig:network-parameter-study-pv} right.We also observed that for values of $k_{\WSS}$ within certain bounds, its impact on the network morphology is low. However, when $k_{\WSS}$ is outside the bound, some care is required so that vessel radii do not tend to zero with time.
}}

\subsection{Angiogenesis and tumor growth for the capillary network}
\begin{figure}[!htb]
	\centering
	\includegraphics[clip,width=.7\textwidth]{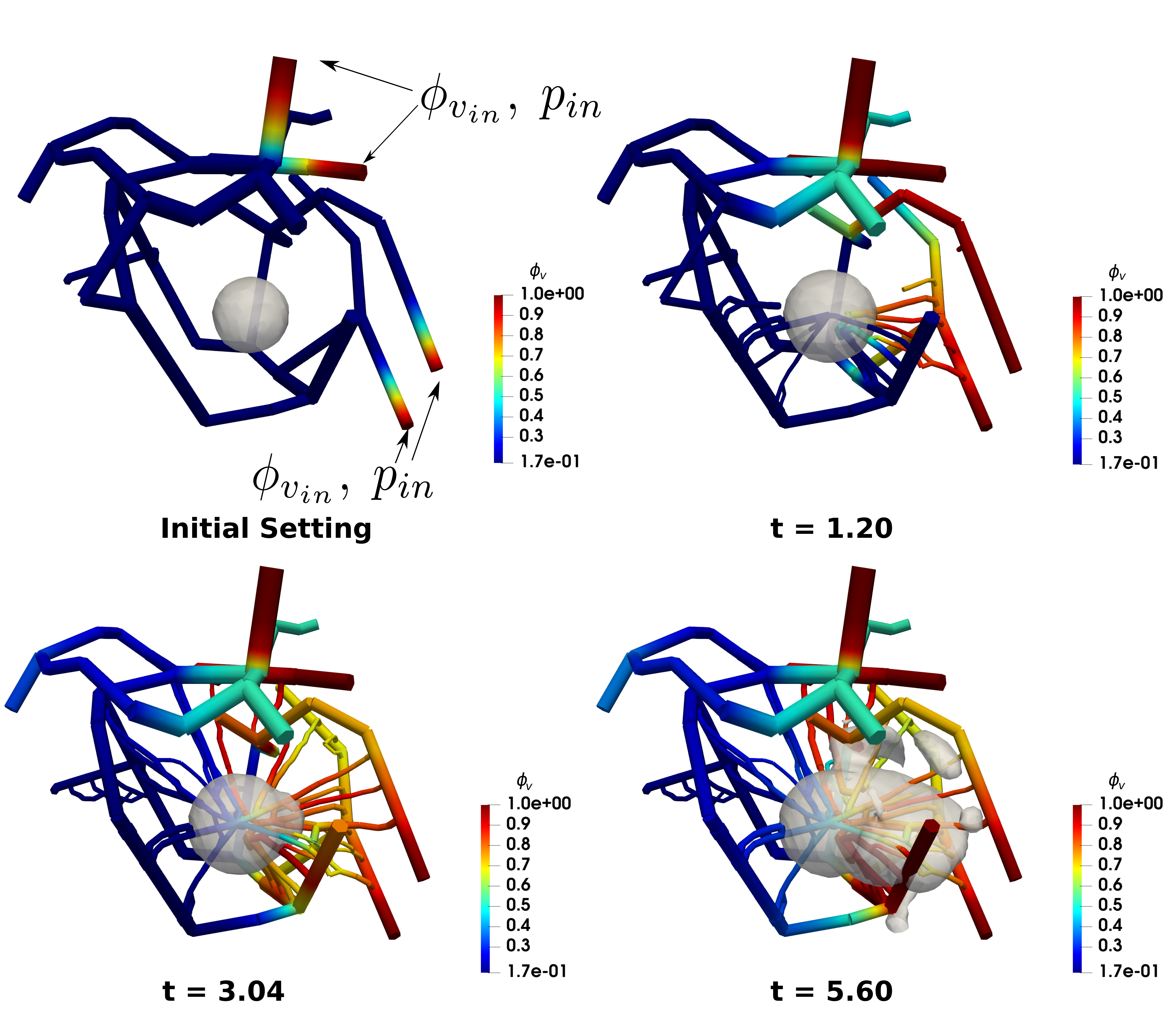}
	\caption{\label{fig:BoundariesNetwork} \textbf{Top left:} Spherical tumor core $\left( \text{Contour plot for }\phi_T = 0.9 \right)$ at $\bx_c = (1.3, 0.9, 0.7)$ with radius $r_c = 0.3$ surrounded by a network of vessels. We identify four inlet ends (red cross-sections) at which the unit nutrient $\phi_v = \phi_{v_{in}} = 1.0$ and pressure $p_v = p_{in} = 8000$ is prescribed as a Dirichlet boundary condition. \textbf{Top right:} Formation of first sprouts at $t=1.20$ growing towards the tumor core. \textbf{Bottom left:} Around $t = 3.04$ a complex vascular network is formed and the tumor starts to grow towards the nutrients. \textbf{Bottom right:} At $t = 5.60$ the tumor is significantly enlarged and creates satellites near the nutrient vessels.}
\end{figure}

Returning to \eqref{eq:ICTumor}, let us consider a smooth spherical tumor core with center at $\bx_c = (1.3, 0.9, 0.7)$ and radius $r_c = 0.3$ in the domain $\Omega = (0,2)^3$. The initial blood vessel network and boundary conditions for pressure and nutrient on the network is described in \cref{fig:BoundariesNetwork}. 

In the simulation, the vessels are uniformly refined two times. We fix $\phi_a = 0$ for $a\in \{H, N, \TAF\}$ and $\phi_\sigma = 0.5$ at $t=0$. The domain is discretized uniformly with a mesh size $h_{3D} = 0.0364$ and the time step size is $\Delta t = 0.0025$. We identify four inlet ends (see Figure \ref{fig:BoundariesNetwork}) at which the unit nutrient $\phi_v = \phi_{v_{in}} = 1$ and pressure $p_v = p_{in} = 8000.0$ is prescribed as the Dirichlet boundary condition. At the remaining boundaries, we prescribe the pressure $p_v = p_{out} = 1000.0$ and apply an upwinding scheme for the nutrients. 

At $t=0$, $p_v$ and $\phi_v$ at internal network nodes are set to zero. Furthermore, we set $L_\sigma = 0.5$, $D_\sigma = 0.05$, $D_{TAF} = 0.1$, $Th_{\TAF} = 0.0075$, $\mu_r = 1.5$, $k_{\WSS} = 0.45$, and $\Delta t_{net} = 10 \Delta t$. All the other parameter values remain unchanged, see \cref{tab:params} and \cref{tab:growth}. 

\begin{figure}[!htb]
	\centering
	\includegraphics[clip,width=0.8\textwidth]{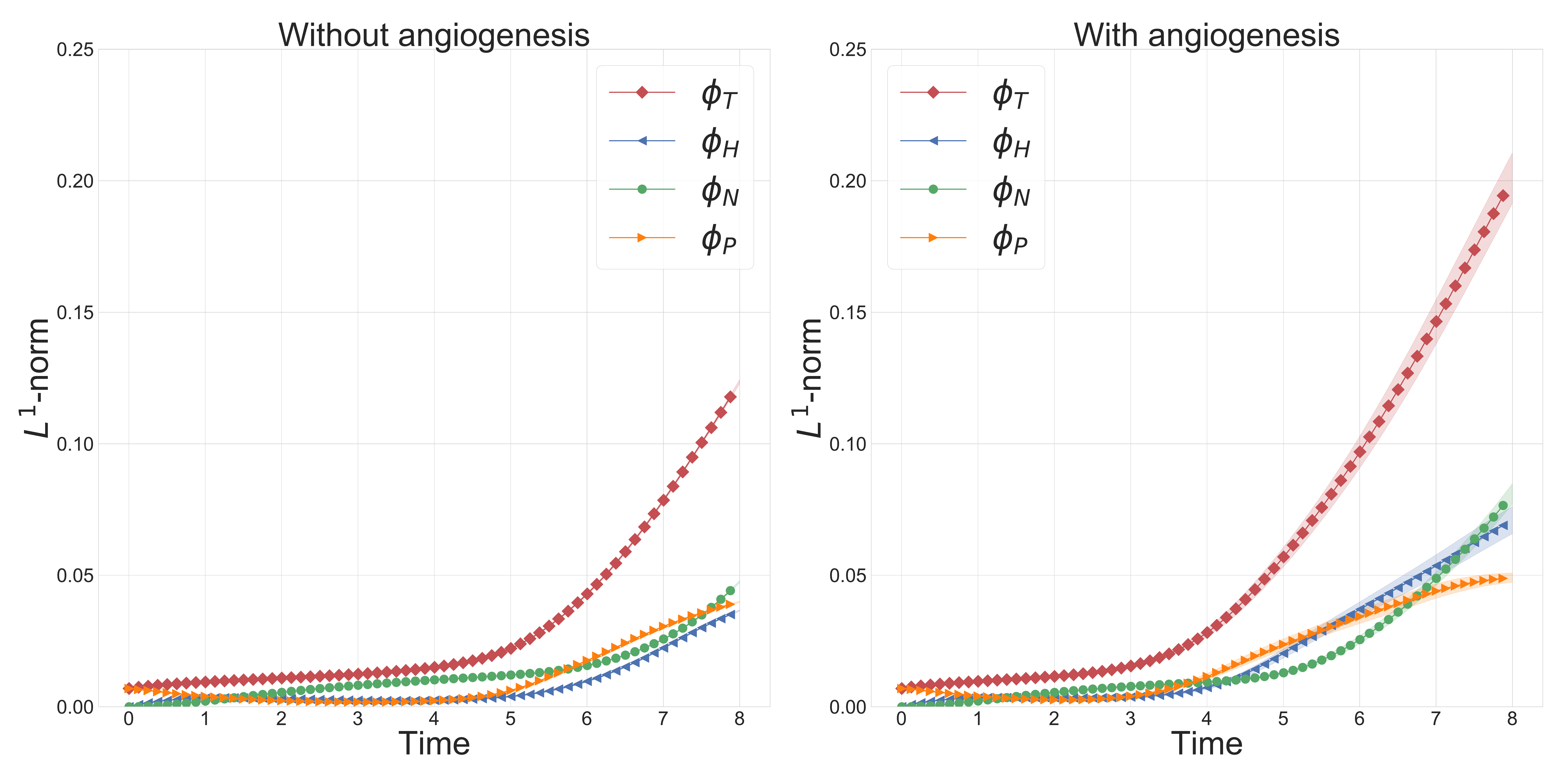}
	\caption{\label{fig:TumorTimeComplex} Quantities of interests (QoIs) related to tumor species over a time interval $[0,8]$ for the capillary network setting. Similar to the two-vessels simulation, we compute the mean and standard deviation using 50 and 10 samples for the case  with and without angiogenesis, respectively. We refer to \cref{fig:TumorTime} for more details on the plots. As in the case of two-vessels setting, the variation in the QoIs are much smaller for the non-angiogenesis case. The mean of the total tumor volume fraction $\left\|\phi_T\right\|_{L^1}$ for the angiogenesis case is about 1.62 times that of the non-angiogenesis case.}
\end{figure} 

\begin{figure}[!htb]
	\centering
	\includegraphics[clip,width=0.9\textwidth]{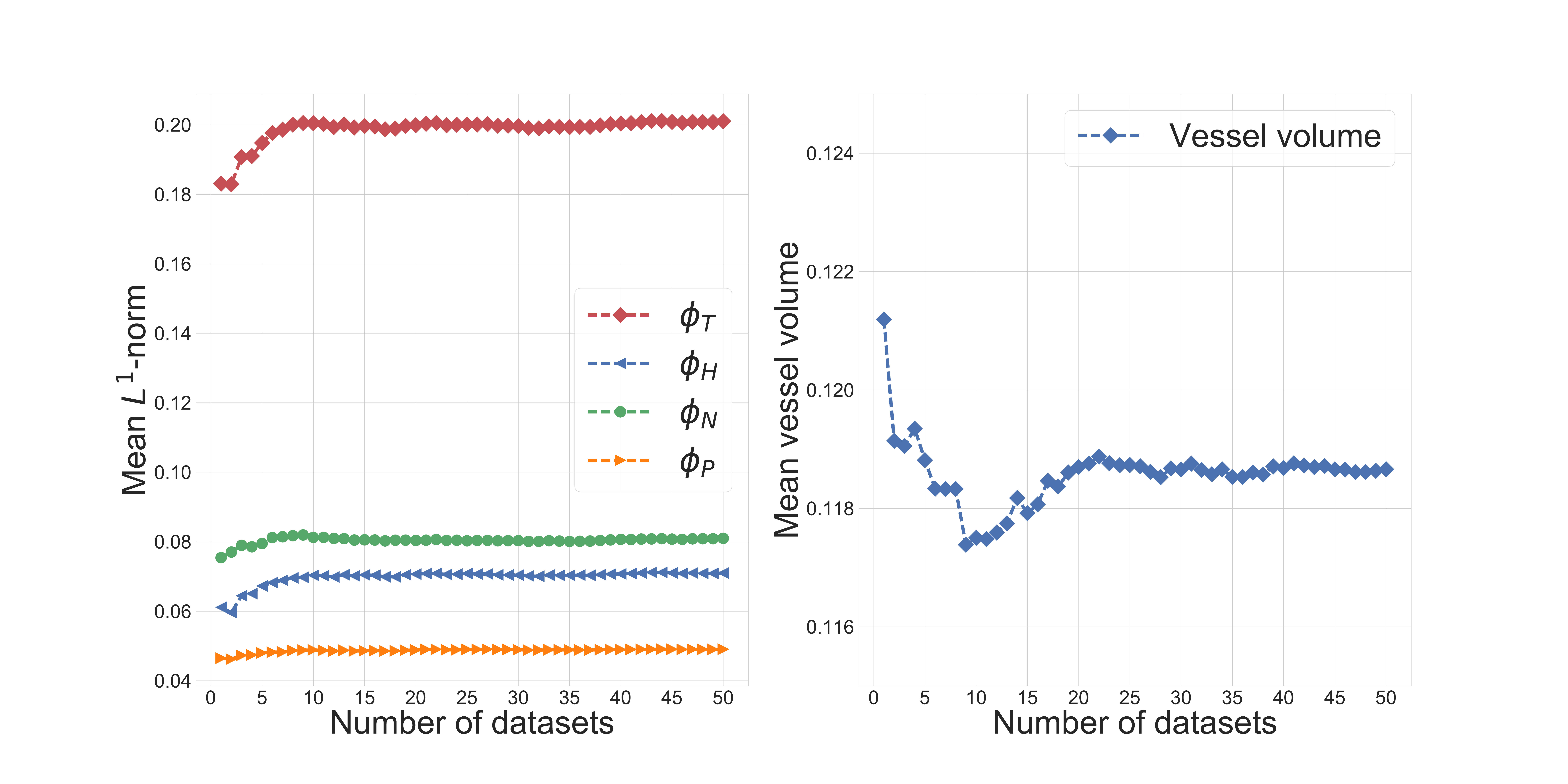}
	\caption{\label{fig:StatisticsComplex} The mean values for the $L^1$-norm of the tumor cell volume fractions $\phi_T, \phi_P, \phi_H, \phi_N$ and the volume of the blood vessel network at time $t = 8$. Results correspond to the capillary network setting.  The mean of the total tumor volume fraction stabilizes with small samples. This agrees with \cref{fig:TumorTimeComplex} that shows that the variations in $L^1$-norm QoIs are overall smaller. The mean of the vessel volume shows some change when samples are small and stabilizes as the size of data set grows. }
\end{figure}

We first compare the tumor volume with and without angiogenesis; see \cref{fig:TumorTimeComplex}. The results are similar to the two-vessel setting. They show that the overall tumor growth is higher with angiogenesis as expected. We also observe that proliferative cells start to grow rapidly at  $t \approx 3.5$ with angiogenesis as compared to $t \approx 4.75$ without angiogenesis. Production of necrotic cells is higher in the non-angiogenesis case until time $t \approx 5$. Compared to \cref{fig:TumorTime} for the two-vessel setting, the variations in the tumor species related QoIs are much smaller in \cref{fig:TumorTimeComplex}. This may be due to the fact that diffusivity of the nutrients in the latter case is much smaller, and that $L_\sigma$ is also smaller in the later case resulting in a smaller exchange of nutrients. Next, we plot the mean of QoIs as we increase the size of data in data sets in \cref{fig:StatisticsComplex}; results show that mean of tumor species related QoIs is stable and can be computed accurately using fewer samples. This relates to the fact that we see smaller variation in the $L^1$-norm of the tumor species $\phi_T$ in \cref{fig:TumorTimeComplex}. The mean of vessel volume shows some variations for smaller data sets and the variations get smaller later on; still the variations are very small and contained in range $[0.117, 0.121]$.

\begin{figure}[!htb]
	\centering
	\includegraphics[clip,width=.70\textwidth]{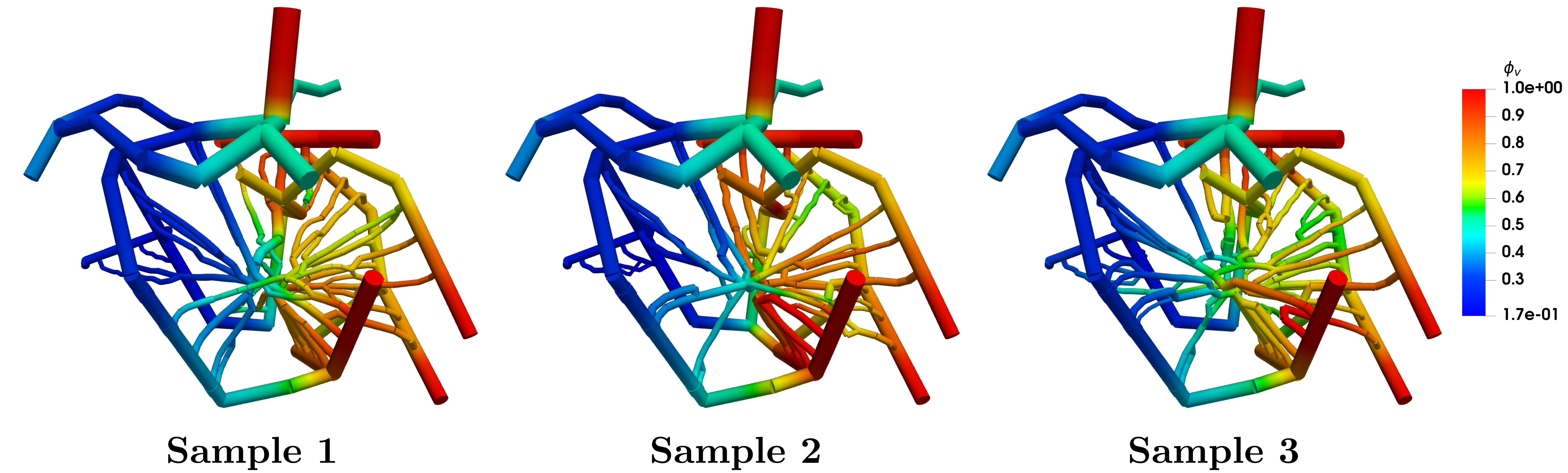}
	\caption{\label{fig:NetworkSamples} Plot of vessel network at $t=8$ from three samples  for the capillary network setting.}
\end{figure}

\begin{figure}[!htb]
	\centering
	\includegraphics[clip,width=.70\textwidth]{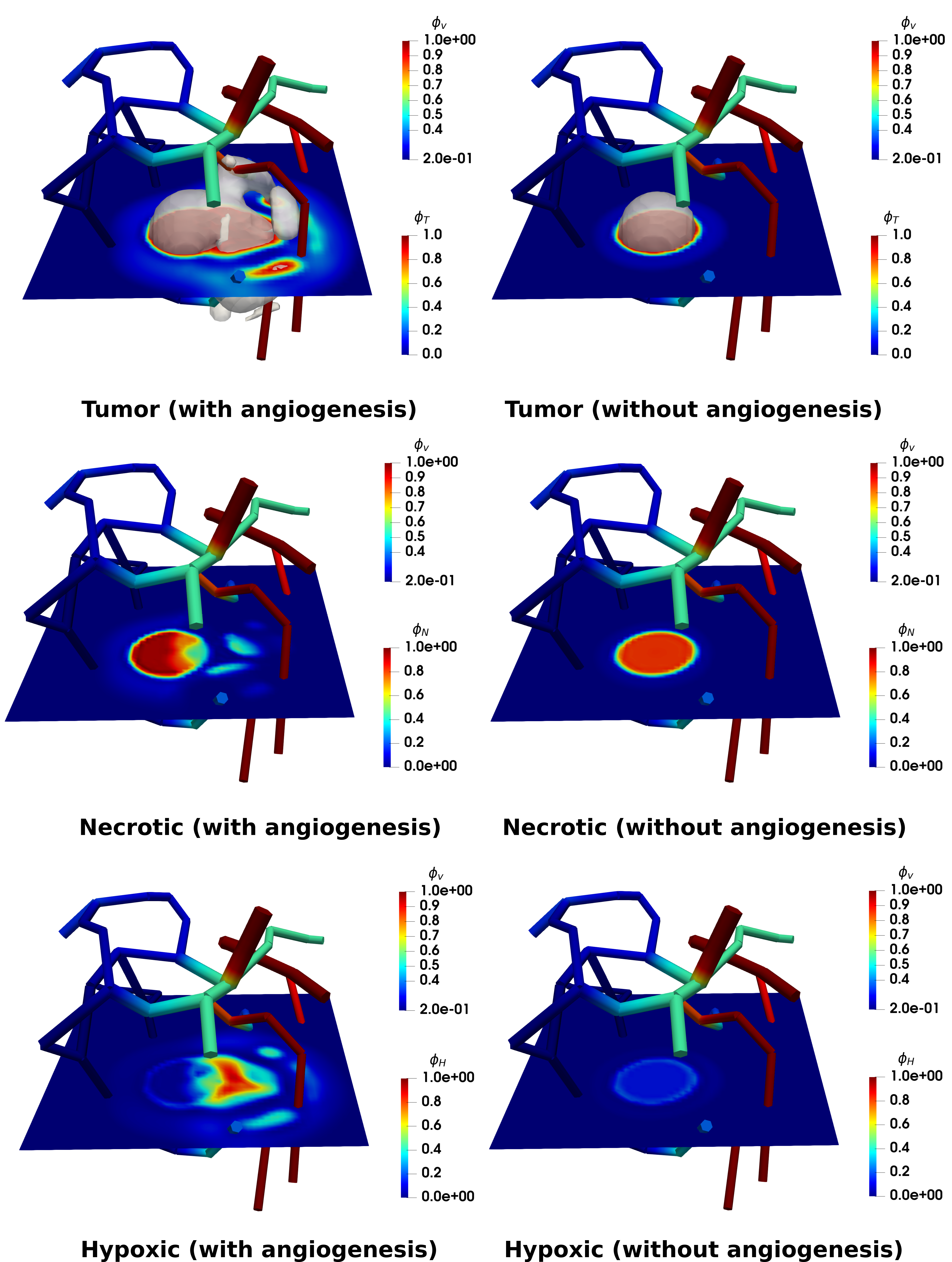}
	\caption{\label{fig:Comparison} Distribution of tumor cells for $t=5.12$. The simulation results for tumor growth supported by angiogenesis are shown on the left hand side, while the results for tumor growth without angiogenesis are presented on the right. The necrotic $\left(\phi_N \right)$ and hypoxic $\left(\phi_H \right)$ volume fractions are visualized in the $z$-plane at $z=0.7$. For $\phi_T$, in both cases a contour plot for $\phi_T = 0.9$  is shown.}
\end{figure}

In Figures \ref{fig:BoundariesNetwork} and \ref{fig:Comparison}, some results for the capillary network are summarized, where in \cref{fig:BoundariesNetwork} the growing network is shown and in the \cref{fig:NetworkSamples} the vessel network as a result of angiogenesis is shown at the final simulation time $t=8$ for three samples. In \cref{fig:Comparison}, we compare the the tumor species at time $t=5.12$ for the angiogenesis and non-angiogenesis case. As in the two-vessel case, the tumor starts to grow faster after it is vascularized. Apparently, the tumor cells tend to migrate towards the nutrient rich part of the computational domain despite the fact that they have to move against the direction of flow which is induced by the pressure gradient. Not surprisingly, the volume fraction of the necrotic cells is larger in the part that is facing away from the nutrient rich part and related to the whole tumor it remains relatively small.

It is interesting to observe that, as in the two-vessel case, the contour plot of $\phi_T$ for $\phi_T=0.9$ exhibits a secondary structure, while in the simulation without angiogenesis, this effect cannot be seen. Moreover, as in the two-vessel experiment, the tumor contains a large necrotic kernel if there is no angiogenesis, indicating that the tumor has almost died. This simulation portrays once again that angiogenesis can play a crucial role in the evolution of tumor growth.


\FloatBarrier

\section{Summary and outlook}
\label{Sec:Conclusion}
In this work, we presented a stochastic model for tumor growth characterized by a coupled system of nonlinear PDEs of Cahn-Hilliard type coupled with a model of an evolving vascular network. A 3D-1D model is developed to simulate flow and nutrient transport within both the network and porous tissue so as to depict the phenomena of angiogenesis. In this model, the blood vessel network is given by a 1D graph-like structure coupling the flow and transport processes in the network and tissue. Furthermore, the model facilitates the handling of a growing network structure with bifurcation of growing vessels which is crucial for the simulation of angiogenesis. The angiogenesis process is simulated by an iterative algorithm starting from a single artery and a vein or a given network. The blood vessel network supplying the tumor employs Murray's law to determine the radii at a bifurcation of network capillaries. The choice of {{radii and lengths of new vessels as well as bifurcations are}} governed by stochastic algorithms. The direction of growth of the vessels is determined by the gradient of the local TAF concentration. 

{{We demonstrate that the model is capable of simulating the development of satellite of tumor concentrations in nutrient rich vessels near necrotic cores in agreement with some experimental observations. Also, as expected, rapid growth of solid tumor mass accompanies increased supply of nutrient through angiogenesis.}}

We believe that our model can serve as a starting point for important predictive simulations of cancer therapy{{; in particular the effect of anti-angiogenic drugs could be studied using models of these types.}} 
{{However, models of these kind require experimental data such as MR imaging data that inform the vasculature in the tissue as well as the parameters in the tumor growth model and vasculature flow model. }} 
In the present work, the adaption of the vessel radii is related to the wall shear stress. However, other effects can influence the vessel radii that could be included, such as metabolic hematocrit-related stimulus, which may also lead to a significant pruning and restructuring of the network. {{Further work on refining and improving computational algorithms is also needed, such as the development of efficient distributed solvers for the 3D-1D systems. We hope to address these issues and other extensions in future work.}}

\section*{Acknowledgements}
The authors gratefully acknowledge the support of the Deutsche Forschungsgemeinschaft (DFG) through TUM International Graduate School of Science and Engineering (IGSSE), GSC 81.
The work of PKJ and JTO was supported by the U.S. Department of Energy, Office of Science, Office of Advanced Scientific Computing Research, Mathematical Multifaceted Integrated Capability Centers (MMICCS), under Award Number DE-SC0019303.


\begin{thebibliography}{10}

\bibitem{ambrosi2005review}
{\sc D.~Ambrosi, F.~Bussolino, and L.~Preziosi}, {\em A review of
  vasculogenesis models}, Journal of Theoretical Medicine, 6 (2005), pp.~1--19.

\bibitem{ambrosi2009cell}
{\sc D.~Ambrosi and L.~Preziosi}, {\em Cell adhesion mechanisms and stress
  relaxation in the mechanics of tumours}, Biomechanics and Modeling in
  Mechanobiology, 8 (2009), pp.~397--413.

\bibitem{anderson1998continuous}
{\sc A.~R. Anderson and M.~A.~J. Chaplain}, {\em Continuous and discrete
  mathematical models of tumor-induced angiogenesis}, Bulletin of Mathematical
  Biology, 60 (1998), pp.~857--899.

\bibitem{antonopoulou2019numerical}
{\sc D.~Antonopoulou, {\'L}.~Ba{\v{n}}as, R.~N{\"u}rnberg, and A.~Prohl}, {\em
  Numerical approximation of the stochastic {C}ahn--{H}illiard equation near
  the sharp interface limit}, Numerische Mathematik, 147 (2021), pp.~505--551.

\bibitem{bellomo2008foundations}
{\sc N.~Bellomo, N.~Li, and P.~K. Maini}, {\em On the foundations of cancer
  modelling: {S}elected topics, speculations, and perspectives}, Mathematical
  Models and Methods in Applied Sciences, 18 (2008), pp.~593--646.

\bibitem{bellomo2000modelling}
{\sc N.~Bellomo and L.~Preziosi}, {\em Modelling and mathematical problems
  related to tumor evolution and its interaction with the immune system},
  Mathematical and Computer Modelling, 32 (2000), pp.~413--452.

\bibitem{byrne2003modelling}
{\sc H.~Byrne and L.~Preziosi}, {\em Modelling solid tumour growth using the
  theory of mixtures}, Mathematical Medicine and Biology: A Journal of the IMA,
  20 (2003), pp.~341--366.

\bibitem{carmeliet2011molecular}
{\sc P.~Carmeliet and R.~K. Jain}, {\em Molecular mechanisms and clinical
  applications of angiogenesis}, Nature, 473 (2011), pp.~298--307.

\bibitem{chai2018conforming}
{\sc S.~Chai, Y.~Cao, Y.~Zou, and W.~Zhao}, {\em Conforming finite element
  methods for the stochastic {C}ahn--{H}illiard--{C}ook equation}, Applied
  Numerical Mathematics, 124 (2018), pp.~44--56.

\bibitem{chaplain2011mathematical}
{\sc M.~A. Chaplain, M.~Lachowicz, Z.~Szyma{\'n}ska, and D.~Wrzosek}, {\em
  Mathematical modelling of cancer invasion: {T}he importance of cell-cell
  adhesion and cell-matrix adhesion}, Mathematical Models and Methods in
  Applied Sciences, 21 (2011), pp.~719--743.

\bibitem{chaplain2005mathematical}
{\sc M.~A. Chaplain and G.~Lolas}, {\em Mathematical modelling of cancer cell
  invasion of tissue: The role of the urokinase plasminogen activation system},
  Mathematical Models and Methods in Applied Sciences, 15 (2005),
  pp.~1685--1734.

\bibitem{cristini2009nonlinear}
{\sc V.~Cristini, X.~Li, J.~S. Lowengrub, and S.~M. Wise}, {\em Nonlinear
  simulations of solid tumor growth using a mixture model: invasion and
  branching}, Journal of Mathematical Biology, 58:723 (2009).

\bibitem{cristini2010multiscale}
{\sc V.~Cristini and J.~Lowengrub}, {\em Multiscale Modeling of Cancer: An
  Integrated Experimental and Mathematical Modeling Approach}, Cambridge
  University Press, 2010.

\bibitem{da1996stochastic}
{\sc G.~Da~Prato and A.~Debussche}, {\em Stochastic {C}ahn--{H}illiard
  equation}, Nonlinear Analysis: Theory, Methods \& Applications, 26 (1996),
  pp.~241--263.

\bibitem{dorraki2020angiogenic}
{\sc M.~Dorraki, A.~Fouladzadeh, A.~Allison, C.~S. Bonder, and D.~Abbott}, {\em
  Angiogenic networks in tumors—insights via mathematical modeling}, IEEE
  Access, 8 (2020), pp.~43215--43228.

\bibitem{eilken2010dynamics}
{\sc H.~M. Eilken and R.~H. Adams}, {\em Dynamics of endothelial cell behavior
  in sprouting angiogenesis}, Current Opinion in Cell Biology, 22 (2010),
  pp.~617--625.

\bibitem{engwer2017structured}
{\sc C.~Engwer, C.~Stinner, and C.~Surulescu}, {\em On a structured multiscale
  model for acid-mediated tumor invasion: The effects of adhesion and
  proliferation}, Mathematical Models and Methods in Applied Sciences, 27
  (2017), pp.~1355--1390.

\bibitem{eyre1998unconditionally}
{\sc D.~J. Eyre}, {\em Unconditionally gradient stable time marching the
  {C}ahn--{H}illiard equation}, in Materials Research Society Symposium
  Proceedings, vol.~529, Materials Research Society, 1998, pp.~39--46.

\bibitem{frieboes2010three}
{\sc H.~B. Frieboes, F.~Jin, Y.-L. Chuang, S.~M. Wise, J.~S. Lowengrub, and
  V.~Cristini}, {\em Three-dimensional multispecies nonlinear tumor growth --
  {II}: {T}umor invasion and angiogenesis}, Journal of Theoretical Biology, 264
  (2010), pp.~1254--1278.

\bibitem{frigeri2018on}
{\sc S.~Frigeri, K.~F. Lam, E.~Rocca, and G.~Schimperna}, {\em On a
  multi-species {C}ahn--{H}illiard--{D}arcy tumor grwoth model with singular
  potentials}, Communications in Mathematical Sciences, 16 (2018),
  pp.~821--856.

\bibitem{fritz2020analysis}
{\sc M.~Fritz, P.~K. Jha, T.~K{\"o}ppl, J.~T. Oden, and B.~Wohlmuth}, {\em
  Analysis of a new multispecies tumor growth model coupling {3D} phase-fields
  with a {1D} vascular network}, Nonlinear Analysis: Real World Applications,
  61 (2021), p.~103331.

\bibitem{fritz2019local}
{\sc M.~Fritz, E.~Lima, V.~Nikolic, J.~T. Oden, and B.~Wohlmuth}, {\em Local
  and nonlocal phase-field models of tumor growth and invasion due to {ECM}
  degradation}, Mathematical Models and Methods in Applied Sciences, 29 (2019),
  pp.~2433--2468.

\bibitem{fritz2019unsteady}
{\sc M.~Fritz, E.~Lima, J.~T. Oden, and B.~Wohlmuth}, {\em On the unsteady
  {D}arcy--{F}orchheimer--{B}rinkman equation in local and nonlocal tumor
  growth models}, Mathematical Models and Methods in Applied Sciences, 29
  (2019), pp.~1691--1731.

\bibitem{garcke2018multiphase}
{\sc H.~Garcke, K.~F. Lam, R.~N{\"u}rnberg, and E.~Sitka}, {\em A multiphase
  {C}ahn--{H}illiard--{D}arcy model for tumour growth with necrosis},
  Mathematical Models and Methods in Applied Sciences, 28 (2018), pp.~525--577.

\bibitem{gerisch2008mathematical}
{\sc A.~Gerisch and M.~Chaplain}, {\em Mathematical modelling of cancer cell
  invasion of tissue: {L}ocal and non-local models and the effect of adhesion},
  Journal of Theoretical Biology, 250 (2008), pp.~684--704.

\bibitem{ginzburg1963frictional}
{\sc B.~Ginzburg and A.~Katchalsky}, {\em The frictional coefficients of the
  flows of non-electrolytes through artificial membranes}, The Journal of
  General Physiology, 47 (1963), pp.~403--418.

\bibitem{hanahan2011hallmarks}
{\sc D.~Hanahan and R.~A. Weinberg}, {\em Hallmarks of cancer: {T}he next
  generation}, Cell, 144 (2011), pp.~646--674.

\bibitem{hawkins2012numerical}
{\sc A.~Hawkins-Daarud, K.~G. van~der Zee, and J.~T. Oden}, {\em Numerical
  simulation of a thermodynamically consistent four-species tumor growth
  model}, International Journal for Numerical Methods in Biomedical
  Engineering, 28 (2012), pp.~3--24.

\bibitem{hillen2013convergence}
{\sc T.~Hillen, K.~J. Painter, and M.~Winkler}, {\em Convergence of a cancer
  invasion model to a logistic chemotaxis model}, Mathematical Models and
  Methods in Applied Sciences, 23 (2013), pp.~165--198.

\bibitem{hodneland2020well}
{\sc E.~Hodneland, X.~Hu, and J.~Nordbotten}, {\em Well-posedness,
  discretization and preconditioners for a class of models for
  mixed-dimensional problems with high dimensional gap}, arXiv preprint
  arXiv:2006.12273,  (2020).

\bibitem{holmgren1995dormancy}
{\sc L.~Holmgren, M.~S. O'Reilly, and J.~Folkman}, {\em Dormancy of
  micrometastases: balanced proliferation and apoptosis in the presence of
  angiogenesis suppression}, Nature Medicine, 1 (1995), pp.~149--153.

\bibitem{koch2020modeling}
{\sc T.~Koch, M.~Schneider, R.~Helmig, and P.~Jenny}, {\em Modeling tissue
  perfusion in terms of 1d-3d embedded mixed-dimension coupled problems with
  distributed sources}, Journal of Computational Physics, 410:100050 (2020).

\bibitem{koppl20203d}
{\sc T.~K{\"o}ppl, E.~Vidotto, and B.~Wohlmuth}, {\em A {3D}-{1D} coupled blood
  flow and oxygen transport model to generate microvascular networks},
  International Journal for Numerical Methods in Biomedical Engineering, e3386
  (2020).

\bibitem{koumoutsakos2013fluid}
{\sc P.~Koumoutsakos, I.~Pivkin, and F.~Milde}, {\em The fluid mechanics of
  cancer and its therapy}, Annual Review of Fluid Mechanics, 45 (2013),
  pp.~325--355.

\bibitem{kunkel2001inhibition}
{\sc P.~Kunkel, U.~Ulbricht, P.~Bohlen, M.~Brockmann, R.~Fillbrandt,
  D.~Stavrou, M.~Westphal, and K.~Lamszus}, {\em Inhibition of glioma
  angiogenesis and growth in vivo by systemic treatment with a monoclonal
  antibody against vascular endothelial growth factor receptor-2}, Cancer
  Research, 61 (2001), pp.~6624--6628.

\bibitem{lima2014hybrid}
{\sc E.~Lima, J.~T. Oden, and R.~Almeida}, {\em A hybrid ten-species
  phase-field model of tumor growth}, Mathematical Models and Methods in
  Applied Sciences, 24 (2014), pp.~2569--2599.

\bibitem{mcdougall2002mathematical}
{\sc S.~R. McDougall, A.~Anderson, M.~Chaplain, and J.~Sherratt}, {\em
  Mathematical modelling of flow through vascular networks: implications for
  tumour-induced angiogenesis and chemotherapy strategies}, Bulletin of
  Mathematical Biology, 64 (2002), pp.~673--702.

\bibitem{mcdougall2006mathematical}
{\sc S.~R. McDougall, A.~R. Anderson, and M.~A. Chaplain}, {\em Mathematical
  modelling of dynamic adaptive tumour-induced angiogenesis: clinical
  implications and therapeutic targeting strategies}, Journal of theoretical
  biology, 241 (2006), pp.~564--589.

\bibitem{murray1926physiological2}
{\sc C.~Murray}, {\em The physiological principle of minimum work applied to
  the angle of branching of arteries}, The Journal of General Physiology, 9
  (1926), pp.~835--841.

\bibitem{murray1926physiological}
\leavevmode\vrule height 2pt depth -1.6pt width 23pt, {\em The physiological
  principle of minimum work: I. the vascular system and the cost of blood
  volume}, Proceedings of the National Academy of Sciences of the United States
  of America, 12 (1926), pp.~207--214.

\bibitem{nargis2016effects}
{\sc N.~Nargis and R.~Aldredge}, {\em Effects of matrix metalloproteinase on
  tumour growth and morphology via haptotaxis}, J. Bioengineer. \& Biomedical
  Sci., 6:1000207 (2016).

\bibitem{nishida2006angiogenesis}
{\sc N.~Nishida, H.~Yano, T.~Nishida, T.~Kamura, and M.~Kojiro}, {\em
  Angiogenesis in cancer}, Vascular Health and Risk Management, 2 (2006),
  pp.~213--219.

\bibitem{oden2016toward}
{\sc J.~T. Oden, E.~Lima, R.~C. Almeida, Y.~Feng, M.~N. Rylander, D.~Fuentes,
  D.~Faghihi, M.~M. Rahman, M.~DeWitt, M.~Gadde, et~al.}, {\em Toward
  predictive multiscale modeling of vascular tumor growth}, Archives of
  Computational Methods in Engineering, 23 (2016), pp.~735--779.

\bibitem{orrierioptimal}
{\sc {Orrieri, Carlo}, {Rocca, Elisabetta}, and {Scarpa, Luca}}, {\em Optimal
  control of stochastic phase-field models related to tumor growth}, ESAIM:
  COCV, 26 (2020), p.~104.

\bibitem{owen2009angiogenesis}
{\sc M.~R. Owen, T.~Alarc{\'o}n, P.~K. Maini, and H.~M. Byrne}, {\em
  Angiogenesis and vascular remodelling in normal and cancerous tissues},
  Journal of Mathematical Biology, 58:689 (2009).

\bibitem{parangi1996antiangiogenic}
{\sc S.~Parangi, M.~O'Reilly, G.~Christofori, L.~Holmgren, J.~Grosfeld,
  J.~Folkman, and D.~Hanahan}, {\em Antiangiogenic therapy of transgenic mice
  impairs de novo tumor growth}, Proceedings of the National Academy of
  Sciences, 93 (1996), pp.~2002--2007.

\bibitem{patsch2015generation}
{\sc C.~Patsch, L.~Challet-Meylan, E.~C. Thoma, E.~Urich, T.~Heckel, J.~F.
  O’Sullivan, S.~J. Grainger, F.~G. Kapp, L.~Sun, K.~Christensen, et~al.},
  {\em Generation of vascular endothelial and smooth muscle cells from human
  pluripotent stem cells}, Nature Cell Biology, 17 (2015), pp.~994--1003.

\bibitem{phillips2020hybrid}
{\sc C.~M. Phillips, E.~A. Lima, R.~T. Woodall, A.~Brock, and T.~E. Yankeelov},
  {\em A hybrid model of tumor growth and angiogenesis: In silico experiments},
  PLoS One, 15 (2020), p.~e0231137.

\bibitem{preziosi2003cancer}
{\sc L.~Preziosi}, {\em Cancer modelling and simulation}, CRC Press, 2003.

\bibitem{pries2001structural}
{\sc A.~Pries, B.~Reglin, and T.~Secomb}, {\em Structural adaptation of
  vascular networks: role of the pressure response}, Hypertension, 38 (2001),
  pp.~1476--1479.

\bibitem{pries2001structural2}
{\sc A.~Pries, B.~Reglin, and T.~W. Secomb}, {\em Structural adaptation of
  microvascular networks: functional roles of adaptive responses}, American
  Journal of Physiology-Heart and Circulatory Physiology, 281 (2001),
  pp.~H1015--H1025.

\bibitem{pries1998structural}
{\sc A.~Pries, T.~Secomb, and P.~Gaehtgens}, {\em Structural adaptation and
  stability of microvascular networks: theory and simulations}, American
  Journal of Physiology-Heart and Circulatory Physiology, 275 (1998),
  pp.~H349--H360.

\bibitem{reichold2009vascular}
{\sc J.~Reichold, M.~Stampanoni, A.~L. Keller, A.~Buck, P.~Jenny, and
  B.~Weber}, {\em Vascular graph model to simulate the cerebral blood flow in
  realistic vascular networks}, Journal of Cerebral Blood Flow \& Metabolism,
  29 (2009), pp.~1429--1443.

\bibitem{ribatti2012sprouting}
{\sc D.~Ribatti and E.~Crivellato}, {\em “sprouting angiogenesis”, a
  reappraisal}, Developmental biology, 372 (2012), pp.~157--165.

\bibitem{rubenstein2000anti}
{\sc J.~Rubenstein, J.~Kim, T.~Ozawa, M.~Zhang, M.~Westphal, D.~Deen, and
  M.~Shuman}, {\em Anti-{VEGF} antibody treatment of glioblastoma prolongs
  survival but results in increased vascular cooption}, Neoplasia, 2 (2000),
  pp.~306--314.

\bibitem{salathe1976mathematical}
{\sc E.~P. Salathe and K.-N. An}, {\em A mathematical analysis of fluid
  movement across capillary walls}, Microvascular Research, 11 (1976),
  pp.~1--23.

\bibitem{schneider2012tissue}
{\sc M.~Schneider, J.~Reichold, B.~Weber, G.~Sz{\'e}kely, and S.~Hirsch}, {\em
  Tissue metabolism driven arterial tree generation}, Medical Image Analysis,
  16 (2012), pp.~1397--1414.

\bibitem{secomb2013angiogenesis}
{\sc T.~Secomb, J.~Alberding, R.~Hsu, M.~Dewhirst, and A.~Pries}, {\em
  Angiogenesis: an adaptive dynamic biological patterning problem}, PLoS
  Computational Biology, 9:e1002983 (2013).

\bibitem{stephanou2005mathematical}
{\sc A.~Stephanou, S.~R. McDougall, A.~R. Anderson, and M.~A. Chaplain}, {\em
  Mathematical modelling of flow in 2d and 3d vascular networks: applications
  to anti-angiogenic and chemotherapeutic drug strategies}, Mathematical and
  Computer Modelling, 41 (2005), pp.~1137--1156.

\bibitem{stephanou2006mathematical}
{\sc A.~St{\'e}phanou, S.~R. McDougall, A.~R. Anderson, and M.~A. Chaplain},
  {\em Mathematical modelling of the influence of blood rheological properties
  upon adaptative tumour-induced angiogenesis}, Mathematical and Computer
  Modelling, 44 (2006), pp.~96--123.

\bibitem{tao2011chemotaxis}
{\sc Y.~Tao and M.~Winkler}, {\em A chemotaxis-haptotaxis model: the roles of
  nonlinear diffusion and logistic source}, SIAM Journal on Mathematical
  Analysis, 43 (2011), pp.~685--704.

\bibitem{travasso2011tumor}
{\sc R.~D. Travasso, E.~C. Poir{\'e}, M.~Castro, J.~C. Rodrguez-Manzaneque, and
  A.~Hern{\'a}ndez-Machado}, {\em Tumor angiogenesis and vascular patterning: a
  mathematical model}, PloS One, 6:e19989 (2011).

\bibitem{vidotto2019hybrid}
{\sc E.~Vidotto, T.~Koch, T.~K\"oppl, R.~Helmig, and B.~Wohlmuth}, {\em Hybrid
  models for simulating blood flow in microvascular networks}, Multiscale
  Modeling \& Simulation, 17 (2019), pp.~1076--1102.

\bibitem{vilanova2018computational}
{\sc G.~Vilanova, M.~Bur{\'e}s, I.~Colominas, and H.~Gomez}, {\em Computational
  modelling suggests complex interactions between interstitial flow and tumour
  angiogenesis}, Journal of The Royal Society Interface, 15:20180415 (2018).

\bibitem{wise2008three}
{\sc S.~M. Wise, J.~S. Lowengrub, H.~B. Frieboes, and V.~Cristini}, {\em
  Three-dimensional multispecies nonlinear tumor growth -- {I}: {M}odel and
  numerical method}, Journal of Theoretical Biology, 253 (2008), pp.~524--543.

\bibitem{wu2020patient}
{\sc C.~Wu, D.~A. Hormuth, T.~A. Oliver, F.~Pineda, G.~Lorenzo, G.~S. Karczmar,
  R.~D. Moser, and T.~E. Yankeelov}, {\em Patient-specific characterization of
  breast cancer hemodynamics using image-guided computational fluid dynamics},
  IEEE Transactions on Medical Imaging,  (2020).

\bibitem{xu2016mathematical}
{\sc J.~Xu, G.~Vilanova, and H.~Gomez}, {\em A mathematical model coupling
  tumor growth and angiogenesis}, PloS One, 11 (2016).

\bibitem{xu2017full}
\leavevmode\vrule height 2pt depth -1.6pt width 23pt, {\em Full-scale,
  three-dimensional simulation of early-stage tumor growth: The onset of
  malignancy}, Computer Methods in Applied Mechanics and Engineering, 314
  (2017), pp.~126--146.

\bibitem{zheng2005nonlinear}
{\sc X.~Zheng, S.~Wise, and V.~Cristini}, {\em Nonlinear simulation of tumor
  necrosis, neo-vascularization and tissue invasion via an adaptive
  finite-element/level-set method}, Bulletin of Mathematical Biology, 67:211
  (2005).

\end{thebibliography}
\bibliographystyle{siam}

\end{document}